\documentclass[showpacs,preprintnumbers,amsmath,amssymb]{revtex4}

\usepackage{graphicx}

\usepackage{amssymb}
\usepackage{bm}
\usepackage{amsmath}
\usepackage[dvips]{color}
\usepackage{epsfig}
\usepackage{epstopdf}

\begin{document}

\title{A full Eulerian finite difference approach for solving fluid-structure coupling problems}

\author{Kazuyasu Sugiyama$^{1}$, 
Satoshi Ii$^{1}$
Shintaro Takeuchi$^{1}$,
Shu Takagi$^{2,1}$ \& Yoichiro Matsumoto$^{1}$\\
$^{1}$ Department of Mechanical Engineering, School of Engineering, The University of Tokyo,\\ 7-3-1 Hongo Bunkyo-Ku, Tokyo 113-8656, Japan\\
$^{2}$ Organ and Body Scale Team, CSRP, Riken, 2-1, Hirosawa, Wako-shi, Saitama 351-0198, Japan}

\begin{abstract}
A new simulation method for solving fluid-structure coupling problems has been developed. 
All the basic equations are numerically solved
 on a fixed Cartesian grid using a finite difference scheme. 
A volume-of-fluid formulation
 (Hirt \& Nichols (1981, J. Comput. Phys., {\bf 39}, 201)),
 which has been widely used for multiphase flow simulations,
 is applied to describing the multi-component geometry. 
The temporal change in the solid deformation 
 is described in the Eulerian frame 
 by updating a left Cauchy-Green deformation tensor, 
 which is used to express constitutive equations 
 for nonlinear Mooney-Rivlin materials.
In this paper, various verifications and validations
 of the present full Eulerian method, 
 which solves the fluid and solid motions on a fixed grid, 
 are demonstrated, 
 and the numerical accuracy involved in the fluid-structure coupling problems
 is examined.
\end{abstract}


\maketitle

\section{Introduction}

Fluid-Structure Interaction (FSI) phenomena appear in many places, 
 e.g., biological systems, and industrial processes.
In life science, the numerical prediction of the FSI problems
 involves twofold significance: fundamental and practical aspects.
For example, it would allow us to gain insight into how life is sustained,
 and support surgical planning in medical treatments.
Conventionally, the computational fluid dynamics is modelled in an Eulerian way,
 while the computational structure dynamics is normally treated in a Lagrangian way. 
The coupling of the fluid and structure dynamics is a formidable task
 due to such a difference in the numerical framework
 as well as its multi-physics nature.

In a creeping (zero Reynolds number limit) flow involving a biological membrane, 
 once boundary conditions on the membrane surface are determined
 via constitutive laws for e.g. an in-plane stress and a bending moment,
 the bulk flow field is found using Green's function.
In such a situation, no volumetric mesh is needed in the bulk region.
A boundary element method is applicable
 to predicting capsule \cite{poz2001} and red blood cell \cite{poz2005}
 motions interacting with the fluid flow.

For non-zero Reynolds number flows, on the other hand,
 it is necessary to set out the computational mesh over the entire domain.
There are currently several major approaches 
 classified with respect to the computational treatment
 how the kinematic and dynamic interactions
 are coupled on the moving interface. 

The most accurate approach is raised
 as Arbitrary Lagrangian Eulerian (ALE) 
 \cite{hir1974,bel1980,hug1981,hue1988b,nit1993} 
 or Deforming-Spatial-Domain/Space-Time (DSD/ST)
 \cite{tez1992a,tez1992b,hug1996} technique,
 in which the body-fitted mesh is used.
The latter method enables one to arbitrarily choose
 spatiotemporal nodes, and facilitates to simulate
 the moving particle and FSI problems.
These approaches are referred to as an interface-tracking approach,
 in which the surface mesh is accommodated to be shared between 
 both the fluid and solid phases,
 and thus to automatically satisfy the kinematic condition.
As the flow is resolved along the moving/deforming object surface, 
 the boundary layer is expected to be highly resolved, 
 and the dynamic interaction is accurately coupled through iterative procedures. 
Once the body-fitted mesh is provided,
 the state-of-the-art of the interface-tracking approach is satisfactory
 for achieving accurate predictions, 
 including the applications of moving rigid particles \cite{joh1997,hu1996},
 moving hyperelastic particles \cite{gao2009}, 
 parachute aerodynamics \cite{ste2001},
 blood flows \cite{tay1998, tor2001, baz2006, fig2006, tor2007, tor2008}
 and heart flows \cite{zha2001,wat2004}.
However, the whole computational domain has to be re-meshed as the
 object moves/deforms, which is computationally expensive.

An alternative to the interface-tracking approach is
 an Eulerian-Lagrangian approach, 
 in which the fluid and solid phases are separately
 formulated on the fixed Eulerian and Lagrangian grids, respectively.
A noticeable contribution is the development of the Immersed Boundary (IB) method
 by Peskin \cite{pes1972,pes2002},
 who introduces a pseudo delta function 
 for communication between the Eulerian and Lagrangian quantities,
 and demonstrated the landmark simulation of the blood flow around heart valves \cite{pes1972,pes2002}.
The Fictious Domain (FD) method \cite{glo1999,glo2001,shi2005,yu2005},
 and PHYSALIS \cite{tak2003,zha2005}
 for specific multiphase flow problems with circular or spherical particle
 are also classified into the Eulerian-Lagrangian approach.
The IB and FD methods have been applied to a variety of studies,
 for example, moving rigid particles \cite{glo1999,glo2001,kaj2001,kaj2002,yuk2007},
 moving flexible bodies \cite{mor2008,hua2009b,zha2008},
 red blood cell motion \cite{egg1998,mor2008,gon2009},
 and restricted diffusion with permeable interfaces \cite{hua2009a}.
The IB method has also inspired many researchers to propose a number of improved methods.
For example, to facilitate the application to medical problems,
 the immersed finite element method, 
 in which a new kernel function instead of the pseudo delta function
 is introduced for determining the cut-off region
 around the interface on a non-uniform mesh system,
 is proposed in  \cite{zha2004,liu2006,zha2007,zha2008b}.
Also, the immersed interface treatment \cite{lev1994,li2001,li2006,tan2009}
 improves the sharpness of the interface by incorporating the jump in
 the stress and velocities across the interface. 
Recently, a conservative momentum exchange method \cite{tak2010}
 is proposed to facilitate the simulation of
 the fluid flow involving a number of elastic particles
 by combining the finite difference and finite element approaches.

In the above-mentioned methods for
 predicting the motion/deformation of hyperelastic material, 
 the solid displacement is temporally updated in the Lagrangian way.
In general, the hyperelastic constitutive law is 
 expressed as a function of the deformation gradient tensor
 ${\bm F}=\partial {\bm x}/\partial {\bm X}$, 
 here  ${\bm x}$ denotes the current coordinates, 
 and ${\bm X}$ the reference coordinates \cite{bon2008}.
The label of each Lagrangian node links the reference configuration
 and the set of the current node positions representing the current configuration.
Therefore, the Lagrangian description has been preferably employed.
However, the re-meshing procedure at each time step leads to
 intensive computation if system involves complicated geometry of solid
 and/or a large number of objects \cite{sug2010}. 

Let us consider patient-specific simulations in a medical field.
The multi-component geometry of the inner side of the human body
 is provided as voxel data, 
 which are converted from the medical image of
 a Computed Tomography (CT) or a Magnetic Resonance Imaging (MRI).
The basic idea of the medical image-based simulation is found in \cite{tay1998,tor2001},
 in which the voxel data are converted into
 the finite element mesh before starting the computation,
 and the mesh is reconstructed with time advancement.
As pointed out in \cite{mat2002,yok2005,nod2006},
 since technical knowledge and expertise are required
 for the mesh generation and reconstruction, 
 an alternative simulation method without mesh generation procedure would
 be desirable to extend the FSI simulation 
 to certain additional classes of problems in the medical field.
Motivated by such a practical background, 
 the full Eulerian finite difference methods,
 which directly access the voxel data
 to describe the {\it rigid} boundary on the fixed Cartesian mesh
 and avoid difficulty in mesh generation and reconstruction, 
 have been developed in \cite{mat2002,yok2005}. 
The application includes the prediction of
 a coil embolization for aneurysms \cite{nod2006},
 which has been used in practice to support surgical planning.

\begin{figure}[h]
\begin{center}

\epsfig{file=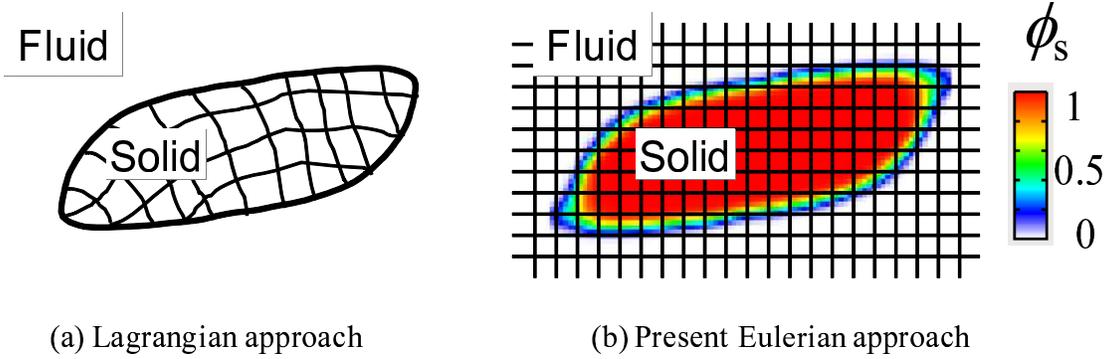,scale=.5,angle=0}

\end{center}

\caption{
Schematic figure  explaining the difference 
 between the interface recognitions of the Lagrangian and Eulerian approaches. 
In the Lagrangian method, 
 the body-fit mesh distinguishes between fluid and solid phases, 
 while in the present Eulerian method, 
 the solid volume fraction $\phi_s$ does. 
The contour at $\phi_s=1/2$ indicates the interface.
}
\label{fig:schem_interface}
\end{figure}

\begin{figure}[h]
\begin{center}

\epsfig{file=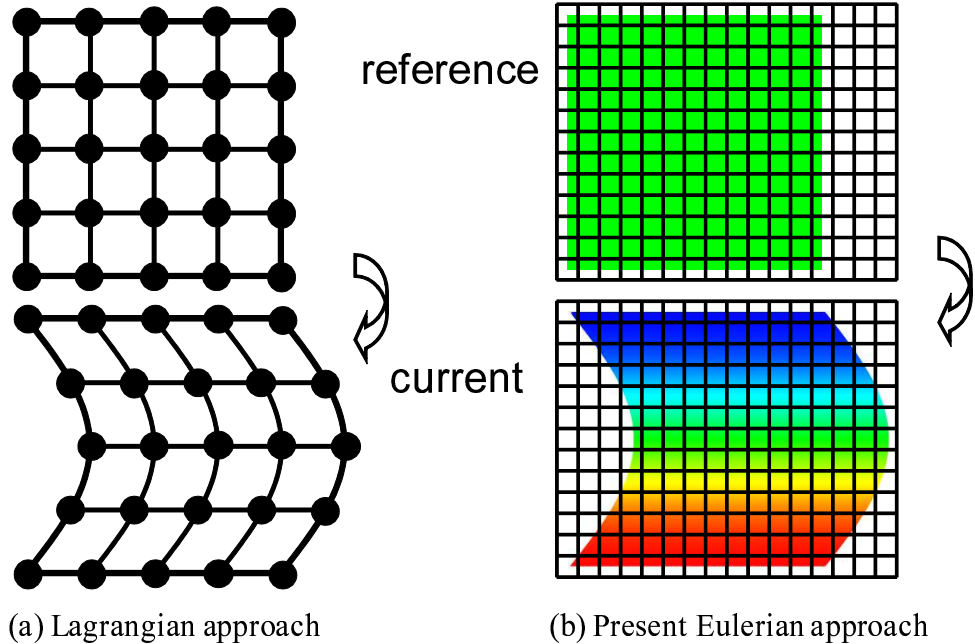,scale=.5,angle=0}

\end{center}

\caption{
Schematic figure explaining the difference between 
 the solid deformation descriptions of the Lagrangian and Eulerian approaches. 
In the Lagrangian method, 
 the relative displacement between adjacent material points 
 from the reference to current configurations
 quantifies the deformation level. 
In the present Eulerian method, to quantify the deformation, 
 the left Cauchy-Green deformation tensor ${\bm B}$ is 
 introduced in the Eulerian frame, and temporally updated.
}
\label{fig:schem_deformation}
\end{figure}

We further develop a full Eulerian method
 for fluid-structure interactions
 involving flexible hyperelastic material.
In the interface-capturing methods for multiphase flow simulations,
 for instance, Volume-Of-Fluid (VOF) \cite{hir1981},
 level set \cite{sus1994,cha1996,osh2001,osh2003,set2003},
 and phase field \cite{whe1992,jac1999} approaches,
 one set of governing equations for the whole flow field, 
 referred to as a one-fluid formulation \cite{try2007}, is often employed.
We follow such an idea,
 and write all the basic equations on a fixed Cartesian grid
 in a finite difference form.
Several Eulerian solvers for modelling the solid deformation
 have been proposed for linear elastic materials 
\cite{xia1999, mat2002}, 
 for elasto-plastic materials 
 \cite{uda2003,oka2007}
 for neo-Hookean materials 
 \cite{vho1991,liu2001,dun2006}
 and for more general hyperelastic materials \cite{cot2008}.
We treat a Mooney-Rivlin hyperelastic material
 \cite{moo1940,riv1948}, 
 and consider
 the nonlinearity of the Cauchy stress with respect to 
 a left Cauchy-Green deformation tensor \cite{bon2008,ham1999}.
Our fluid-structure coupling approach is characterized by the feasibility 
 in implementing the hyperelastic constitutive law
 into the standard incompressible fluid flow solvers,
 and also by the treatments
 in capturing the fluid-structure interface and the solid deformation. 
In the Lagrangian method, 
 the body-fit finite elements automatically 
 distinguish between the fluid and solid phases.
In the present Eulerian method, 
 considering that the voxel data contain the volume
 fraction of fluid and solid, 
 we apply the VOF formulation \cite{hir1981}
 to describing the multi-component geometry 
 (see figure \ref{fig:schem_interface}). 
The large deformation is usually described
 by using the Piola-Kirchhoff stress tensor
 as a function of the deformation gradient, 
 which is suited to the Lagrangian formulation
 as mentioned above.
By contrast, 
 the Eulerian formulation lacks of the material points
 to link between the reference and current configurations.
Therefore, we must devise a method to 
 quantify the level of deformation.
To this end, 
 we introduce the left Cauchy-Green deformation tensor
 ${\bm B}(={\bm F}\cdot{\bm F}^T)$
 defined on each grid point,
 and temporally update it 
 (see figure \ref{fig:schem_deformation}). 
We will devote several test computations to the verification and validation,
 and investigate the numerical accuracy involved in the fluid-structure coupling.

The paper is organized as follows. 
In \S \ref{sec:basic_equation},
 we present the details of the basic equations
 of the system consisting of Newtonian fluid and hyperelastic material,
 and formulate constitutive equations suited
 to the full Eulerian FSI simulation. 
In \S \ref{sec:simulation_method}, 
 we explain the simulation methods in terms of 
 the time-stepping algorithm and the finite difference descriptions.
In \S \ref{sec:validation},
 to explore the validity of the advocated numerical procedure,
 we perform three kinds of tests.
Firstly, we address a one-dimensional problem   
 of the oscillatory parallel fluid-solid layers.
Secondly, we make comparisons with available simulations.
Thirdly, we examine how the shape reversibility
  of the hyperelastic materials is reproduced.
In \S \ref{sec:conclusion},
 we provide some perspectives to conclude the paper. 

\section{Basic equations}
\label{sec:basic_equation}

\subsection{Governing equations and fluid-structure mixture representations}

\begin{figure}[h]
\begin{center}

\epsfig{file=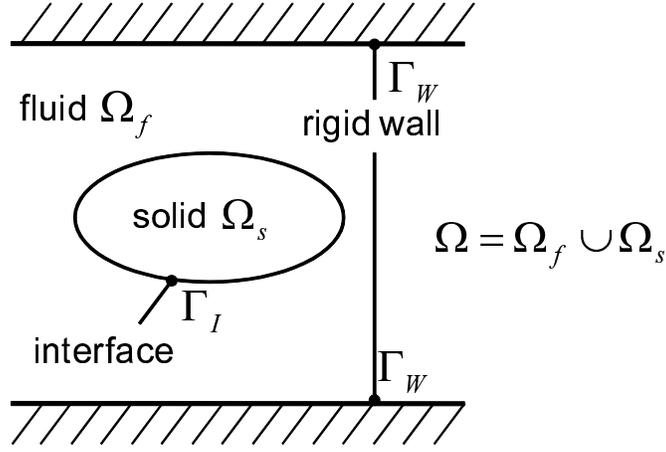,width=9cm,angle=0}
\end{center}
\caption{
Abstract setting for the FSI problem considered
 in the present study. 
}
\label{fig:schem_domain_boundary}
\end{figure}

Figure \ref{fig:schem_domain_boundary} shows 
 the notation of the fluid-structure systems to be addressed. 
Let us consider an incompressible hyperelastic domain
 $\Omega_s$ submerged in an incompressible fluid domain $\Omega_f$,
 which is bounded with rigid flat walls.
Hereafter, the suffices $f$ and $s$ stand for the fluid and solid phases, respectively.
We focus on the system, where the walls are in contact with only fluid
 at the boundary $\Gamma_W$,
 and the moving wall drives the fluid and solid motions.
Both fluid and solid are homogeneous, i.e., 
 the material properties are uniform inside each phase.
We shall restrict our attention to 
 the kinematic and dynamic interactions
 at the fluid-structure interface $\Gamma_I$.
The fluid and solid densities
 are assumed to be identical ($\rho_f=\rho_s=\rho$)
 as in many analyses for biological systems (e.g. \cite{tor2008,zha2008}),
 and no external body force (${\bm b}=0$) is to be exerted on the continua.
The extensions to the non-identical density
 and the non-zero body force would be readily achieved. 

For incompressible fluid and solid, the governing equations consist of 
 the mass and momentum conservations:
\begin{equation}
\begin{split}
\nabla\cdot{\bm v}_f=0,\ \ \ {\bm x}\in \Omega_f,\\
\nabla\cdot{\bm v}_s=0,\ \ \ {\bm x}\in \Omega_s,
\label{eq:cont}
\end{split}
\end{equation}
\begin{equation}
\begin{split}
\rho(\partial_t {\bm v}_f+{\bm v}_f\cdot\nabla{\bm v}_f)=&
 \nabla\cdot {\bm \sigma}_f,\ \ \ {\bm x}\in \Omega_f,\\
\rho(\partial_t {\bm v}_s+{\bm v}_s\cdot\nabla{\bm v}_s)=&
 \nabla\cdot {\bm \sigma}_s,\ \ \ {\bm x}\in \Omega_s,
\label{eq:mom}
\end{split}
\end{equation}
where ${\bm v}$ denotes the velocity vector, 
$t$ the time, 
$\rho$ the density, 
and 
${\bm \sigma}$ the Cauchy stress tensor.

The no-slip condition is imposed on the fluid-wall boundary, namely 
\begin{equation}
{\bm v}_f={\bm V}_W,
\ \ \ {\bm x}\in \Gamma_W,
\end{equation}
where ${\bm V}_W$ denotes the wall velocity.
The kinematic and dynamic interactions between the fluid and solid phases
 are determined by continuity in the velocity and in the traction force
 at the fluid-structure interface, namely, 
\begin{equation}
{\bm v}_f={\bm v}_s,
\ \ \ {\bm x}\in \Gamma_I,
\label{eq:cont_u01}
\end{equation}
\begin{equation}
{\bm \sigma}_f\cdot{\bm n}={\bm \sigma}_s\cdot{\bm n},
\ \ \ {\bm x}\in \Gamma_I,
\label{eq:cont_sig01}
\end{equation}
where ${\bm n}$ denotes the unit normal vector at the interface.

In the practical numerical procedure based on the full Eulerian perspective,
 instead of separately partitioned two velocity fields ${\bm v}_f$ and ${\bm v_s}$
 respectively in $\Omega_f$ and in $\Omega_s$, 
 it is convenient to introduce a monolithic velocity vector ${\bm v}$ 
 applied to the entire domain $\Omega (= \Omega_f \bigcup \Omega_s)$.
In multiphase flow simulations, 
 one set of governing equations for the whole flow field, 
 known as a one-fluid formulation \cite{try2007}, is often employed
 to be discretized on a fixed grid.
In the present study, 
 such an idea is applied to the fluid-structure system by using ${\bm v}$, 
 that is here referred to as a {\it one-continuum} formulation.
The one-continuum formulation would immediately satisfy (\ref{eq:cont_u01})
 because ${\bm v}$ is supposed to be continuous across the interface $\Gamma_I$. 
Following the volume-averaging procedure \cite{tak2010},
 we establish the velocity field ${\bm v}$ as
\begin{equation}
{\bm v}=(1-\phi_s){\bm v}_f + \phi_s {\bm v}_s,
\end{equation}
where $\phi_s$ is the volume fraction of solid inside a computational cell:
\begin{equation}
\phi_s(x,y,z)=
\frac{1}{\Delta_x\Delta_y\Delta_z}
\int_{-\Delta_x/2}^{\Delta_x/2}\!\!\!{\rm d}\hat{x}
\int_{-\Delta_y/2}^{\Delta_y/2}\!\!\!{\rm d}\hat{y}
\int_{-\Delta_z/2}^{\Delta_z/2}\!\!\!{\rm d}\hat{z}
\ I_s(x+\hat{x},y+\hat{y},z+\hat{z}),
\end{equation}
where $\Delta$ denotes the grid size,
 the suffices $x$, $y$ and $z$ stand for the respective directions,
 and $I_s$ the indicator function defined by
\begin{equation}
I_s({\bm x})=\left\{
\begin{array}{ll}
1&{\rm for}\ {\bm x}\in \Omega_s,\\
0&{\rm for}\ {\bm x}\in \Omega_f.
\end{array}
\right.
\end{equation}
We may regard the volume fraction $\phi_s$ as 
 a smoothed Heaviside function at the grid scale. 
The distribution of the volume fraction reveals $\phi_s=0$ in fluid,
 $\phi_s=1$ in solid, and $0<\phi_s<1$ for the grid involving the
 fluid-solid interface. 
As explained in \S\ref{sec:zhao}, 
 the modulus of its gradient $|\nabla\phi_s|$ is regarded as
 a smoothed one-dimensional delta function at the grid scale $\Delta_x$. 
Namely, for the grid involving only fluid or solid, $|\nabla \phi_s|$ is zero, 
 whereas for the grid involving the interface, 
 $|\nabla\phi_s|$ reveals a peak of the order of $\Delta_x^{-1}$.
The isoline at $\phi_s=1/2$ represents the interface $\Gamma_I$
 (see figure \ref{fig:schem_interface}(b)). 
In several approaches for multiphase flow simulations
(e.g., VOF \cite{hir1981} and level set \cite{sus1994,cha1996} methods),
 the viscous stress is often written in a mixture form:
 i.e., the smoothed Heaviside function $H$ smears out the viscosity
 such as $\mu_{\rm mix}=(1-H) \mu_1+H \mu_2$ to remove the discontinuity
 of the stress across the interface $\Gamma_I$. 
The mixture representation is employed in the present study. 
For incompressible continua,
 the pressure $p$ may be regarded as
 of a Lagrangian multiplier imposing the solenoidal condition
 over the whole velocity field. 
The Poisson equation will be solved
 to find the pressure field $p$, written in the one-continuum form,
 over the entire domain $\Omega$.  
Taking the weighted average with respect to $\phi_s$,
 we write the mixture stress ${\bm \sigma}$ as
\begin{equation}
{\bm \sigma}=
-p{\bm I}+(1-\phi_s){\bm \sigma}_f^\prime+\phi_s{\bm \sigma}_s^\prime,
\ \ \ {\bm x}\in \Omega,
\label{eq:mix_stress01}
\end{equation}
where ${\bm I}$ denotes the unit tensor,
 and the prime on the second-order tensor stands for
 the deviatoric tensor, 
 e.g. ${\bm T}^\prime={\bm T}-\frac{1}{3}{\rm tr}({\bm T}){\bm I}$
 for a tensor ${\bm T}$.
Since $\phi_s$ is smoothed at the grid scale
 and ${\bm \sigma}$ is supposed to be smoothly distributed over the entire domain,
 the expression (\ref{eq:mix_stress01}) at $\phi_s=1/2$
 would satisfy the continuity of the traction force (\ref{eq:cont_sig01}).
To advect the scalar field $\phi_s$,
 the purely Eulerian interface method is employed.
Throughout the paper, 
 the fluid component is assumed to be Newtonian,
 and thus the deviatoric stress of fluid is given by
\begin{equation}
{\bm \sigma}_f^\prime= 2\mu_f {\bm D}^\prime,
\label{eq:sigf}
\end{equation}
where  $\mu_f$ denotes the dynamic viscosity of fluid, 
 and ${\bm D}(=\frac{1}{2}(\nabla{\bm v}+\nabla{\bm v}^T))$ the strain rate tensor. 
Instead of (\ref{eq:cont}) and (\ref{eq:mom})
 with (\ref{eq:cont_u01}), (\ref{eq:cont_sig01}) and (\ref{eq:mix_stress01}),
 we solve the following equations 
 in the one-continuum form
 over the entire domain:
\begin{equation}
\nabla\cdot{\bm v}=0,\ \ \ {\bm x}\in \Omega,
\label{eq:cont2}
\end{equation}
\begin{equation}
\rho(\partial_t {\bm v}+{\bm v}\cdot\nabla{\bm v})=
-\nabla p+
\nabla\cdot\{
2(1-\phi_s)\mu_f{\bm D}^\prime+\phi_s{\bm \sigma}_s^\prime
\},\ \ \ {\bm x}\in \Omega,
\label{eq:mom2}
\end{equation}
\begin{equation}
\partial_t\phi_s+{\bm v}\cdot\nabla\phi_s=0,\ \ \ {\bm x}\in \Omega.
\label{eq:transphi}
\end{equation}
The deviatoric stress ${\bm \sigma}_s^\prime$ of solid
 is dependent on the constitutive law.
The incompressible Mooney-Rivlin law
 \cite{moo1940,riv1948,ham1999}
 involving a nonlinearity 
 with respect to ${\bm B}$
 (here ${\bm B}={\bm F}\cdot{\bm F}^T$ denotes
 the left Cauchy-Green deformation tensor,
 ${\bm F}=\partial {\bm x}/\partial {\bm X}$
 the deformation gradient, 
 ${\bm x}$ the current coordinates, 
 and ${\bm X}$ the reference coordinates \cite{bon2008})
 is incorporated into the finite difference method.
The constitutive equations and
 the solution algorithm will be presented in the following sections.

\subsection{Constitutive equations for solid}

In most of finite element computations,
 the constitutive equations of hyperelastic material
 are written over the reference configuration.
The hyperelastic constitutive law is usually provided
 by using the first or second Piola-Kirchhoff stress tensor,
 which is differentiated with respect to
 the reference coordinates ${\bm X}$ in the momentum equation.
It is suited to the Lagrangian frame.
By contrast, in the Eulerian approach, 
 the equation set is written over the current configuration. 
Therefore, the constitutive equations are
 needed to be temporally updated on the fixed grid
 without using the Lagrangian mesh.
In this section, 
 we describe the constitutive law in the Cauchy stress form, 
 and the transport of the left Cauchy-Green deformation tensor field, 
 which are the core elements of the present approach.

\subsubsection{Cauchy stress expression of the incompressible
 Mooney-Rivlin law involving nonlinearity up to $O({\bm B}^2)$}

We consider incompressible visco-hyperelastic materials undergoing
 only the isochoric motion. 
The deviatoric Cauchy stress of solid is expressed as
\begin{equation}
{\bm \sigma}_s^\prime=
2\mu_s{\bm D}^\prime + {\bm \sigma}_{sh}^\prime,
\end{equation}
where the first term on the right-hand-side
 corresponds to the viscous contribution with dynamic viscosity $\mu_s$.
The second term ${\bm \sigma}_{sh}^\prime$ corresponds 
 to the hyperelastic contribution to be derived below.

To formulate the constitutive equation, 
 we refer to the general theories \cite{tra1971,gur1973,sim1985}
 of constrained material. 
Choosing the Mooney-Rivlin expression \cite{moo1940,riv1948}, 
 and considering the nonlinearity up to $O({\bm B}^2)$ in the deviatoric
 Cauchy stress,
 we write the hyperelastic strain energy potential $W$ as
\begin{equation}
W=c_1(\widetilde{\rm I}_{C}-3)+c_2(\widetilde{\rm II}_{C}-3)+c_3(\widetilde{\rm I}_{C}-3)^2,
\end{equation}
where $\widetilde{\rm I}_{C}$ and $\widetilde{\rm II}_{C}$ 
 denote reduced invariants \cite{ham1999}
 of the right Cauchy-Green deformation tensor ${\bm C}={\bm F}^T\cdot{\bm F}$ 
 defined by 
\begin{equation}
\widetilde{\rm I}_{C}=\frac{{\rm I}_{C}}{{\rm III}_{C}^{1/3}},\ \ \ 
\widetilde{\rm II}_{C}=\frac{{\rm II}_{C}}{{\rm III}_{C}^{2/3}},
\end{equation}
where 
$$
{\rm I}_{C}={\rm tr}({\bm C}),\ \ \ 
{\rm II}_{C}=\frac{1}{2}\{{\rm I}_{C}^2-{\rm tr}({\bm C}\cdot{\bm C})\},\ \ \ 
{\rm III}_{C}={\rm det}({\bm C}).
$$
Utilizing the equivalence of the invariants
 between the left and right Cauchy-Green deformation tensors \cite{his1992},
 we write the deviatoric Cauchy stress tensor as
\begin{equation}
\begin{split}
{\bm \sigma}_{sh}^\prime=&
\frac{1}{{\rm det}({\bm F})}
{\bm F}\cdot\biggl\{
2\frac{\partial W}{\partial {\rm   I}_{C}}\frac{\partial {\rm   I}_{C}}{\partial {\bm C}}+
2\frac{\partial W}{\partial {\rm  II}_{C}}\frac{\partial {\rm  II}_{C}}{\partial {\bm C}}+
2\frac{\partial W}{\partial {\rm III}_{C}}\frac{\partial {\rm III}_{C}}{\partial {\bm C}}
\biggr\}\cdot{\bm F}^T
\\=&
\bigl(
2c_1{\bm B}+2c_2({\rm tr}({\bm B}){\bm B}-{\bm B}\cdot{\bm B})
+4c_3({\rm tr}({\bm B})-3){\bm B}
\bigr)^\prime.
\label{eq:sigs_b01}
\end{split}
\end{equation}
We will give several demonstrations afterward
 for some specific cases
 based on the linear Mooney-Rivlin, neo-Hookean, and incompressible
 Saint Venant-Kirchhoff materials.
Note that all these materials obey (\ref{eq:sigs_b01}).
For {\it linear} Mooney-Rivlin material ($c_3=0$) \cite{his1992,moo1940,riv1948},
 which is often used as the biological material, 
 Cauchy stress becomes
\begin{equation}
{\bm \sigma}_{sh}^\prime=
2c_1{\bm B}^\prime+2c_2({\rm tr}({\bm B}){\bm B}-{\bm B}\cdot{\bm B})^\prime.
\label{eq:ce_lin_mr}
\end{equation}
The neo-Hookean material is a particular case 
 of the linear Mooney-Rivlin material
 with the coefficients 
$$
c_1=\frac{G}{2}, \ \ \ c_2=c_3=0,
$$
where $G$ denotes the modulus of transverse elasticity. 
The corresponding Cauchy stress is 
\begin{equation}
{\bm \sigma}_{sh}^\prime=G{\bm B}^\prime.
\label{eq:ce_nh}
\end{equation}

As another typical hyperelastic material,
 we consider Saint Venant-Kirchhoff material \cite{bon2008,gil2006,saw2007},
 which often models a thin but finite volume membrane.
The constitutive equation is expressed as
 a simple extension of Hooke's law, 
 as defined by 
$$
{\bm S}=
\lambda_{\mbox{\tiny Lam\'e}}^s{\rm tr}({\bm E}){\bm I}
+  2\mu_{\mbox{\tiny Lam\'e}}^s{\bm E},
$$
where ${\bm S}$ is the second Piola-Kirchhoff stress, 
 $\lambda_{\mbox{\tiny Lam\'e}}^s$, $\mu_{\mbox{\tiny Lam\'e}}^s$
 are the Lam\'{e} constants, 
 and ${\bm E}=\frac{1}{2}({\bm C}-{\bm I})$ is the Green-Lagrange strain tensor.
Although the Saint Venant-Kirchhoff material
 is usually referred to as dilatable,
 we implement the {\it incompressible} Saint Venant-Kirchhoff
 model, as defined in \cite{gul2003}.
The deviatoric Cauchy stress of the incompressible 
 Saint Venant-Kirchhoff material is expressed as
\begin{equation}
{\bm \sigma}_{sh}^\prime=
\left(\frac{{\bm F}\cdot {\bm S}\cdot {\bm F}^T}{{\rm det}({\bm F})}\right)^\prime
=\frac{\lambda_{\mbox{\tiny Lam\'e}}^s}{2}({\rm tr}({\bm B})-3){\bm B}^\prime
+\mu_{\mbox{\tiny Lam\'e}}^s({\bm B}\cdot{\bm B}-{\bm B})^\prime.
\label{eq:ce_svk}
\end{equation}
Substituting the relations
$$
c_1=\mu_{\mbox{\tiny Lam\'e}}^s,\ \ \ 
c_2=-\frac{\mu_{\mbox{\tiny Lam\'e}}^s}{2},\ \ \ 
c_3=\frac{\lambda_{\mbox{\tiny Lam\'e}}^s+2\mu_{\mbox{\tiny
 Lam\'e}}^s}{8},
$$
into (\ref{eq:sigs_b01}), we can readily realize that 
 the constitutive law (\ref{eq:ce_svk})
 falls within the class of nonlinear Mooney-Rivlin laws.

\subsubsection{Transport of left Cauchy-Green deformation tensor field in the Eulerian frame}

Once the coefficients $c_1$, $c_2$ and $c_3$ are given,
 the constitutive equation (\ref{eq:sigs_b01})
 is expressed as a function of the left Cauchy-Green deformation tensor ${\bm B}$.
If the tensor field ${\bm B}$ is determined in a purely Eulerian manner,
 all the equations will be closed in the Eulerian form.
Utilizing the fact that the upper convected time derivative of ${\bm B}$
 is identically zero \cite{bon2008}, 
 one may use the following transport equation to update the ${\bm B}$ field:
\begin{equation}
\partial_t{\bm B}
+{\bm v}\cdot\nabla{\bm B}
={\bm L}\cdot{\bm B}
+{\bm B}\cdot{\bm L}^T,
\label{eq:transb01}
\end{equation}
where ${\bm L}(=(\nabla{\bm v})^T)$ denotes the velocity gradient tensor.
For the initially unstressed solid, 
 the initial condition is given by ${\bm B}={\bm I}$. 
It should be noticed that, however,
 it is quite cumbersome to solve (\ref{eq:transb01}) from a numerical viewpoint,
 because ${\bm B}$ exhibits rough distribution
 in the fluid domain $\Omega_f$ \cite{liu2001}.
The fluid element subject to a shearing motion
 is likely to elongate toward the extensional direction. 
Such an elongation causes a temporally exponential growth
 of some components of ${\bm B}$.
To avoid the numerical instability brought by the exponential growth, 
 instead of ${\bm B}$, 
 we update the {\it modified} left Cauchy-Green deformation tensor
 $\tilde{\bm B}=\phi_s^{\alpha}{\bm B}$ ($\alpha>0$),
 which yields
\begin{equation}
\partial_t\tilde{\bm B}
+{\bm v}\cdot\nabla\tilde{\bm B}
={\bm L}\cdot\tilde{\bm B}
+\tilde{\bm B}\cdot {\bm L}^T,
\label{eq:transb02}
\end{equation}
with the initial condition $\tilde{\bm B}=\phi_s^\alpha {\bm I}$. 
Because of $\tilde{\bm B}=0$ for $\phi_s=0$, 
 the numerical instability is evaded in the fluid domain $\Omega_f$.
In the hyperelastic constitutive law (\ref{eq:ce_lin_mr}), 
 we consider the term up to $O({\bm B}^2)$, 
 of which the contribution to the phase averaged stress
 $\phi_s {\bm \sigma}_s$ in (\ref{eq:mix_stress01})
 is proportional to 
 $\phi_s {\bm B}^2 = \phi_s^{1-2\alpha} \tilde{\bm B}^2$.
In order to avoid division by zero in the fluid region $\phi_s=0$, 
 the exponent $1-2\alpha$ on $\phi_s$ should be non-negative,
 indicating $\alpha\leq 1/2$.
In the present study, we choose the largest value $\alpha=1/2$.
Further, to avoid the inevitable exponential growth 
 at the cell near the interface $\Gamma_I$ containing the fluid-solid mixture,
 and to obtain a viable compromise
 between the numerical consistency and stability,
 we introduce a threshold $\phi_{\rm min}$ and enforce $\tilde{\bm B}=0$ where $\phi_s< \phi_{\rm min}$.
In the present study, we set $\phi_{\rm min}$ between $0.01$ and $0.1$.

\subsubsection{Characteristics of the present approach in treating the solid stress}

From (\ref{eq:sigs_b01}),
 the resulting deviatoric stress of solid multiplied by $\phi_s$, 
 which is involved in (\ref{eq:mom2}), is expressed as
\begin{equation}
\begin{split}
\phi_s{\bm \sigma}_{s}^\prime=&
2\phi_s\mu_s{\bm D}^\prime+
\bigl(
2c_1\phi_s^{1/2}\tilde{\bm B}
\\&
+2c_2({\rm tr}(\tilde{\bm B})\tilde{\bm B}-\tilde{\bm B}\cdot\tilde{\bm B})
+4c_3({\rm tr}(\tilde{\bm B})-3\phi_s^{1/2})\tilde{\bm B}
\bigr)^\prime,
\label{eq:sigs_b02}
\end{split}
\end{equation}
which can be evaluated together with the temporally updated $\tilde{\bm B}$
 from (\ref{eq:transb02}).

It should be noted that 
 when the material points are tracked in the Lagrangian way, 
 the relation (\ref{eq:transb01}) is identically satisfied 
 via the change in the relative position of the adjacent material points
 (see figure \ref{fig:schem_deformation}(a)).
Thus, in the pure Lagrangian approach, 
 it is not necessary to introduce a transport equation 
 describing the deformation level such as ${\bm B}$.
On the other hand, in the Eulerian approach, 
 which does not rely on the material point,
 we must introduce the deformation level on the fixed mesh.
The present approach
 is characterized by the introduction
 of the transport equation of $\tilde{\bm B}$,
 which is temporally updated in the Eulerian frame
 (see figure \ref{fig:schem_deformation}(b)).

\section{Simulation methods}
\label{sec:simulation_method}

\subsection{Overview}

Once the initial field of the solid volume fraction $\phi_{s0}$
 is given over the entire domain $\Omega$,
 the present Eulerian method enables one to carry out 
 the FSI simulation without mesh generation procedure.
In order to prescribe the $\phi_{s0}$ field, 
 it is only required to numerically compute
 the ratio of the occupied solid to each control volume
 from the initial geometry as a preprocess.
If the system is initially at rest and unstressed, 
 one can launch the simulation
 by setting the initial conditions of the velocity vector, pressure, and 
 modified left Cauchy-Green deformation tensor 
 to ${\bm v}=0$, $p=0$, and $\tilde{\bm B}=\phi_s^{1/2}{\bm I}$,
 respectively.

\begin{figure}[h]
\begin{center}

\epsfig{file=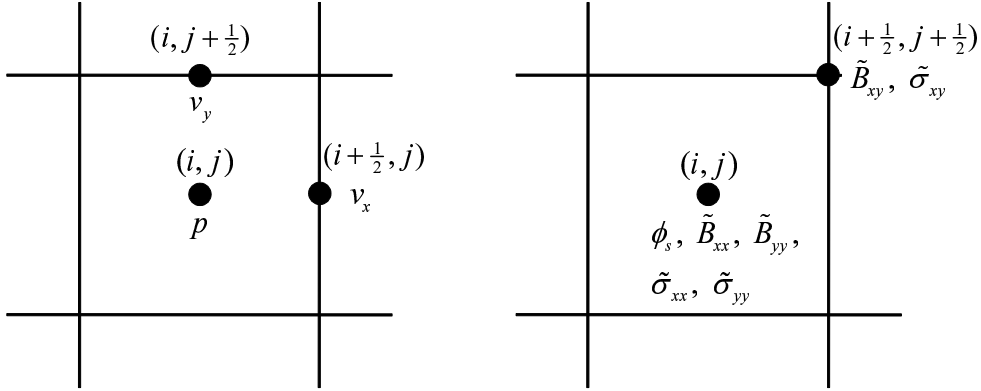,width=13cm,angle=0}

\end{center}
\caption{
Schematic figure of computational grid with the mesh size of $\Delta_x\times \Delta_y$, 
here $\Delta_x=\Delta_y$.
(a) left panel: 
Definition points of the velocity components $v_x$, $v_y$, and the pressure $p$. 
(b) right panel: 
Definition points of 
the solid volume fraction $\phi_s$, 
the stress components $\tilde{\sigma}_{xx}$, $\tilde{\sigma}_{yy}$, $\tilde{\sigma}_{xy}$, 
and the modified left Cauchy-Green deformation components
$\tilde{B}_{xx}$, $\tilde{B}_{yy}$, $\tilde{B}_{xy}$.
}
\label{fig:schem_staggered}
\end{figure}

Instead of the {\it deviatoric} stress ${\bm \sigma}^\prime$
 (with a prime),
 we may use the following {\it pseudo} stress
 to make some discretized expressions for the individual stress components simple:
\begin{equation}
\tilde{\bm \sigma}=(\mu_f+(\mu_s-\mu_f)\phi_s)\left(\nabla{\bm v}+\nabla{\bm v}^{T}\right)
+\phi_s\tilde{\bm\sigma}_{sh},
\end{equation}
where 
\begin{equation}
\phi_s\tilde{\bm\sigma}_{sh}=
2c_1\phi_s^{1/2}\tilde{\bm B}
+2c_2({\rm tr}(\tilde{\bm B})\tilde{\bm B}-\tilde{\bm B}\cdot\tilde{\bm B})
+4c_3({\rm tr}(\tilde{\bm B})-3\phi_s^{1/2})\tilde{\bm B}.
\label{eq:sigs_b03}
\end{equation}
Introducing a {\it pseudo} pressure $\tilde{p}$, instead of (\ref{eq:mom2}), we solve
\begin{equation}
\rho\left(\partial_t{\bm v}+{\bm v}\cdot\nabla{\bm v}\right)=
-\nabla \tilde{p}
+\nabla\cdot \tilde{\bm\sigma}.
\label{eq:mom3}
\end{equation}
The actual pressure and deviatoric stress read
$$
p=\tilde{p}-\frac{{\rm tr}(\tilde{\bm\sigma})}{3},
\ \ \ 
{\bm\sigma}^\prime=
\tilde{\bm\sigma}-\frac{{\rm tr}(\tilde{\bm\sigma})}{3}{\bm I}.
$$
Hereafter, the formulation will be given on
 the two-dimensional (2D) system. 
The extension to the three-dimensional (3D) system is straightforward \cite{sug2010}.
The basic equations
 (\ref{eq:cont2}), (\ref{eq:mom2}), (\ref{eq:transphi}), 
 (\ref{eq:transb02}) and (\ref{eq:sigs_b02}) 
 are solved by a finite difference method on a fixed Cartesian grid. 
The quantities ${\bm v}$, $p$, $\phi_s$, and $\tilde{\bm B}$ are temporally
 updated in the Eulerian frame. 
The entire domain is discretized by a uniform square mesh.
We follow a conventional staggered Marker-And-Cell (MAC) cell arrangement \cite{har1965},
 where the velocity component is located on the cell face,
 and the pressure at the cell center
(see figure \ref{fig:schem_staggered}(a)). 
The symmetry in the tensor $\tilde{\bm B}$ 
 and the two-dimensionality of the system imply $\tilde{B}_{zz}=\phi_s^{\alpha}$,
 $\tilde{B}_{xz}=\tilde{B}_{zx}=\tilde{B}_{yz}=\tilde{B}_{zy}=0$,
 and $\tilde{B}_{yx}=\tilde{B}_{xy}$.
Hence, (\ref{eq:transb02}) is solved for 
 the three independent components $\tilde{B}_{xx}$, $\tilde{B}_{yy}$ and $\tilde{B}_{xy}$.
The diagonal components of $\tilde{\bm B}$
 and the solid volume fraction $\phi_s$ are defined on the cell center,
 while the non-diagonal components of $\tilde{\bm B}$ are on the cell apex 
 (see figure \ref{fig:schem_staggered}(b)). 

As opposed to usual methods of computational structure dynamics,
 for instance, the GMRES approach \cite{saa1986}
 and the weak compressibility approach \cite{hue1988a,che1996},
 the pressure Poisson equation is solved
 to exactly satisfy the solenoidal condition (\ref{eq:cont2})
 over the entire domain $\Omega$
 likewise widely-used incompressible fluid flow algorithms.
In solving the discretized Poisson equation, 
 the fast Fourier transform is used to ensure high accuracy
 as well as high efficiency.

\subsection{Time-stepping algorithm}

Here, the time-stepping algorithm
 to update the variables at the $(n+1)$-th time level
 from the $n$-th time level is briefly explained.
Following the Simplified MAC method \cite{ams1970},
 corresponding to a standard incompressible fluid flow algorithm,
 with an incremental pressure correction applied to the finite difference scheme, 
 we decompose the time-stepping into three steps. 

In the first step, the second-order Adams-Bashforth scheme 
\cite{can1988} is applied
 to explicitly updating the volume fraction $\phi_s^{(n+1)}$
 and the modified left Cauchy-Green deformation tensor  $\tilde{\bm B}^{(n+1)}$:
\begin{equation}
\phi_s^{(n+1)}=\phi_s^{(n)}-(\Delta t)\left(
\frac{3}{2}{\bm v}^{(n  )}\cdot\nabla\phi_s^{(n  )}-
\frac{1}{2}{\bm v}^{(n-1)}\cdot\nabla\phi_s^{(n-1)}\right),
\end{equation}
\begin{equation}
\begin{split}
\tilde{\bm B}^{(n+1)}&=\tilde{\bm B}^{(n)}-(\Delta t)\biggl\{
 \frac{3}{2}{\bm v}^{(n  )}\cdot\nabla\tilde{\bm B}^{(n  )}
-\frac{1}{2}{\bm v}^{(n-1)}\cdot\nabla\tilde{\bm B}^{(n-1)}
\\&
+\frac{3}{2}\left(
-{\bm L}^{(n)}\cdot\tilde{\bm B}^{(n)}
-\tilde{\bm B}^{(n)}\cdot{\bm L}^{T(n)}\right)
\\&
-\frac{1}{2}\left(
-{\bm L}^{(n-1)}\cdot\tilde{\bm B}^{(n-1)}
-\tilde{\bm B}^{(n-1)}\cdot{\bm L}^{T(n-1)}\right)
\biggr\},
\end{split}
\end{equation}
where 
$(\Delta t)$ denotes the time increment,
 and the superscript $(n)$ stands for the $n$-th time level. 
It should be noted that $(\Delta t)$ is chosen
 such that the Courant-Friedrichs-Lewy (CFL)
 condition is satisfied i.e. the CFL number 
 based on the larger velocity scale between the maximum advection speed
 ${\rm max}(|v_x|,|v_y|)$ and the linear elastic wave speed $\sqrt{2(c_1+c_2)/\rho}$
 is $0.1$ or less for all the present computations.

In the second step, 
the second-order Adams-Bashforth and Crank-Nicolson 
 schemes \cite{can1988} are applied to iteratively calculating
 the unprojected velocity vector ${\bm v}^*$
 and stress tensor $\tilde{\bm \sigma}^{*}$:
\begin{equation}
\begin{split}
{\bm v}^*&={\bm v}^{(n  )}-\frac{(\Delta t)\nabla \tilde{p}^{(n)}}{\rho}
-(\Delta t)\biggl(
\frac{3}{2}{\bm v}^{(n  )}\cdot\nabla{\bm v}^{(n  )}-
\frac{1}{2}{\bm v}^{(n-1)}\cdot\nabla{\bm v}^{(n-1)}
\biggr)
\\&
+\frac{(\Delta t)}{\rho}
\left(
\frac{1}{2}\nabla\cdot \tilde{\bm \sigma}^{*}
+\frac{1}{2}\nabla\cdot\tilde{\bm \sigma}^{(n)}
\right),
\end{split}
\end{equation}
\begin{equation}
\begin{split}
\tilde{\bm \sigma}^{*}&=
(\mu_f+(\mu_s-\mu_f)\phi_s^{(n+1)})
\left(\nabla{\bm v}^*+\nabla{\bm v}^{* T}\right)
+(\phi_s\tilde{\bm\sigma}_{sh})^{(n+1)},
\end{split}
\end{equation}
where the solid stress is given by (\ref{eq:sigs_b03}). 

Finally, pressure, solenoidal velocity vector, and stress tensor
 are updated during the projection step:
\begin{equation}
\tilde{p}^{(n+1)}=\tilde{p}^{(n)}+\varphi,
\label{eq:update_p}
\end{equation}
\begin{equation}
{\bm v}^{(n+1)}={\bm v}^{*}-(\Delta t)\nabla\varphi,
\label{eq:update_v}
\end{equation}
\begin{equation}
\tilde{\bm \sigma}^{(n+1)}=\tilde{\bm \sigma}^{*}
-2(\Delta t)
(\mu_f+(\mu_s-\mu_f)\phi_s^{(n+1)})
(\nabla\nabla\varphi),
\label{eq:update_sig}
\end{equation}
where the incremental pressure $\varphi$ is determined by solving the Poisson equation
\begin{equation}
\nabla^2\varphi = \frac{\rho\nabla\cdot{\bm v}^*}{(\Delta t)}.
\label{eq:pres_eq}
\end{equation}

\subsection{Spatial discretizations}

The spatial derivatives are approximated by the second-order central differences,
 except for those of the advection terms in (\ref{eq:transphi}) and (\ref{eq:transb02}),
 to which the fifth-order WENO scheme \cite{liu1994, jia1996} is applied. 
For the momentum equation, following the spirit in \cite{kaj1994,ike2007}, 
 we discretize the advection terms to satisfy
 the identity $\nabla\cdot({\bm v}{\bm v}) = ({\bm v}\cdot \nabla){\bm v}+{\bm v}(\nabla\cdot{\bm v})$
 in the discretized space,
 that would make the energy highly conserved.
The finite difference descriptions are detailed in Appendix \ref{sec:fdm}. 

We here focus on the discretization of the right-hand-side of (\ref{eq:transb02}),
 that is important to accurately describe the isochoric solid deformation. 
We take care of the difference between the definition points of 
 the diagonal $\tilde{B}_{xx}$, $\tilde{B}_{yy}$
 and non-diagonal $\tilde{B}_{xy}$ components
 as illustrated in figure \ref{fig:schem_staggered}(b).
The incompressibility ${\rm det}({\bm B})=1$ 
 implies $\tilde{B}_{xx}\tilde{B}_{yy}-\tilde{B}_{xy}^2=\phi_s^{2\alpha}$. 
Since we choose $\alpha=1/2$, we find 
\begin{equation}
\frac{{\rm d}}{{\rm d}t}
\int\!\!\!\int_{\Omega}\!{\rm d}^2{\bm x}\ 
(\tilde{B}_{xx}\tilde{B}_{yy}-\tilde{B}_{xy}^2)
=
\frac{{\rm d}}{{\rm d}t}
\int\!\!\!\int_{\Omega}\!{\rm d}^2{\bm x}\ 
\phi_s
=0,
\label{eq:bb_b}
\end{equation} 
In consideration of the time derivative of each component of $\tilde{\bm B}$
 in (\ref{eq:transb02}), 
 the left-hand-side of (\ref{eq:bb_b}) is given by
\begin{equation}
\begin{split}
&
\int\!\!\!\int_{\Omega}\!{\rm d}^2{\bm x}\ 
\left\{
(\partial_t \tilde{B}_{xx})
\tilde{B}_{yy}
+
\tilde{B}_{xx}
(\partial_t \tilde{B}_{yy})
-2(\partial_t \tilde{B}_{xy})\tilde{B}_{xy}
\right\}\\
=&
\int\!\!\!\int_{\Omega}\!{\rm d}^2{\bm x}\ 
\bigl\{
(-{\bm v}\cdot \nabla \tilde{B}_{xx})\tilde{B}_{yy}
+
\tilde{B}_{xx}(-{\bm v}\cdot \nabla\tilde{B}_{yy})
-2(-{\bm v}\cdot\nabla\tilde{B}_{xy})\tilde{B}_{xy}
\bigr\}
\\+&
\int\!\!\!\int_{\Omega}\!{\rm d}^2{\bm x}\ 
\bigl\{
{\bm e}_x\cdot ({\bm L}\cdot\tilde{\bm B}+\tilde{\bm B}\cdot {\bm L}^T)\cdot {\bm e}_x
\tilde{B}_{yy}
+
{\bm e}_y\cdot ({\bm L}\cdot\tilde{\bm B}+\tilde{\bm B}\cdot {\bm L}^T)\cdot {\bm e}_y
\tilde{B}_{xx}
\\&
-2{\bm e}_x\cdot ({\bm L}\cdot\tilde{\bm B}+\tilde{\bm B}\cdot {\bm L}^T)\cdot {\bm e}_y
\tilde{B}_{xy}
\bigr\},
\label{eq:constraint_lb0}
\end{split}
\end{equation}
where ${\bm e}$ denotes the unit vector.
Applying Gauss' divergence theorem together with
 the kinematic condition ${\bm n}\cdot{\bm v}=0$ at the rigid wall $\Gamma_w$
 and the solenoidal condition (\ref{eq:cont2}),
 we rewrite the first term in the right-hand-side of (\ref{eq:constraint_lb0}) as
\begin{equation*}
-\oint_{\Gamma_w}\!{\rm d}{\bm x}\ 
\underbrace{{\bm n}\cdot{\bm v}}_{=0}
\bigl\{
\tilde{B}_{xx}\tilde{B}_{yy}
-\tilde{B}_{xy}^2
\bigr\}=0.
\end{equation*}
Hence, in
 view of the solid volume conservation, 
 the following relation should be satisfied:
\begin{equation}
\begin{split}
&\int\!\!\!\int_{\Omega}\!{\rm d}^2{\bm x}\ 
\bigl\{
{\bm e}_x\cdot ({\bm L}\cdot\tilde{\bm B}+\tilde{\bm B}\cdot {\bm L}^T)\cdot {\bm e}_x
\tilde{B}_{yy}
+
{\bm e}_y\cdot ({\bm L}\cdot\tilde{\bm B}+\tilde{\bm B}\cdot {\bm L}^T)\cdot {\bm e}_y
\tilde{B}_{xx}
\\&
-2{\bm e}_x\cdot ({\bm L}\cdot\tilde{\bm B}+\tilde{\bm B}\cdot {\bm L}^T)\cdot {\bm e}_y
\tilde{B}_{xy}
\bigr\}=0.
\label{eq:constraint_lb}
\end{split}
\end{equation}
We choose the interpolation and the finite difference formulae
 to satisfy the integral relation (\ref{eq:constraint_lb}) in a discrete form.
For a quantity $q$,
 let us introduce finite difference operators 
 $\delta_i$ and $\delta_j$, of which the indices $i$ and $j$ correspond
 to discretized coordinates along the respective directions $x$ and $y$,
 such as
\begin{equation}
\delta_i(q)|_{i,j}=q_{i+\frac{1}{2},j}-q_{i-\frac{1}{2},j},\ \ \ 
\delta_j(q)|_{i,j}=q_{i,j+\frac{1}{2}}-q_{i,j-\frac{1}{2}},
\label{eqa:fdop}
\end{equation}
and a four-point interpolation operator denoted by double overline such as
\begin{equation}
\overline{\overline{q}}|_{i,j}=\frac{
q_{i-\frac{1}{2},j-\frac{1}{2}}+
q_{i+\frac{1}{2},j-\frac{1}{2}}+
q_{i-\frac{1}{2},j+\frac{1}{2}}+
q_{i+\frac{1}{2},j+\frac{1}{2}}}{4}.
\end{equation}
The integral $\int_\Omega {\rm d}^2{\bm x}\ fg$ 
 provides the equality in a finite volume representation
\begin{equation}
\sum_{i}\sum_{j}\Delta_x\Delta_y
f_{i,j}\ \overline{\overline{g}}|_{i,j}
=
\sum_{i}\sum_{j}\Delta_x\Delta_y
\overline{\overline{f}}|_{i+\frac{1}{2},j+\frac{1}{2}}
\ g_{i+\frac{1}{2},j+\frac{1}{2}},
\label{eq:sum_fvm}
\end{equation}
where the quantities $f$ and $g$ are defined at the cell center and
 at the cell apex, respectively, 
 and assumed to vanish at the boundary $\Gamma_W$.
 We write each component of ${\bm L}\cdot\tilde{\bm B}+\tilde{\bm B}\cdot{\bm L}^T$
 involved in (\ref{eq:constraint_lb}) as
\begin{align}
&\left({\bm e}_x\cdot ({\bm L}\cdot\tilde{\bm B}+\tilde{\bm B}\cdot{\bm L}^T)\cdot{\bm e}_x
\right)_{i,j}
=2L_{xx,i,j}\tilde{B}_{xx,i,j}+2\overline{\overline{L_{xy}\tilde{B}_{xy}}}\bigr|_{i,j},
\label{eq:exlbblx}
\\&
\left({\bm e}_y\cdot ({\bm L}\cdot\tilde{\bm B}+\tilde{\bm B}\cdot{\bm L}^T)\cdot{\bm e}_y
\right)_{i,j}
=2L_{yy,i,j}\tilde{B}_{yy,i,j}+2\overline{\overline{L_{yx}\tilde{B}_{xy}}}\bigr|_{i,j},
\\&
\left({\bm e}_x\cdot ({\bm L}\cdot\tilde{\bm B}+\tilde{\bm B}\cdot{\bm L}^T)\cdot{\bm e}_{y}
\right)_{i+\frac{1}{2},j+\frac{1}{2}}
\!\!=
\left(
\overline{\overline{L_{xx}}}\bigr|_{i+\frac{1}{2},j+\frac{1}{2}}+
\overline{\overline{L_{yy}}}\bigr|_{i+\frac{1}{2},j+\frac{1}{2}}
\right)\tilde{B}_{xy,i+\frac{1}{2},j+\frac{1}{2}}
\nonumber\\&
\ \ \ \ \ \ \ \ \ \ \ \ \ \ \ \ \ \ \ \ \ \ \ \ \ \ 
+L_{xy,i+\frac{1}{2},j+\frac{1}{2}}\overline{\overline{\tilde{B}_{yy}}}\bigr|_{i+\frac{1}{2},j+\frac{1}{2}}
+L_{yx,i+\frac{1}{2},j+\frac{1}{2}}\overline{\overline{\tilde{B}_{xx}}}\bigr|_{i+\frac{1}{2},j+\frac{1}{2}},
\label{eq:exlbbly}
\end{align}
where 
$$
L_{xx,i,j}=
\frac{\delta_i(v_x)|_{i,j}}{\Delta_x},
\ \ 
L_{yy,i,j}=
\frac{\delta_j(v_y)|_{i,j}}{\Delta_y},
$$
$$
L_{xy,i+\frac{1}{2},j+\frac{1}{2}}=
\frac{\delta_j(v_x)|_{i+\frac{1}{2},j+\frac{1}{2}}}{\Delta_y},
\ \ 
L_{yx,i+\frac{1}{2},j+\frac{1}{2}}=
\frac{\delta_i(v_y)|_{i+\frac{1}{2},j+\frac{1}{2}}}{\Delta_x}.
$$
Substituting (\ref{eq:exlbblx})--(\ref{eq:exlbbly}) into (\ref{eq:constraint_lb})
 together with the relation (\ref{eq:sum_fvm}), 
 and considering the solenoidal condition (\ref{eq:cont2}),
 we confirm that the requirement (\ref{eq:constraint_lb}) would be fulfilled
 in a discrete form as
$$
2\sum_{i}\sum_{j}\Delta_x\Delta_y
\underbrace{(L_{xx,i,j}+L_{yy,i,j})}_{=0}
\left(
\tilde{B}_{xx,i,j}\tilde{B}_{yy,i,j}
-\overline{\overline{\tilde{B}_{xy}^2}}|_{i,j}
\right)=0.
$$

\section{Validation tests}
\label{sec:validation}

For the purpose of addressing several computational issues
 involved in the procedure advocated in \S \ref{sec:basic_equation}
 and \S \ref{sec:simulation_method},
 firstly, we consider a one-dimensional problem
 for the motion of the oscillatory parallel fluid-solid layers.
Though this system is a simple example, 
 it includes the fundamental aspects of the FSI problem
 and the system allows estimate of the numerical accuracy
 in the present full Eulerian approach
 as it has analytical solutions.
In the second example for validation, 
 we make comparisons with
 the available data of the deformable solid motion
 in a lid-driven cavity \cite{zha2008} and
 the particle-particle interaction
 in a Couette flow \cite{gao2009}.
Also, in the third example, 
 a response of hyperelastic material
 to the external shear strain is examined
 to check the reversibility in shape
 as this aspect is vital for a full Eulerian formulation.

\begin{figure}[h]
\begin{center}

\vspace{1em}

\epsfig{file=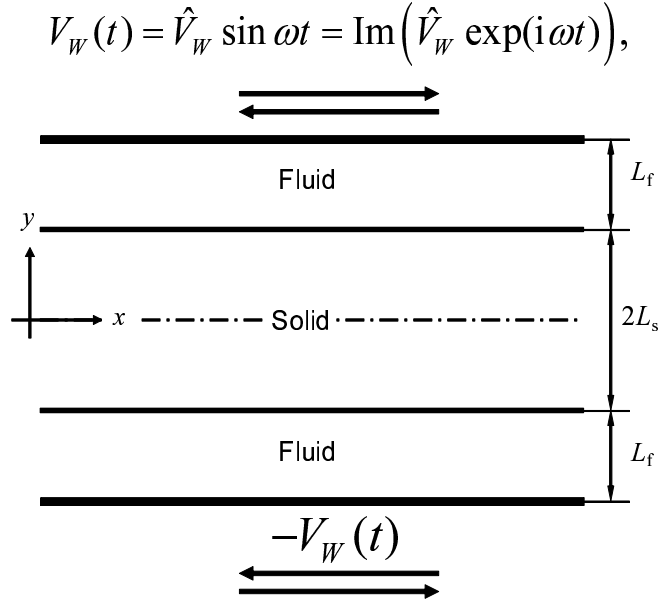,width=9cm,angle=0}

  \end{center}
\caption{
Schematic figure of the parallel fluid-structure layers.
The upper and lower plates move in opposite directions 
 with temporally sinusoidal velocities 
 to drive the fluid and solid motions. 
}
\label{fig:schem_layer}
\end{figure}

\begin{figure}[h]
\begin{center}

\epsfig{file=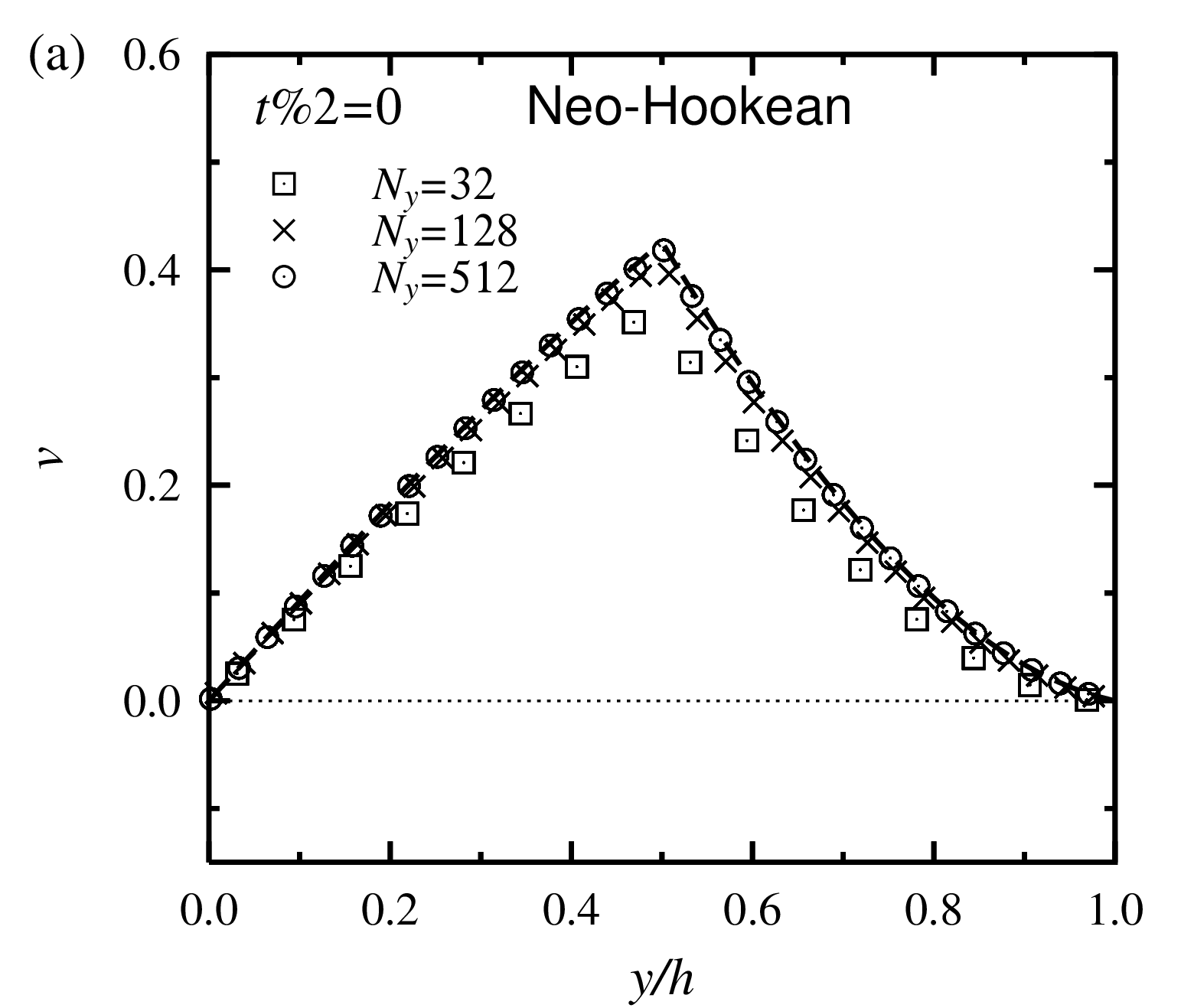,width=5.7cm,angle=270}
\epsfig{file=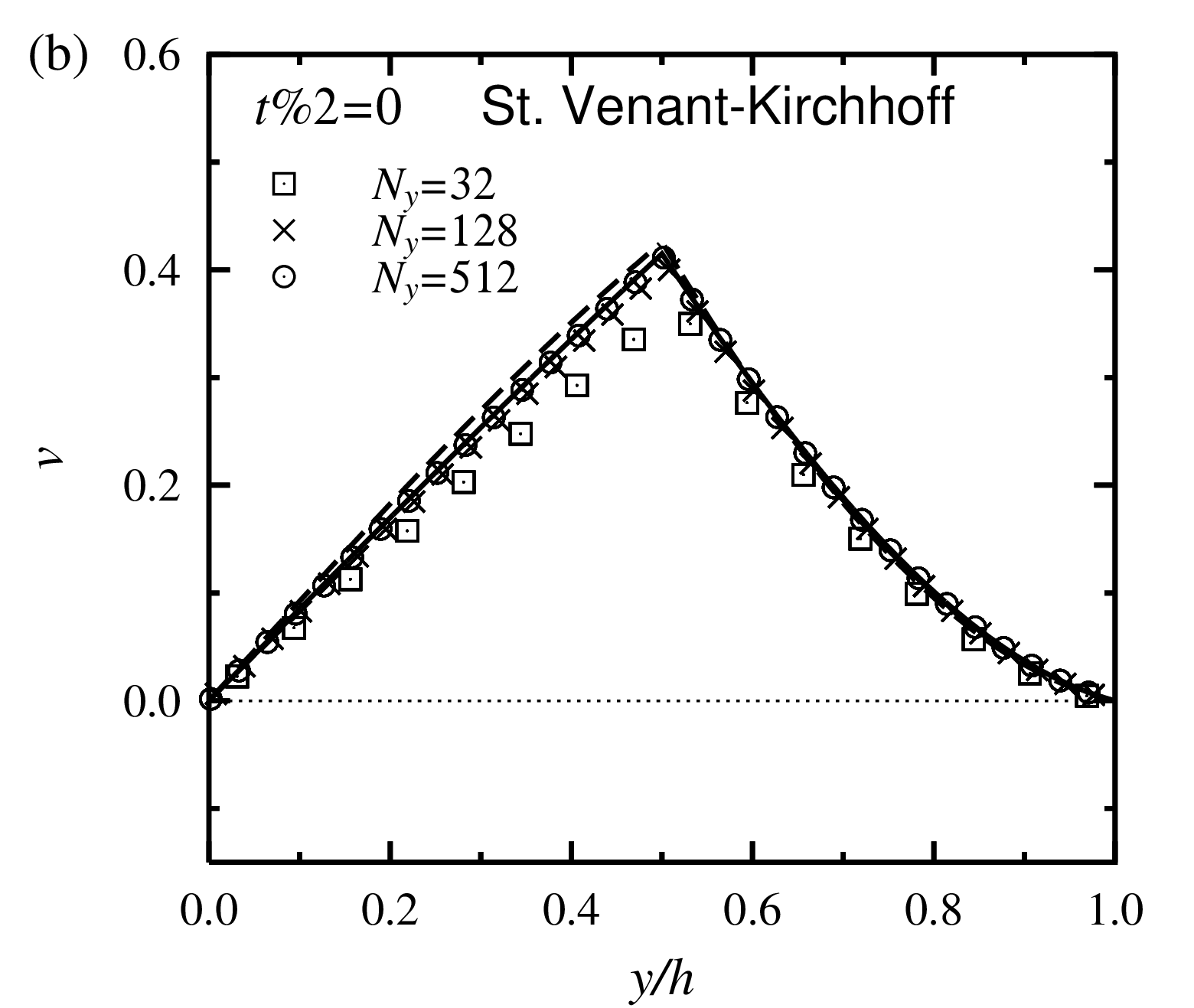,width=5.7cm,angle=270}

\epsfig{file=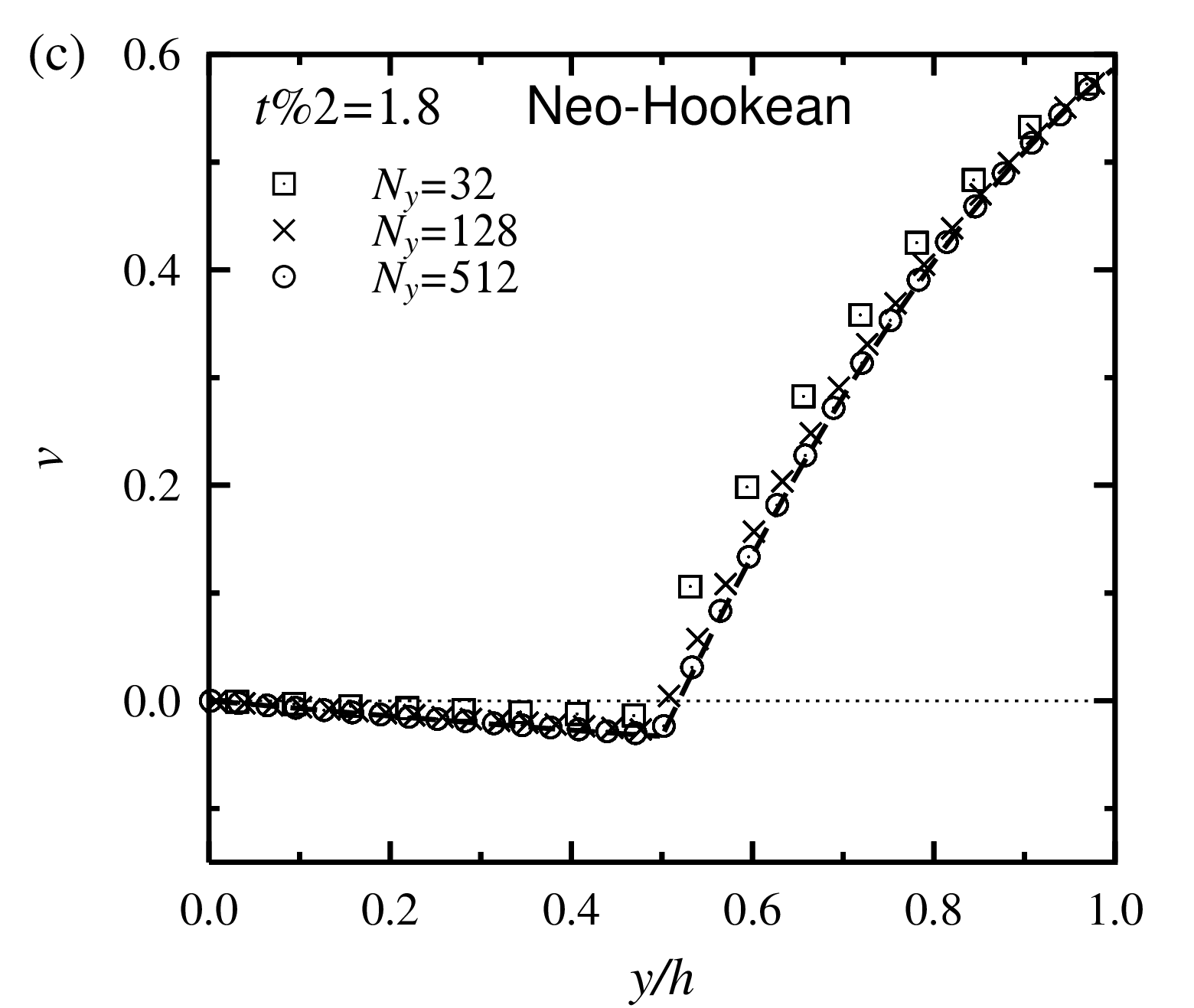,width=5.7cm,angle=270}
\epsfig{file=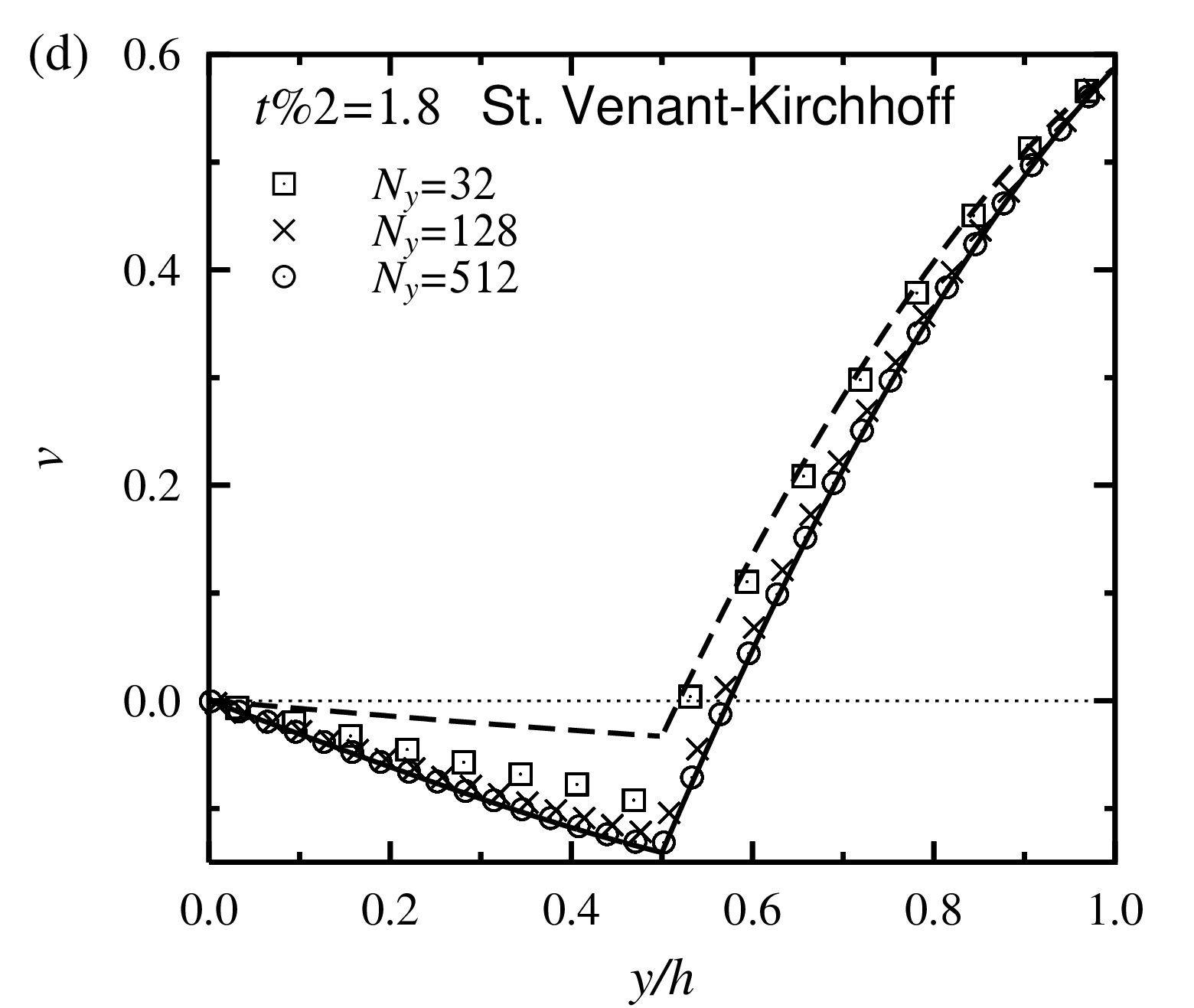,width=5.7cm,angle=270}

\end{center}
\caption{
The velocity profile in the upper half region of 
 the parallel fluid-structure layers
 under the conditions of 
 $\mu_f=1$, $L_s=L_f=0.5$, $\omega=\pi$, and $\hat{V}_w=1$. 
Left panels: 
 the neo-Hookean material with the modulus of transverse elasticity of $G=5$.
Right panels: 
 the incompressible Saint Venant-Kirchhoff material with the Lam\'e constants of
 $\lambda_{\mbox{\tiny Lam\'e}}^s=7.5$ and $\mu_{\mbox{\tiny Lam\'e}}^s=5$.
The upper and lower panels show the results at the temporal phases
 of $t\%2=0$ and  $t\%2=1.8$, respectively.
The dashed and solid curves respectively
 represent the linear and nonlinear solutions
 with the sharp interface,
 which are determined by means of the spectral approach
 (see Appendix \ref{sec:spec}).
The symbols correspond to the present Eulerian simulation results 
 for various number of grid points
 ($N_x\times N_y =8\times 32$, $8\times 128$, $8\times 512$).
}
\label{fig:layer_nh_svk}
\end{figure}

\subsection{Oscillatory response in parallel layers of fluid and solid}
\label{sec:layer}

As schematically illustrated in figure \ref{fig:schem_layer}, 
 we deal with the interactive motions of the three (fluid-solid-fluid) parallel layers 
 bounded with two oscillatory plane walls. 
Here, making comparisons with accurate solutions 
 obtained by means of a sharp interface approach, 
 we examine the validation and verification 
 of the present Eulerian approach, 
 accompanied with a diffuse interface. 
Supposing homogeneity in $x$ direction,
 we may omit the $x$-dependence of any quantity 
 in the theoretical analysis. 
In the numerical simulation, 
 the periodic condition is applied in $x$ direction.
We here treat pure hyperelastic material, i.e., $\mu_s=0$.
The relations between the velocity $v$, 
the displacement $u$ and the shear stress $\sigma$ are given by
\begin{equation}
\partial_t v=\partial_y \sigma,
\label{eq:v_layer}
\end{equation}
\begin{equation}
\partial_t u=v,
\end{equation}
\begin{equation}
\sigma=
\left\{
\begin{array}{ll}
\mu_f\partial_y v&\mbox{  for fluid}\ (L_s< |y|\leq  L_s+L_f),\\
2(c_1+c_2)\partial_y u+
4c_3(\partial_y u)^3&\mbox{  for solid}\ (0\leq |y|< L_s),
\label{eq:constitutive_layer1}
\end{array}
\right.
\end{equation}
with no-slip condition at the upper and lower plates ($y=\pm (L_f+L_s)$)
\begin{equation}
v=\left\{
\begin{array}{ll}
 \hat{V}_W\sin\omega t&{\rm at }\ y= L_f+L_s,\\
-\hat{V}_W\sin\omega t&{\rm at }\ y=-(L_f+L_s).
\end{array}
\right.
\end{equation}
The solid stress expression (\ref{eq:constitutive_layer1})
 indicates that the system involving
 a linear Mooney-Rivlin material with $c_3=0$
 is linear with respect to the displacement $u$, 
 while that involving the Saint Venant-Kirchhoff material
 with $c_3\neq 0$ is nonlinear.

\begin{figure}[t]
\begin{center}

\epsfig{file=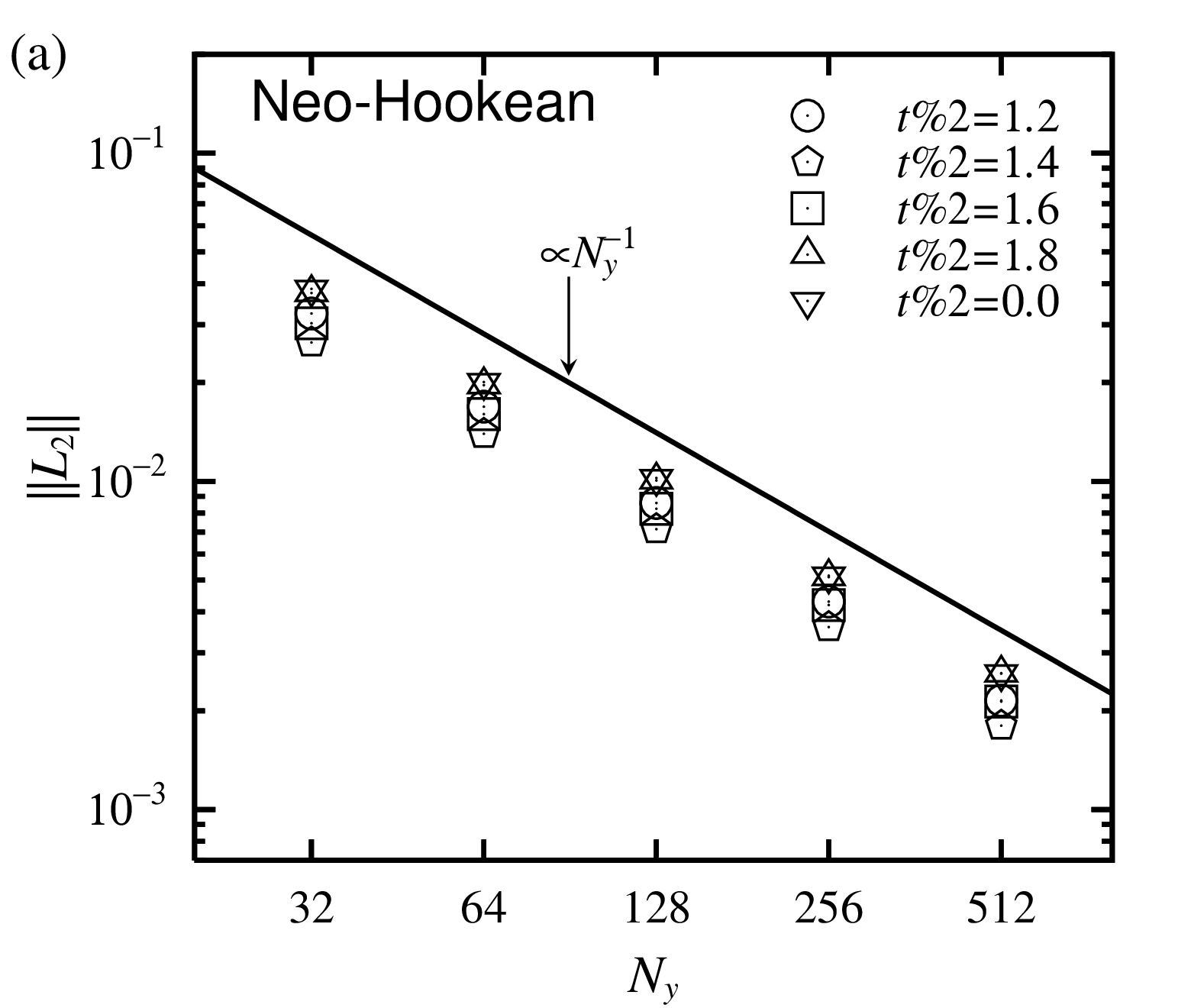,width=5.7cm,angle=270}
\epsfig{file=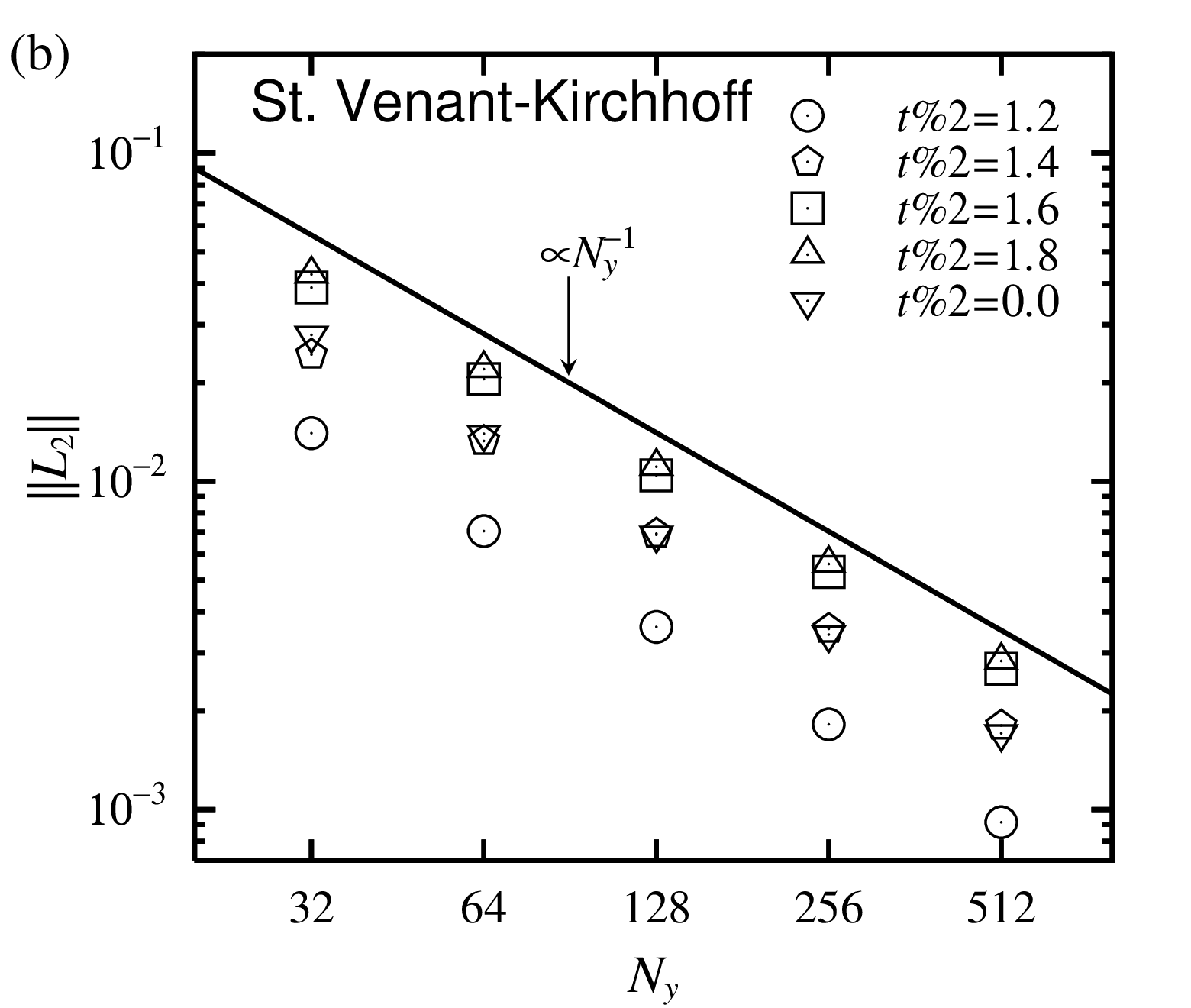,width=5.7cm,angle=270}

\epsfig{file=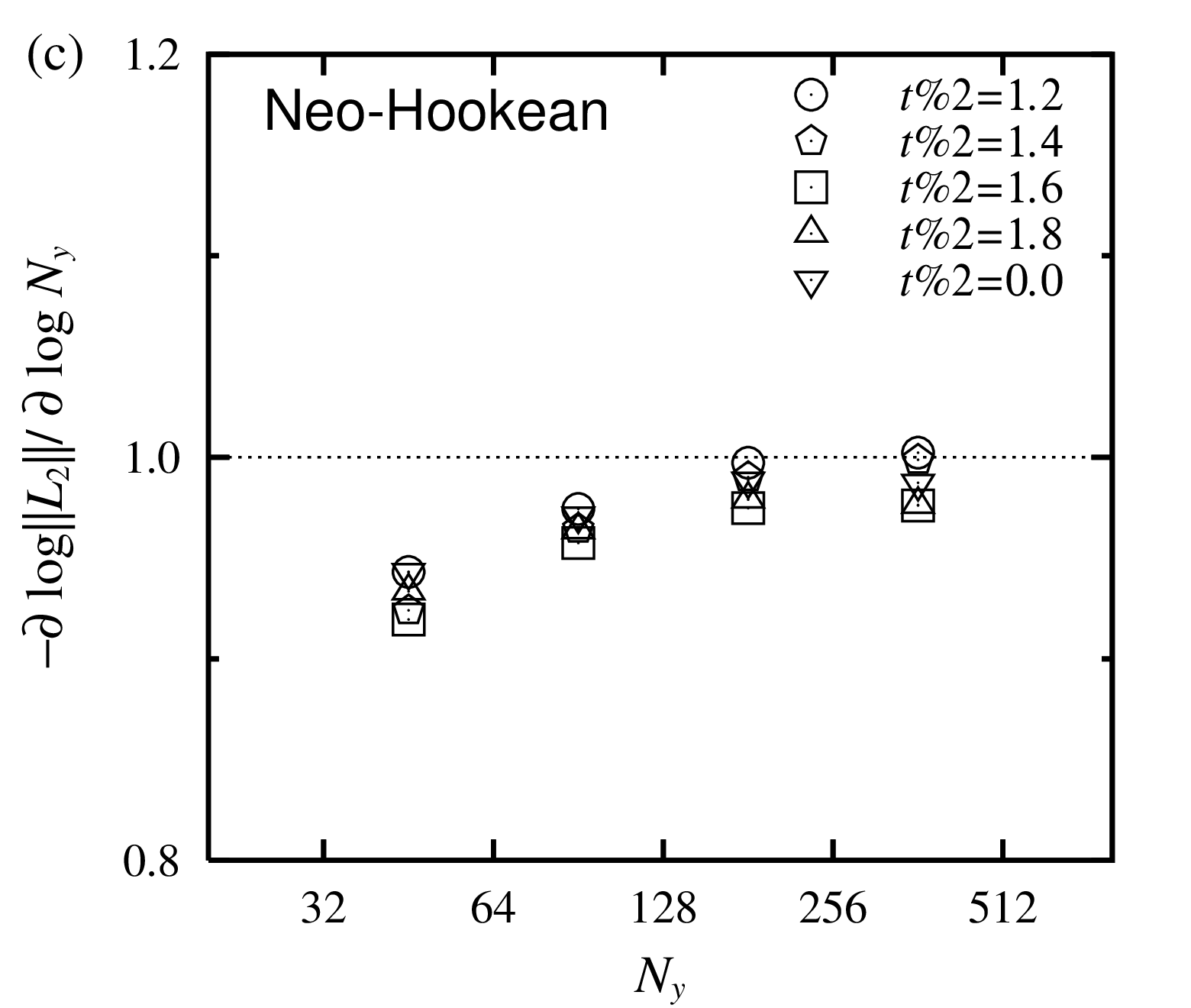,width=5.7cm,angle=270}
\epsfig{file=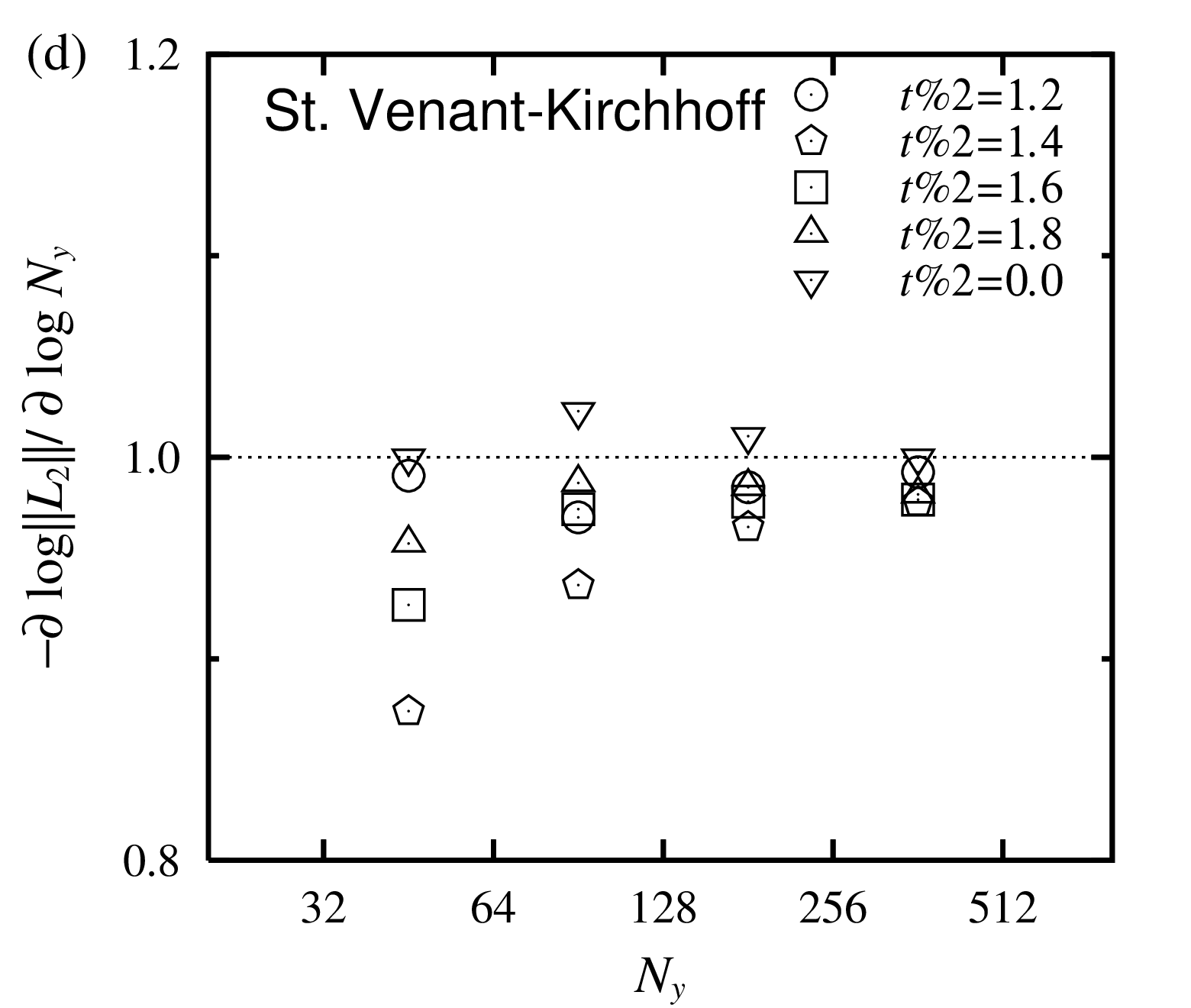,width=5.7cm,angle=270}

\end{center}
\caption{
Upper panels: 
The error of the velocity in L$_2$ norm versus the number $N_y$ of grid points 
 in the vertical ($y$) direction 
 for various temporal phases 
 in the parallel fluid-structure layers. 
The error is determined from the difference 
 between the results of the present Eulerian
 and sharp interface methods based on (\ref{eq:error_layer}). 
Lower panels: 
The absolute slope in the plot of the error versus $N_y$.
The local slope is determined from (\ref{eq:local_slope_error}).
The left and right panels correspond to
 the neo-Hookean and incompressible Saint-Venant Kirchhoff materials, 
 respectively.
The conditions are the same as those of
 figure \ref{fig:layer_nh_svk}.
}
\label{fig:errlayer_l2}
\end{figure}

\begin{figure}[t]
\begin{center}

\epsfig{file=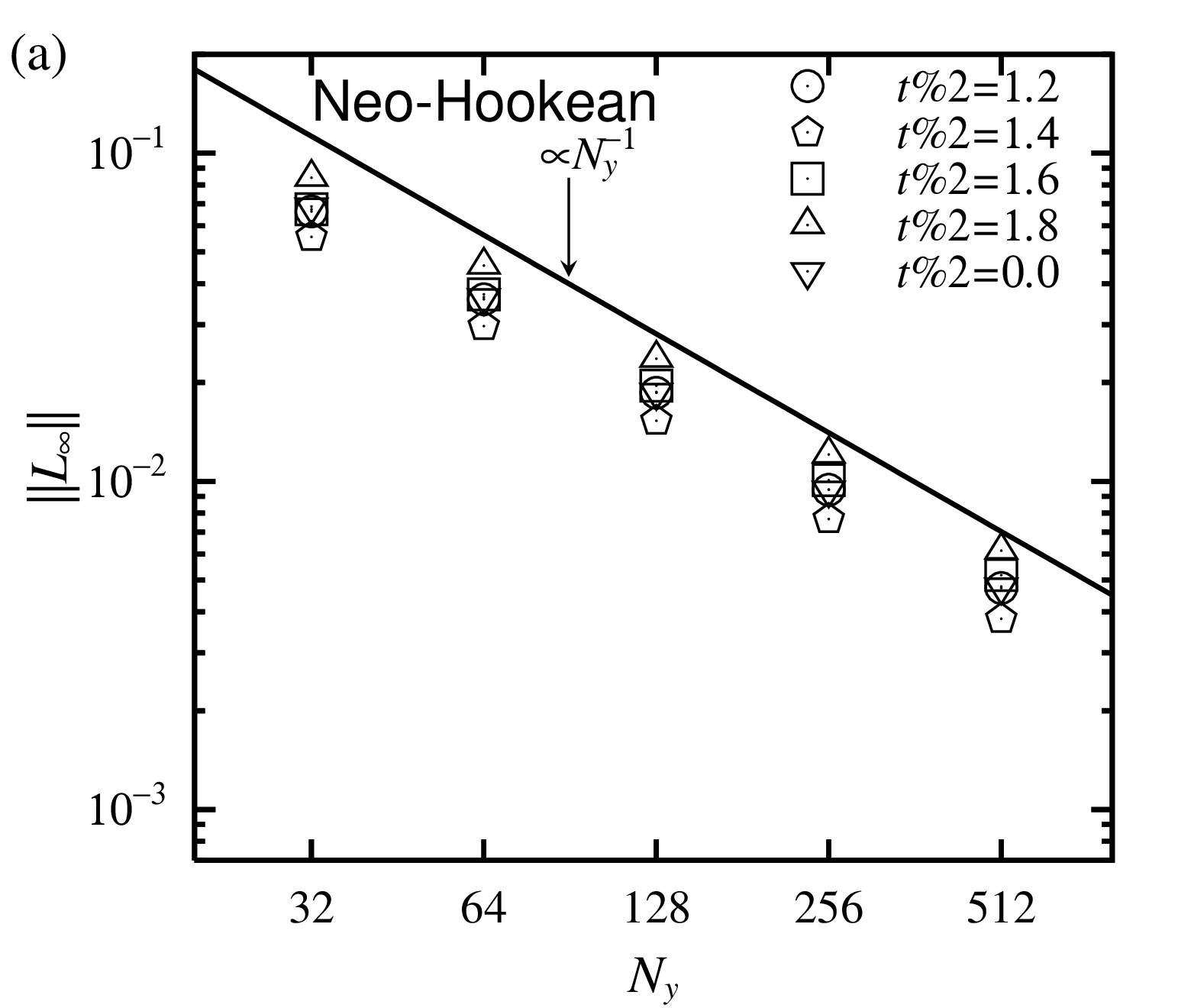,width=5.7cm,angle=270}
\epsfig{file=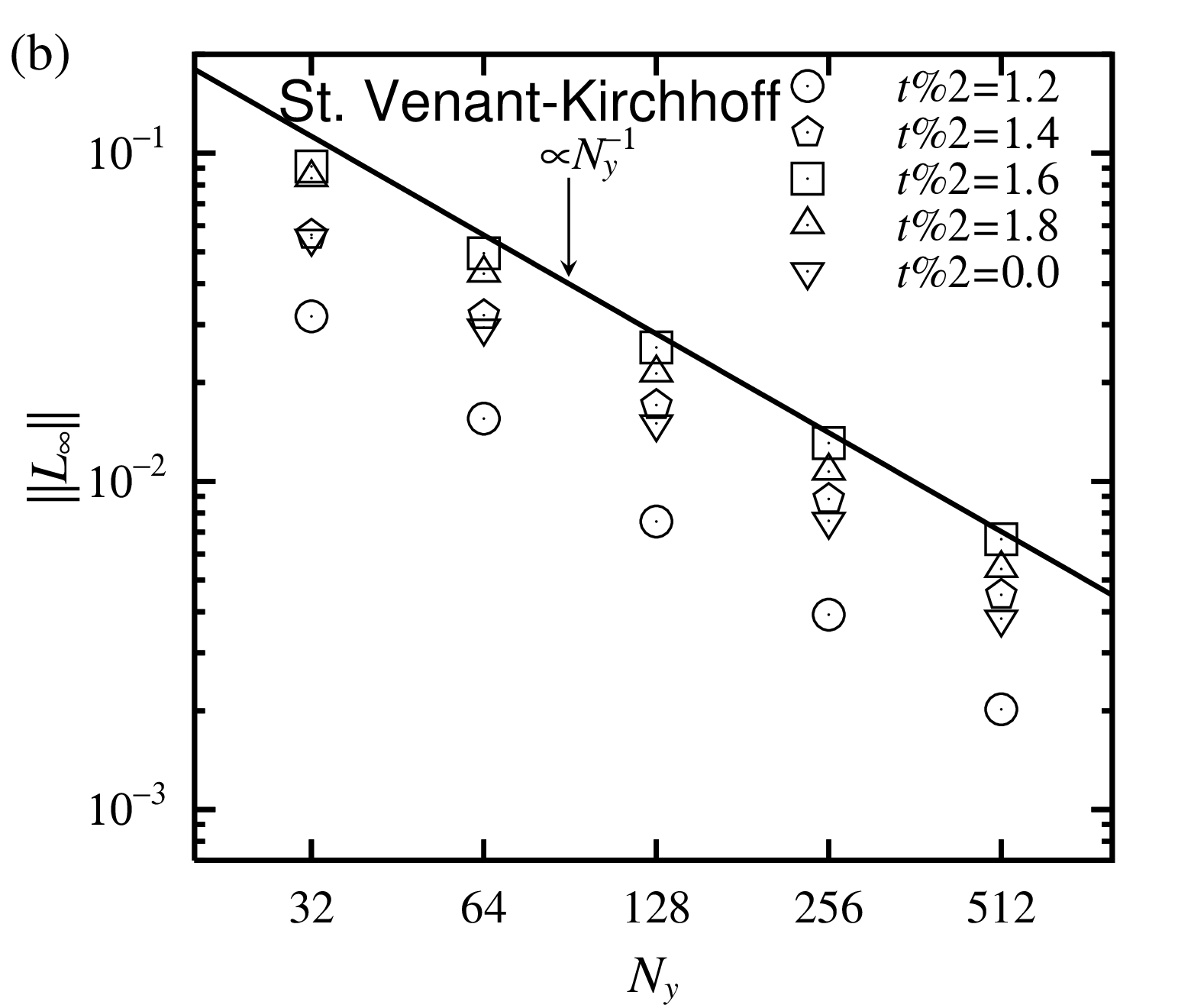,width=5.7cm,angle=270}

\epsfig{file=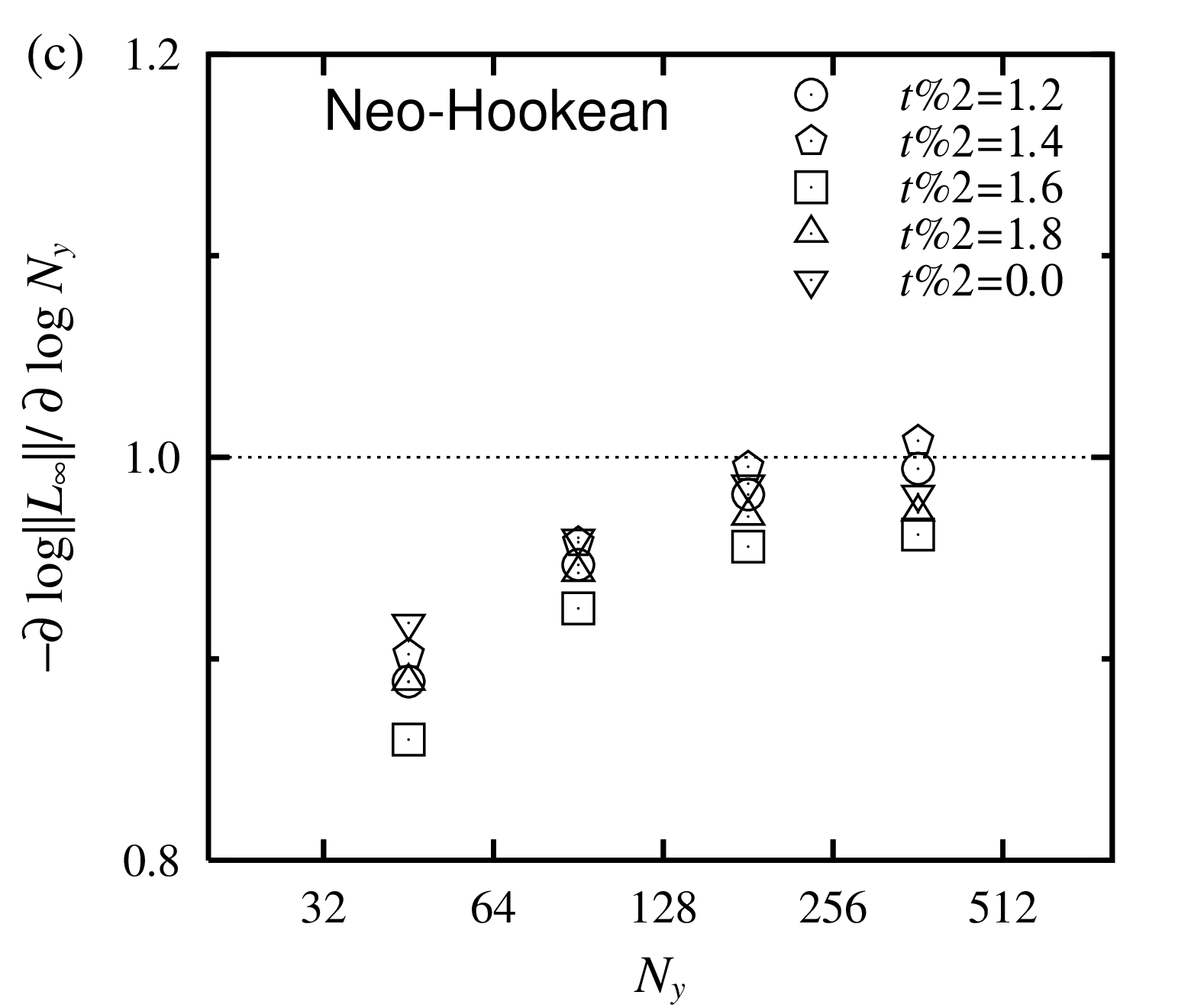,width=5.7cm,angle=270}
\epsfig{file=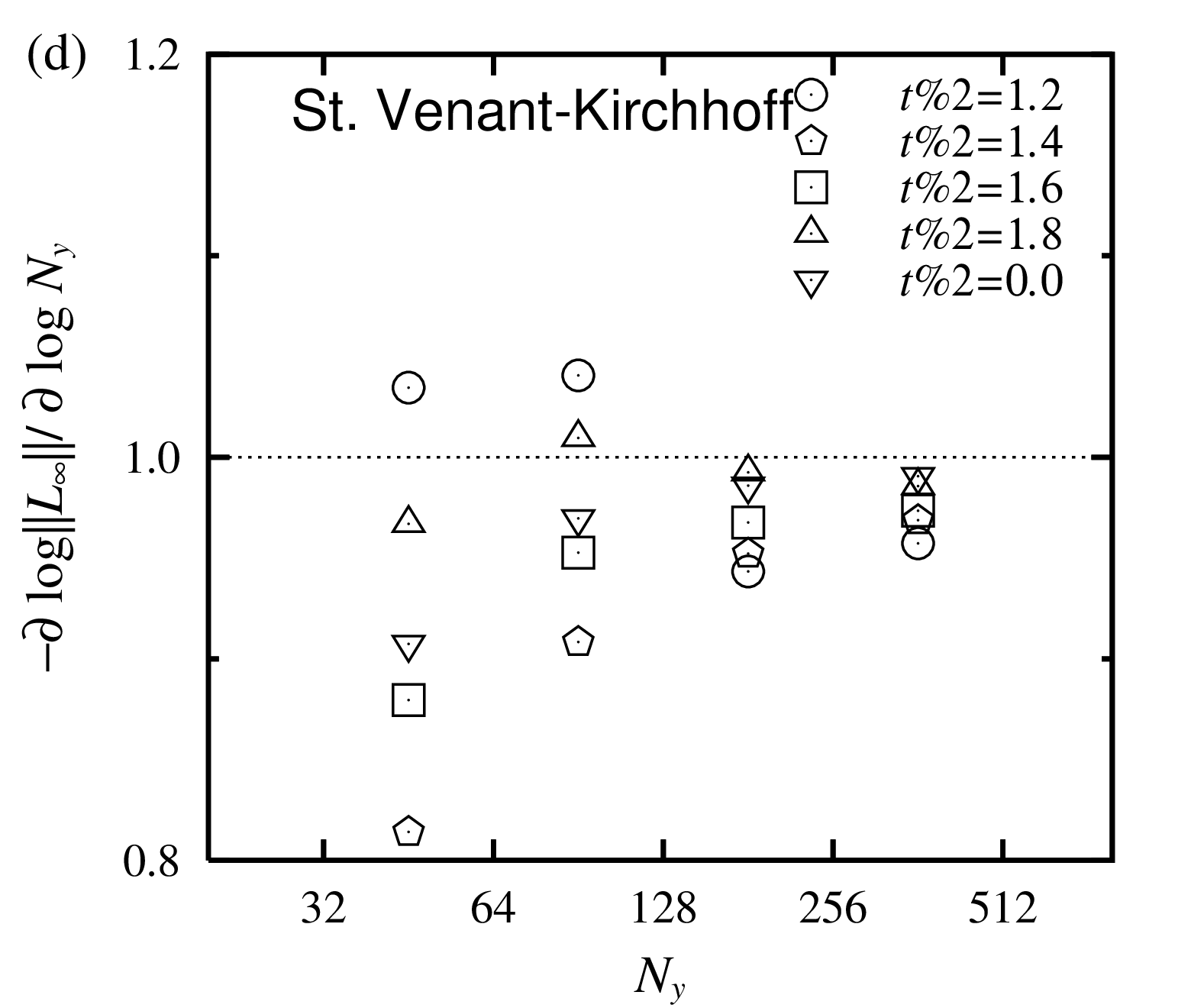,width=5.7cm,angle=270}
\end{center}
\caption{
Same as figure \ref{fig:errlayer_l2}, 
but in L$_\infty$ norm. 
}
\label{fig:errlayer_linf}
\end{figure}

The simulation results based on the present full Eulerian approach
 are compared with the analytical solution obtained 
 by means of the sharp interface approach 
 (see Appendix \ref{sec:spec} for detail). 
We fix the conditions $\rho=1$, $\mu_f=1$, $L_x=8$, $L_s=L_f=0.5$, $\omega=\pi$, 
 and $\hat{V}_w=1$, and vary the number of grid points 
 ($N_x\times N_y =8\times 32$, $8\times 128$, $8\times 512$). 
Initially, the system is at rest.
The total computational period is $t=40$,  corresponding to $20$ cycles, 
 and the sampling is performed within the last one cycle i.e., $t\in (38,40]$.
Figure \ref{fig:layer_nh_svk} shows
 the velocity profiles for the neo-Hookean model 
  ($G=5$, i.e., $c_1=2.5$, $c_2=c_3=0$)
and for the Saint Venant-Kirchhoff one
 ($\lambda_{\mbox{\tiny Lam\'e}}^s=7.5$, 
$\mu_{\mbox{\tiny Lam\'e}}^s=5$, 
i.e., $c_1=5.0$, $c_2=-2.5$, $c_3=2.1875$).
The present simulation results 
 converge to the sharp interface solutions 
 with the higher spatial resolution. 
Figure \ref{fig:layer_nh_svk}(d) shows
 the obvious difference in profile between
 the linear ($c_3=0$) and nonlinear ($c_3\neq 0$) solutions
 at the phase of $t\%2=1.8$ (here $\%$ stands for the remainder). 
The simulation results in figure \ref{fig:layer_nh_svk}(d)
 clearly get closer to the nonlinear solution 
 with increasing the number of grid points, 
 indicating that the nonlinearity in the solid constitutive law
 is reasonably captured in the present Eulerian approach. 

The accuracy in the fluid-structure coupling
 is quantified by the errors in L$_2$ and L$_\infty$ norms, 
 which are respectively defined as 
\begin{equation}
\begin{split}
||L_2||(N_y)=&
\left\{
\frac{1}{N_y}\sum_{j=1}^{N_y}(v_j^{(n)}-v_j^{(a)})^2
\right\}^{\frac{1}{2}},\\
||L_\infty||(N_y)=&
\max_{j\in [1,N_y]}
\left|v_j^{(n)}-v_j^{(a)}\right|,
\label{eq:error_layer}
\end{split}
\end{equation}
where $v_j^{(n)}$ and $v_j^{(a)}$, respectively,
 denote the present result and the sharp interface solution 
 on the node $y_j=(j-\frac{1}{2})/N_y$. 
The log-log plots of the L$_2$ and L$_\infty$ errors 
 versus the number of grid points 
 are shown in figure \ref{fig:errlayer_l2}(a)(b) and figure \ref{fig:errlayer_linf}(a)(b),
respectively. 
The local slopes are obtained therefrom,
 and shown in figure \ref{fig:errlayer_l2}(c)(d) and figure \ref{fig:errlayer_linf}(c)(d).
Here, the local slopes are determined from the following approximations
\begin{equation}
\begin{split}
\frac{\partial \log \left(||L_2||\right)}{\partial \log \left(N_y\right)}
(N_y)
\approx&
\frac{\log\left(||L_2||(\sqrt{2}N_{y})\right)
-\log \left(||L_2||(N_{y}/\sqrt{2})\right)}
{\log 2},
\\
\frac{\partial \log \left(||L_\infty||\right)}{\partial \log \left(N_y\right)}
(N_y)
\approx&
\frac{\log\left(||L_\infty||(\sqrt{2}N_{y})\right)
-\log \left(||L_\infty||(N_{y}/\sqrt{2})\right)}
{\log 2}.
\label{eq:local_slope_error}
\end{split}
\end{equation}
The slope indicates the degree of the accuracy
 in the fluid-structure coupling.
In both the L$_2$ error (corresponding to a global indicator)
 and the L$_\infty$ error (corresponding to a local maximum indicator)
 are nearly proportional to $N_y^{-1}$.
Since the second-order finite difference
 is applied to describing the spatial derivatives,
 this near-first order trend must have resulted from
 the mixture stress expression (\ref{eq:mix_stress01})
 involving the first-order accuracy locally at the interface,
 which dominates the global degree of accuracy. 
Note that we also investigated the grid convergence of
 the shear stress, and confirmed that
 the accuracy at the interface is 
 of the first-order with respect to the grid size.

\begin{figure}[h]
\begin{center}

\epsfig{file=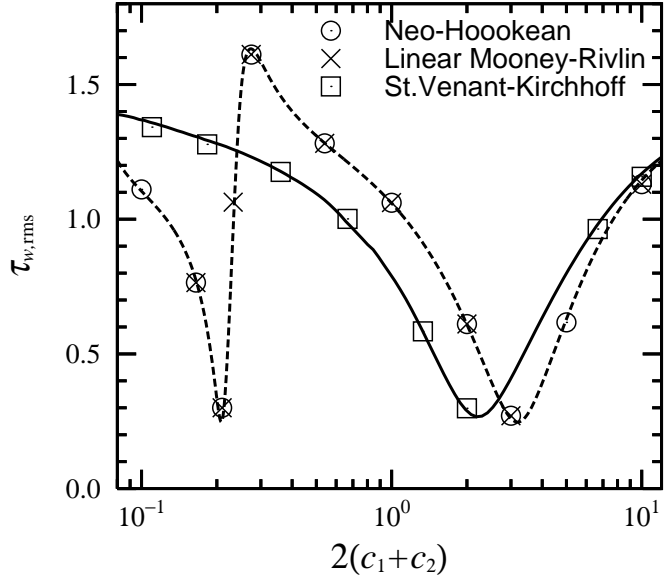,width=8cm,angle=270}
\end{center}
\caption{
Response curve of the skin friction amplitude $\tau_{w,{\rm rms}}$
 in the motion of the
 parallel fluid-structure layers as a function of $2(c_1+c_2)$
 with  $\mu_f=1$, $L_s=L_f=0.5$, $\omega=\pi$, and $\hat{V}_w=1$.
The symbols correspond to 
 the results based on the present Eulerian method with the $8\times 256$ mesh.
The circles, crosses, and squares represent 
 results of the neo-Hookean ($c_2=c_3=0$),
 linear Mooney-Rivlin ($c_2=c_1/2$, $c_3=0$) and
 incompressible Saint Venant-Kirchhoff ($c_2=-c_1/2$, $c_3=7c_1/16$)
 models, respectively.
The curves represent the sharp interface solutions,
 which are determined by means of the spectral methods
 (see Appendix \ref{sec:spec}).
The dashed curve represents the linear solution 
 for the neo-Hookean and linear Mooney-Rivlin materials with $c_3=0$, 
 while the solid curve the nonlinear solution 
 for the above-mentioned Saint Venant-Kirchhoff material.
}
\label{fig:comp_tau}
\end{figure}

Figure \ref{fig:comp_tau} shows the sensitivity
 of the wall friction amplitude $\tau_{w,{\rm rms}}$
  on moduli.
We fix the conditions  $\mu_f=1$, $\mu_s=0$, $L_s=L_f=0.5$, $\omega=\pi$, and $\hat{V}_w=1$.
Likewise the computations in figure \ref{fig:layer_nh_svk},
 the total computational period is set to $20$ cycles.
The root-mean-square of $\tau_w$ was sampled
 over the last one cycle. 
The results of the linear Mooney-Rivlin model with ($c_2=c_1/2$, $c_3=0$)
 as well as the neo-Hookean ($c_2=c_3=0$) and 
 Saint Venant-Kirchhoff ($c_2=-c_1/2$, $c_3=7c_1/16$) models 
 are plotted as a function of $2(c_1+c_2)$.
Because the deformed motion of solid behaves 
 like as the spring-mass system,
 the plot of the wall friction amplitude versus $2(c_1+c_2)$
 reveals the non-monotonous resonant behavior. 
As long as the solid strain is sufficiently small, 
 the nonlinearity involved in the constitutive law is negligible, 
 and therefore the linear assumption is justified.
Indeed, the curve of the nonlinear solution approaches
 the linear solution with increasing $2(c_1+c_2)$
 since the solid strain is suppressed for the stiffer material.
By contrast, for the smaller $2(c_1+c_2)$, 
 the discrepancy between the linear and nonlinear solutions
 becomes more obvious. 
It is because the larger strain makes the nonlinear system
 effectively stiffer as implied by (\ref{eq:constitutive_layer1}). 
All the results of the present Eulerian approach are 
 in good agreement with the sharp interface solution.
The present approach is confirmed to capture
 the resonance behavior resulting from the dynamic interaction between
 the fluid and solid motions,
 and the nonlinearity in the solid constitutive law.

\subsection{Comparison with independently conducted FSI analyses}

We here make comparisons with two well-validated FSI analyses.
In the constitutive law for (visco-)hyperelastic material, 
 one has $\mu_s=\mu_f$, $c_1\neq 0$ and $c_2=c_3=0$,
 and the other has $\mu_s=0$, $c_2\neq 0$ and $c_1=c_3=0$.

\begin{figure}[h]
\begin{center}

\epsfig{file=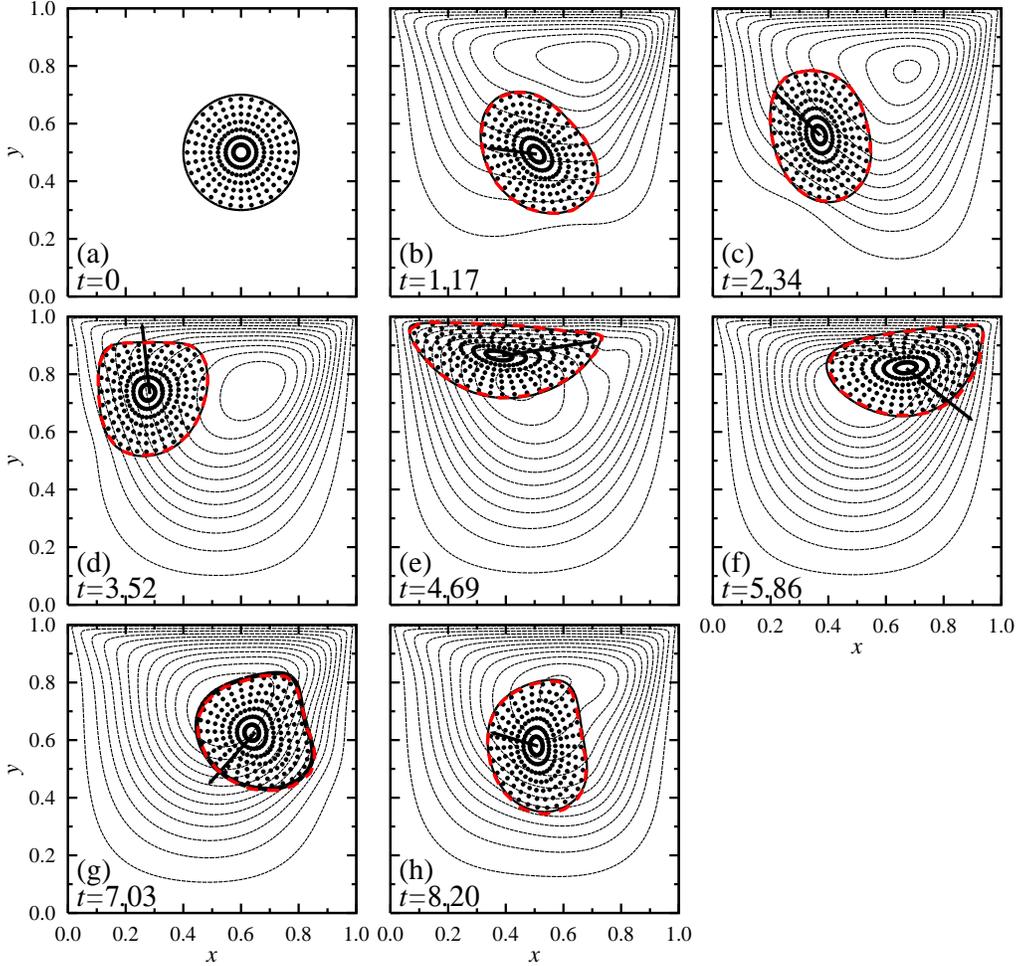,width=13cm,angle=270}
\end{center}
\caption{
Comparison of the solid deformation 
in the lid-driven flow with the simulation result \cite{zha2008}. 
The dashed outline represents the result of Zhao {\em et al}. \cite{zha2008}, 
 in which the Lagrangian tracking approach 
 was employed to describe the solid deformation. 
The solid outline, the dotted material points and the streamlines 
 correspond to the present simulation results 
 based on the full Eulerian approach
 with a mesh $1024\times 1024$. 
}
\label{fig:comp_zhao}
\end{figure}

\subsubsection{A solid motion in a lid-driven cavity flow}
\label{sec:zhao}

We perform full Eulerian simulations
 of deformable solid motion in a lid-driven cavity
 with the same setup and conditions as Zhao {\em et al.}\cite{zha2008},
 who employed mixed Lagrangian and Eulerian approach. 
The initial setup is schematically illustrated
 in figure \ref{fig:comp_zhao}(a).
The size of the cavity is $L_x\times L_y=1\times 1$. 
Initially, the system is at rest. 
The unstressed solid shape is circular
 with a radius of $0.2$, and centered at $(0.6,0.5)$.
At $t=0$, to drive the fluid and solid motions, 
 the top wall starts to move at a speed of $V_W=1$
 in $x$ direction.
The no-slip condition is imposed on the walls. 
The solid component is neo-Hookean material.
The material properties are
 $\rho=1$, $\mu_f=\mu_s=10^{-2}$, $c_1=0.05$ and $c_2=c_3=0$.

Figure \ref{fig:comp_zhao} visualizes the particle deformation
 and the flow field for eight consecutive time instants.
The dashed curve in figure \ref{fig:comp_zhao} 
 represents the outline of the particle obtained 
 by Zhao {\em et al}. \cite{zha2008},
 in which they computed the solid deformation on the Lagrangian mesh.
The solid lines represent the instantaneous particle shapes, 
 corresponding to the isoline at $\phi_s=1/2$,
 obtained by the present full Eulerian simulation. 
The dotted material points are tracked
 just to transfer images of the particle deformation,
 but we did not use these material points
 for computing solid stress and strain. 
The particle moves and deforms driven by the fluid flow,
 and exhibits highly deformed shape when the particle approaches the wall. 
It should be noticed that no special artifact 
 for avoiding a particle-wall overlap
 is implemented into the present method
 because the particle-wall hydrodynamic repulsion is likely to
 be brought by the soft lubrication effect \cite{sko2005}
 due to the geometry change via the particle deformation.
The solid shapes obtained by the present Eulerian simulation 
 are in excellent agreement with the well-validated result \cite{zha2008}. 

\begin{figure}[h]
\begin{center}
\epsfig{file=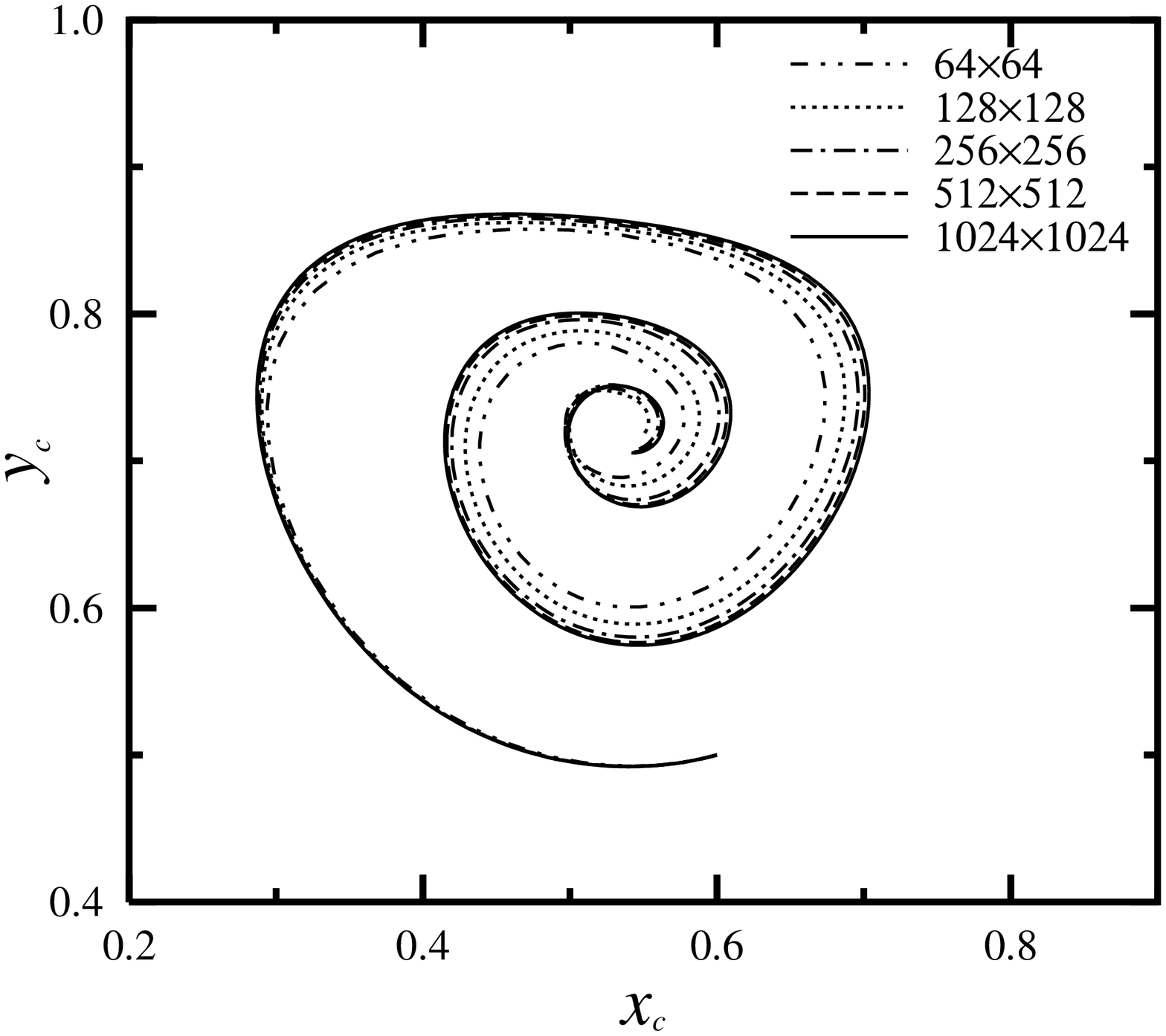,width=8cm,angle=270}
\end{center}
\caption{
Trajectories of the solid centroid 
in the lid-driven flow 
in a time range  $t\in [0,20]$
for various number of grid points.
}
\label{fig:trajectory_zhao}
\end{figure}

\begin{figure}[h]
\begin{center}
\epsfig{file=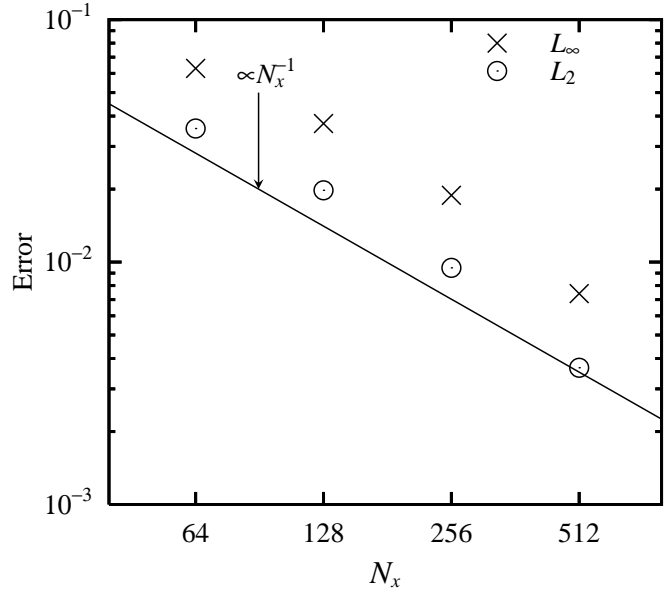,width=8cm,angle=270}
\end{center}
\caption{
The errors of the particle centroid in L$_2$ norm and in L$_\infty$ norm
 versus the number $N_x$ of grid points in the lid-driven flow.
}
\label{fig:err_zhao_xc}
\end{figure}

In addition to the comparative study,
 we address grid convergence issues below.
We trace the centroid ${\bm x}_c=(x_c,y_y)$ of the particle, 
 which is evaluated from the approximation
\begin{equation}
{\bm x}_c(t)
\approx
\frac{\displaystyle\sum_{i=1}^{N_x}
\sum_{j=1}^{N_y}\Delta_x\Delta_y\ {\bm x}_{i,j}\ \phi_{s}({\bm x}_{i,j}, t)
}{\displaystyle
\sum_{i=1}^{N_x}
\sum_{j=1}^{N_y}\Delta_x\Delta_y\ \phi_{s}({\bm x}_{i,j}, t)
}.
\label{eq:xc}
\end{equation}
Figure \ref{fig:trajectory_zhao} 
 shows the trajectory of the centroid ${\bm x}_c$ in a time range of $t\in [0,20]$
 for various number of grid points 
 ($N_x\times N_y = 64\times 64$, $128\times 128$, $256\times 256$, $512\times 512$, $1024\times 1024$). 
The trajectories clearly exhibit a convergent trend to the curve of the highest spatial resolution.

To quantify the grid convergence behavior, 
 the distance of the particle centroid ${\bm x}_c$ with respect to
 that of the highest resolution $N_x = 1024$
 is monitored.
We evaluate the errors in L$_2$ and L$_\infty$ norms respectively from
\begin{equation}
\begin{split}
||L_2||(N_x)=&
\left\{
\frac{1}{T}
\int_0^T\!\!\!{\rm d}t\ 
|{\bm x}_c(t,N_x)-{\bm x}_c(t,N_x=1024)|^2
\right\}^{\frac{1}{2}},\\
||L_\infty||(N_x)=&\max_{t\in [0,T]}
|{\bm x}_c(t,N_x)-{\bm x}_c(t,N_x=1024)|.
\label{eq:error_zhao_xcent}
\end{split}
\end{equation}
Figure \ref{fig:err_zhao_xc} shows
 the L$_2$ and L$_\infty$ errors as a function of $N_x$.
Both the errors are nearly proportional to $N_x^{-1}$, 
 indicating the first-order accuracy.
The first-order accuracy involved
 in the fluid-structure coupling 
 as described in \S \ref{sec:layer}
 is reflected on the slopes in these plots. 

In addition to the particle centroid, 
 to further examine the grid convergence behavior,
 we here perform modal analyses of the particle deformation.
The distance from the particle centroid ${\bm x}_c$ to 
 the interface ${\bm x}_I = (x_I, y_I)$ is written as
\begin{equation}
R(\theta)=|{\bm x}_I-{\bm x}_c|,
\end{equation}
where $\theta$ is found to satisfy the relations
\begin{equation}
\cos\theta = \frac{x_I-x_c}{|{\bm x}_I-{\bm x}_c|},\ \ \ 
\sin\theta = \frac{y_I-y_c}{|{\bm x}_I-{\bm x}_c|}.
\end{equation}
The distance is written in a Fourier series form
\begin{equation}
R(\theta)=R_0+\sum_{n=1}^{\infty}(R_{cn}\cos n\theta +R_{sn}\sin n\theta),
\label{eq:mode1}
\end{equation}
where $R_n$ denotes the $n$-th order deformation mode.
The deformation modes are uniquely determined
 via the orthogonality in the cosine and sine functions
 from definite integrals
\begin{equation}
\begin{split}
R_0=&\frac{1}{2\pi}\int_{0}^{2\pi}\!\!\!{\rm d}\theta\ R(\theta),\\
R_{cn}=&\frac{1}{\pi}\int_{0}^{2\pi}\!\!\!{\rm d}\theta\ R(\theta)\cos n\theta,\ \ \ 
R_{sn}= \frac{1}{\pi}\int_{0}^{2\pi}\!\!\!{\rm d}\theta\ R(\theta)\sin n\theta.
\label{eq:mode2}
\end{split}
\end{equation}
\begin{figure}[h]
\begin{center}
\epsfig{file=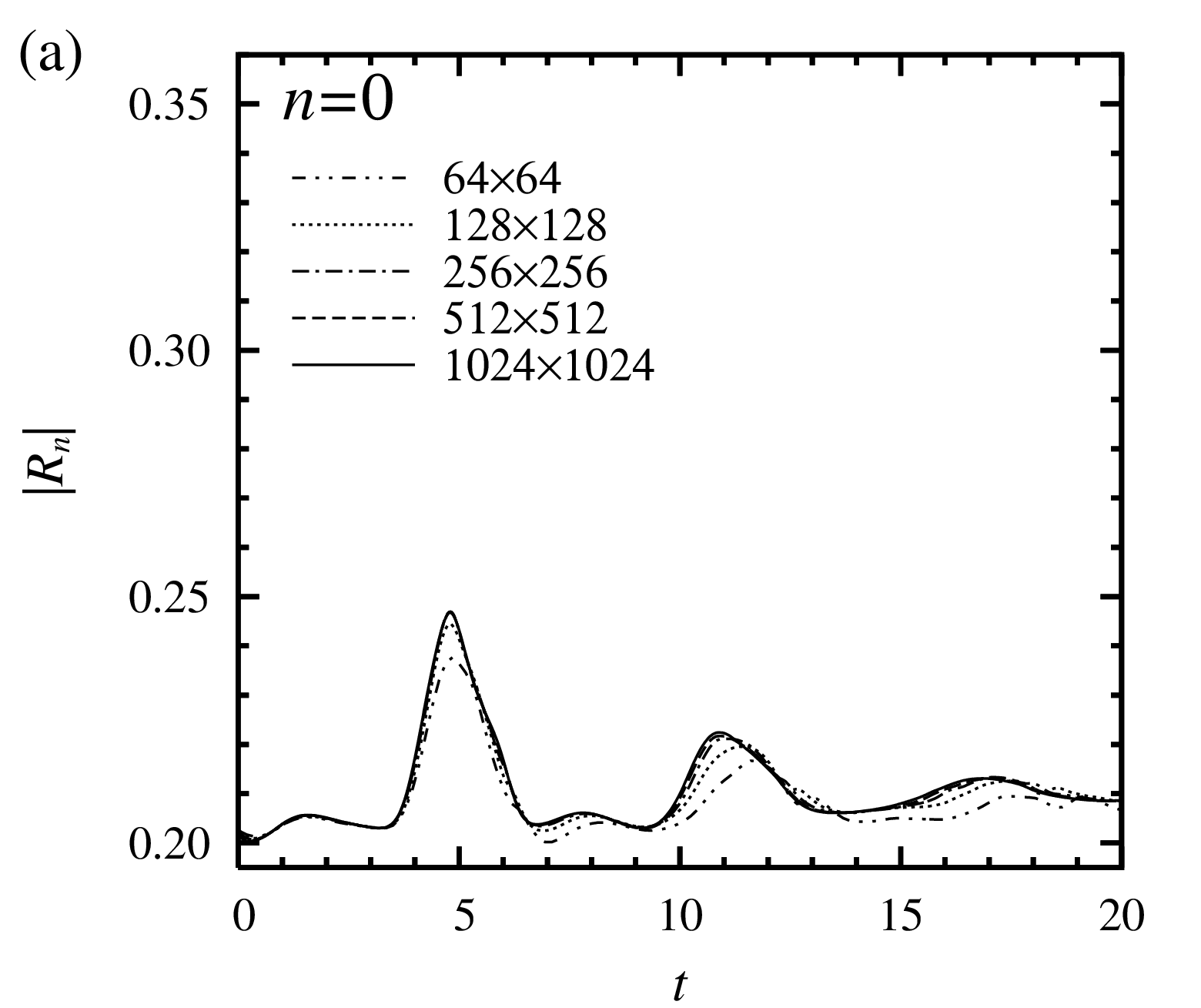,width=5.7cm,angle=270}
\epsfig{file=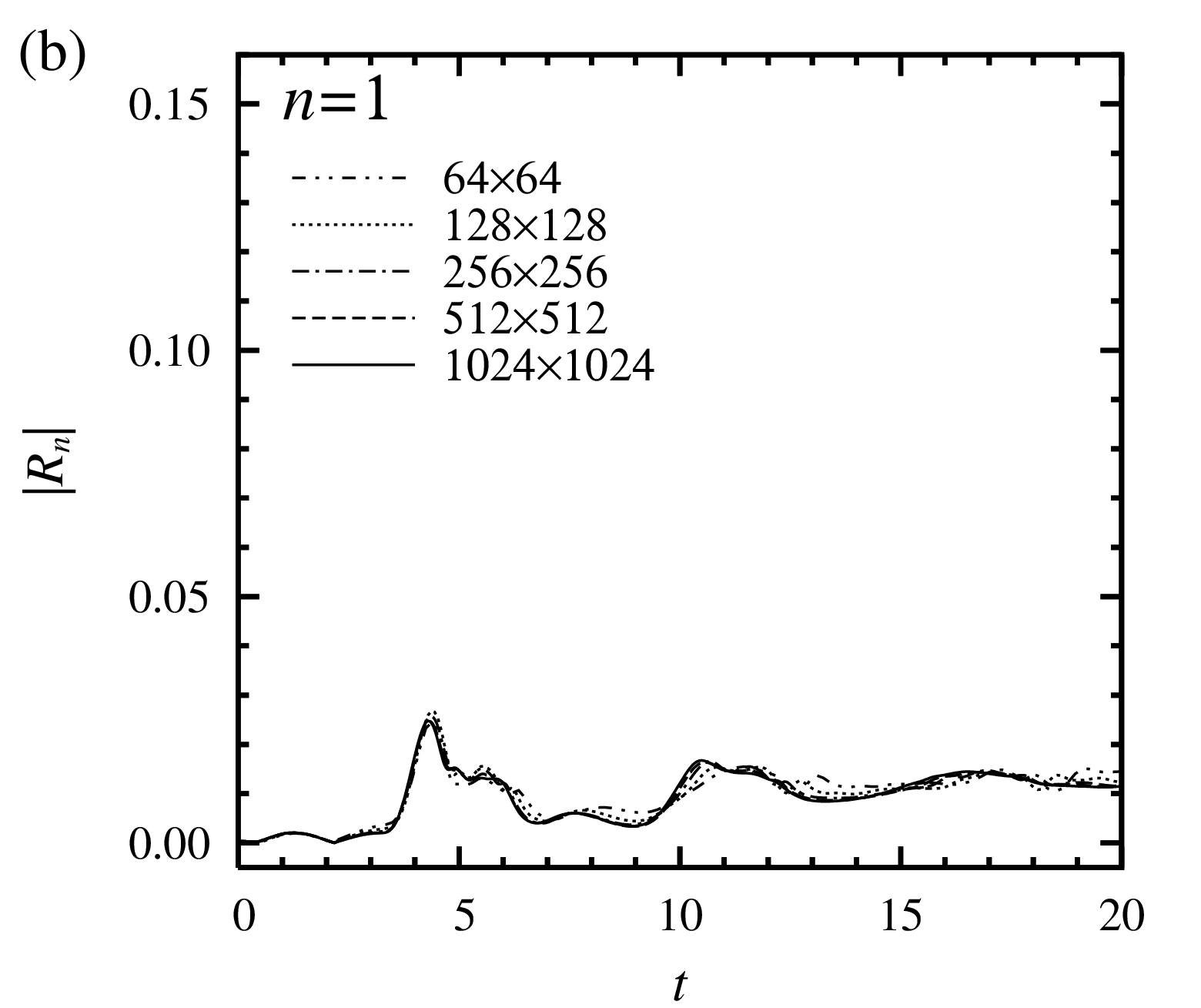,width=5.7cm,angle=270}

\epsfig{file=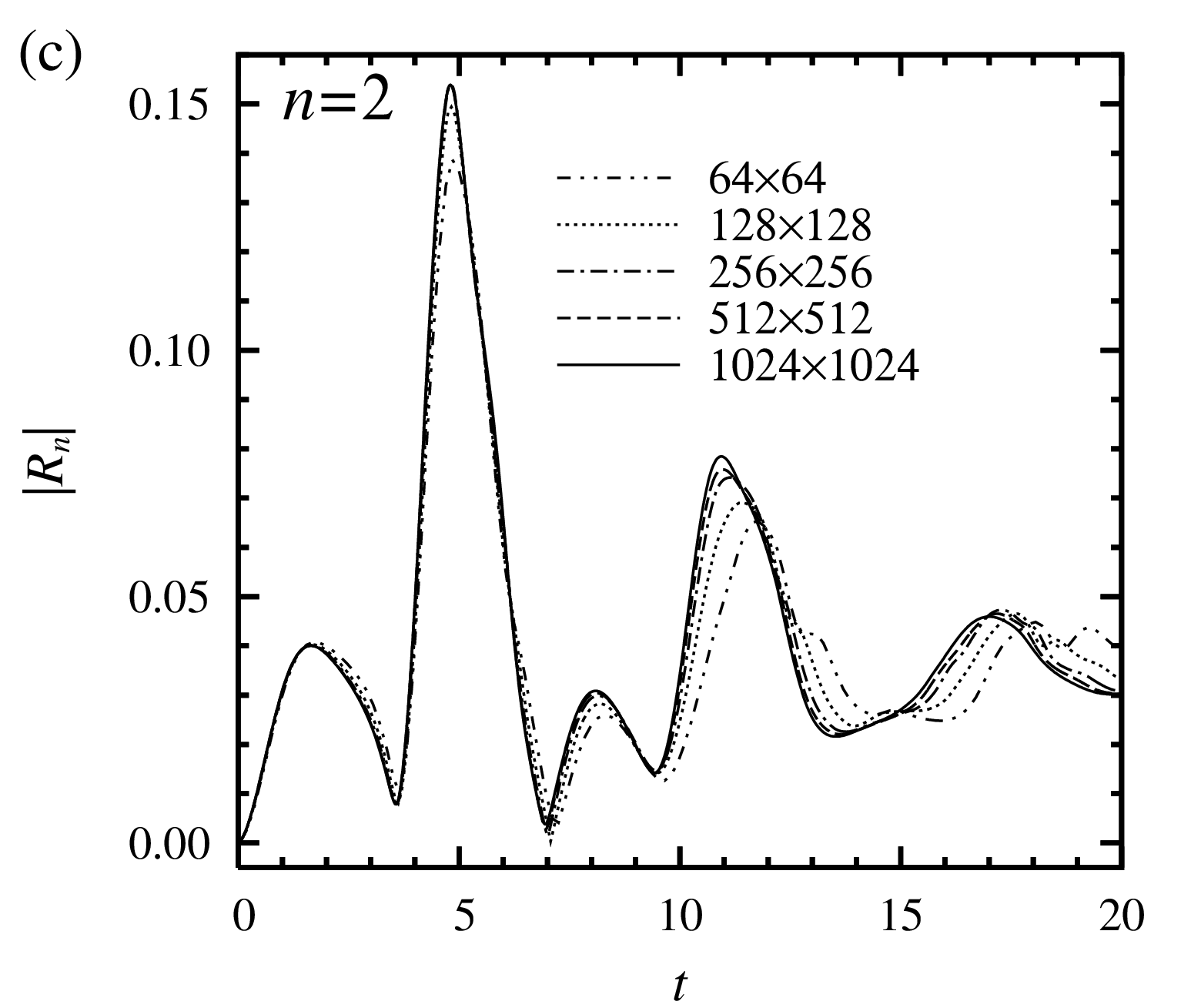,width=5.7cm,angle=270}
\epsfig{file=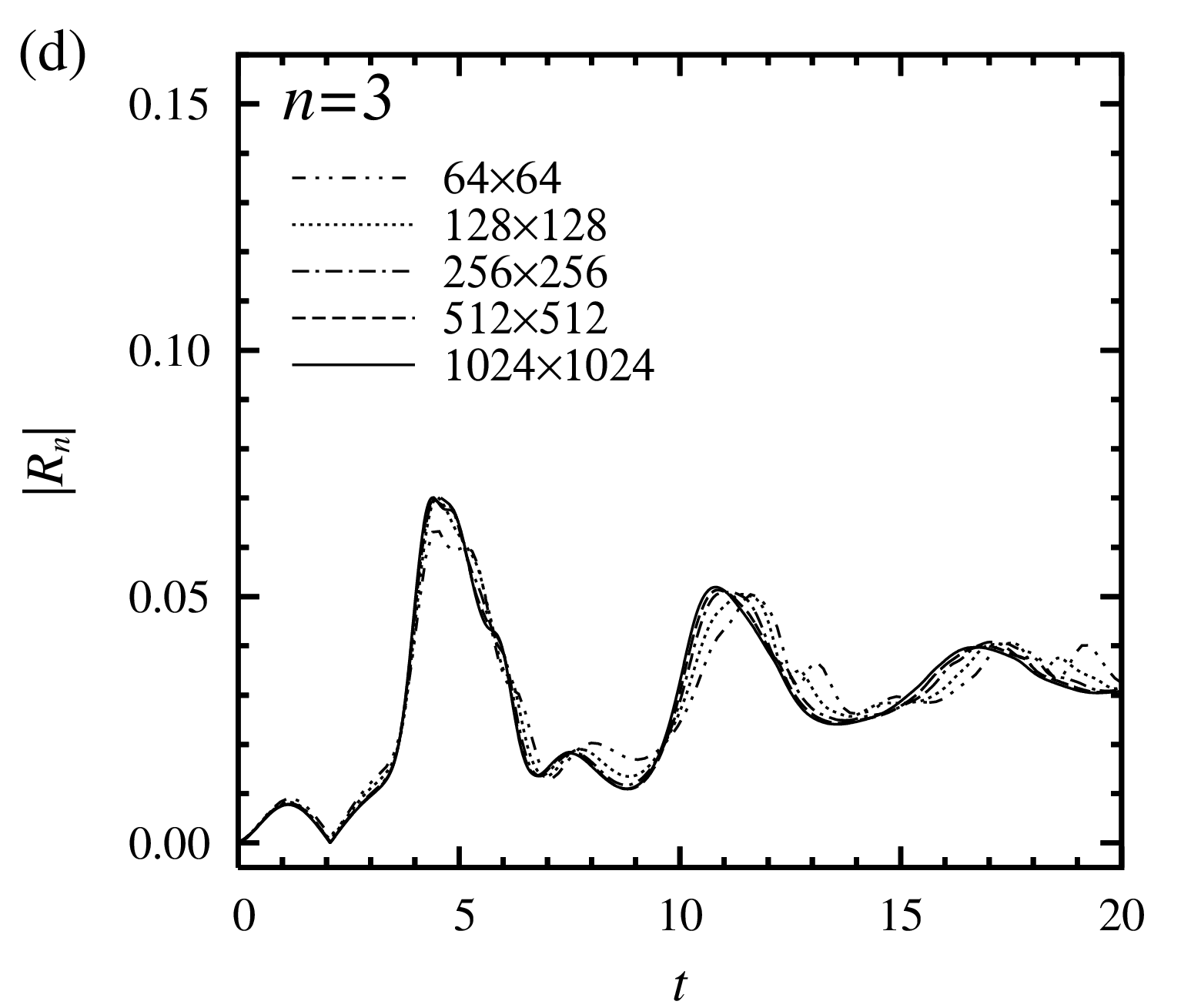,width=5.7cm,angle=270}
\end{center}
\caption{
Time history of the $n$-th order modal amplitude $|R_n|$
 of the particle deformation
 in the lid-driven cavity flow
 for various number of grid points.
(a) $n=0$, (b) $n=1$, (c) $n=2$, and (d) $n=3$.
}
\label{fig:deform_mode_zhao}
\end{figure}

\noindent
In the present Eulerian approach, 
 the fluid and solid phases are distinguished 
 by the solid volume fraction $\phi_s$, 
 and there is no explicit quantity describing the angular profile of $R$.
Instead of the circumferential integral in (\ref{eq:mode2}),
 we will apply the area integral to evaluating the deformation mode.
Let us consider the following relation for a function $f(\theta)$:
\begin{equation}
\int_0^{2\pi}\!\!\!{\rm d}\theta\ R(\theta)f(\theta)
=
\int\!\!\!\int_{\Omega}\!{\rm d}^2{\bm x}\ \delta(|{\bm x}-{\bm x}_I|)\
f(\theta),
\label{eq:mode3}
\end{equation}
where $\delta$ stands for a one-dimensional Dirac's delta function, 
 which is related to the gradient of the solid indicator function $I_s$, namely, 
$$
\nabla I_s = -{\bm n} \delta (|{\bm x}-{\bm x}_I|),
$$
where ${\bm n}$ denotes the unit normal vector pointing towards the fluid
 and is given by 
$$
{\bm n}=-\frac{\nabla I_s}{|{\nabla I_s}|}.
$$
Hence, the delta function used in (\ref{eq:mode3}) is expressed as
$$
\delta(|{\bm x}-{\bm x}_I|)=|{\nabla I_s}|.
$$
Implementing (\ref{eq:mode3}) in the finite difference approach, 
 one must smooth the delta function at the grid scale. 
Upon using the solid volume fraction $\phi_s$, 
 which is regarded as the solid indicator function smoothed at the grid scale,
 we obtain the approximation
\begin{equation}
\delta(|{\bm x}-{\bm x}_I|)
\approx 
|\nabla \phi_s|,
\label{eq:mode4}
\end{equation}
and then evaluate the definite integrals in the summation form
\begin{equation}
\int_0^{2\pi}\!\!\!{\rm d}\theta\ R(\theta) f(\theta)
\approx
\sum_{i=1}^{N_x}
\sum_{j=1}^{N_y}\Delta_x\Delta_y|\nabla \phi_{s,i,j}|f(\theta_{i,j}),
\label{eq:mode5}
\end{equation}
together with (\ref{eq:xc})
to find $\theta_{i,j}=\theta({\bm x}_{I,i,j},{\bm x}_c)$.
For $n\geq 1$, we write the modal amplitude as
$|R_n|=\sqrt{R_{cn}^2+R_{sn}^2}$.

Figure \ref{fig:deform_mode_zhao} shows 
 the temporal evolutions of the modal amplitude
 $|R_n|$ ($n=0$, $1$, $2$ and $3$) of the particle deformation
 for various number of grid points.
The largest elongation ($n=2$ mode) of the particle is observed
 at about $t=5$ when the particle is in the proximity of the moving wall,
 and the synchronized increases in $|R_n|$ are found
 for different modes.
With increasing the number of grid points,
 the profiles for each mode $n$ settle to corresponding convergent curves,
 suggesting the verification with respect to the deformation of the particles.

\begin{figure}[h]
\begin{center}
\epsfig{file=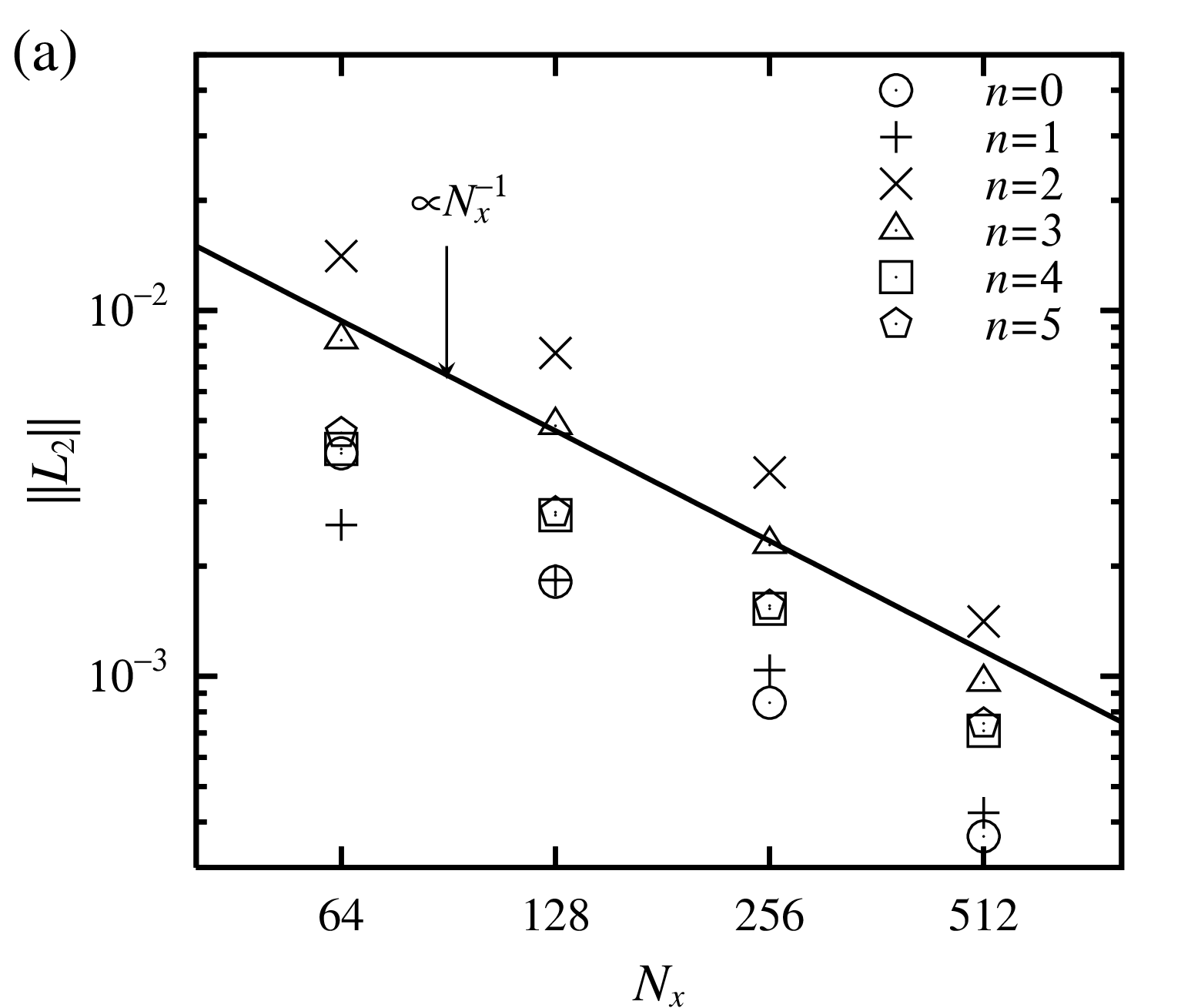,width=5.7cm,angle=270}
\epsfig{file=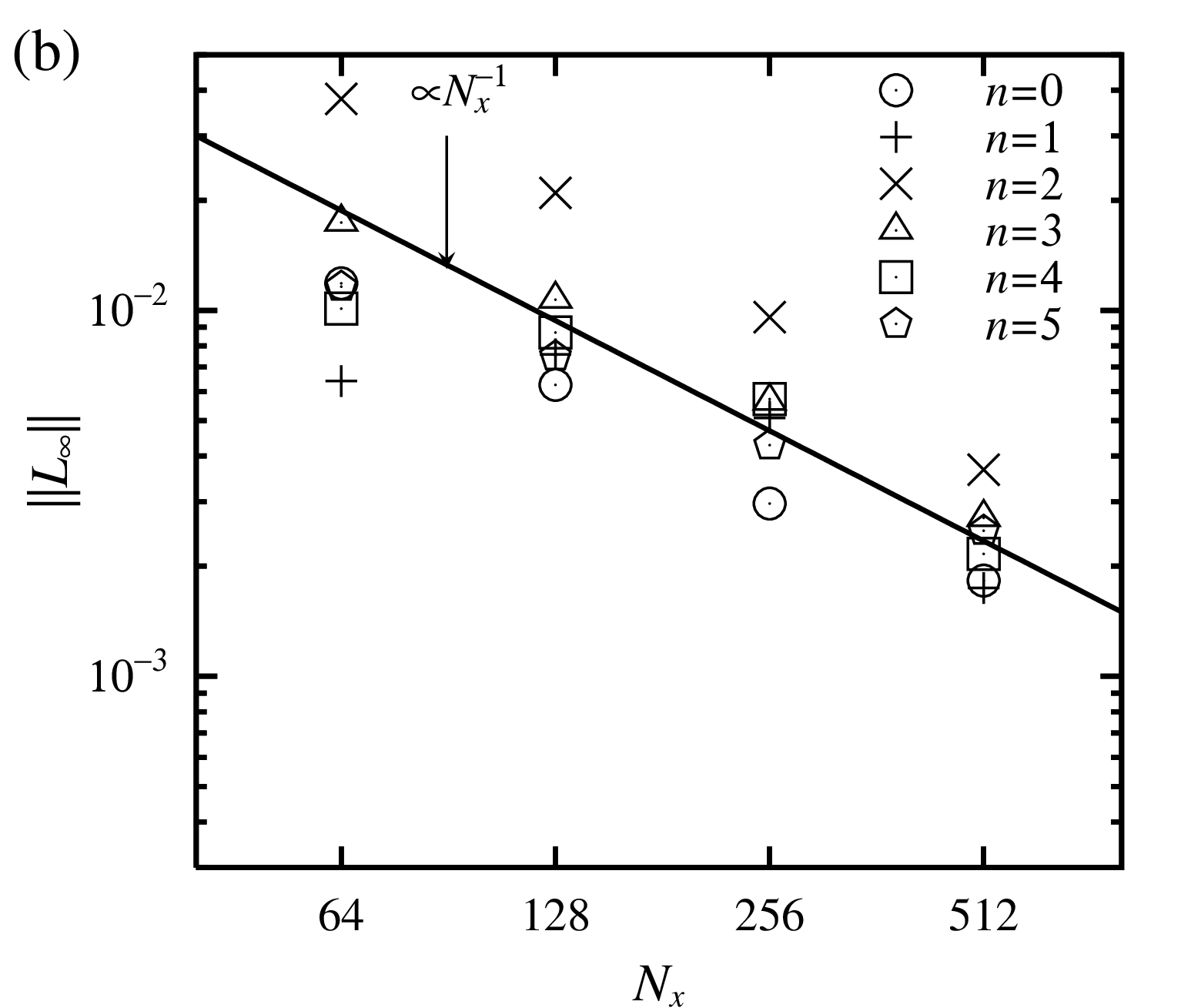,width=5.7cm,angle=270}
\end{center}
\caption{
The errors of the $n$-th order modal amplitude $|R_n|$
 ($n=0$, $1$, $2$, $3$, $4$ and $5$)
 of the particle deformation 
 (a) in L$_2$ norm and (b) in L$_\infty$ norm
 versus the number $N_x$ of grid points in the lid-driven flow.
}
\label{fig:err_deform_mode_zhao}
\end{figure}

In a similar manner to (\ref{eq:error_zhao_xcent}), 
 the  L$_2$ and L$_\infty$ errors
 of the modal amplitude with respect to 
 that of the highest resolution $N_x = 1024$
 are quantified as follows:
\begin{equation}
\begin{split}
||L_2||(N_x)=&
\left\{
\frac{1}{T}
\int_0^T\!\!\!{\rm d}t\ 
\bigl||R_n|(t,N_x)-|R_n|(t,N_x=1024)\bigr|^2
\right\}^{\frac{1}{2}},\\
||L_\infty||(N_x)=&\max_{t\in [0,T]}
\bigl||R_n|(t,N_x)-|R_n|(t,N_x=1024)\bigr|.
\label{eq:error_zhao_def}
\end{split}
\end{equation}
Figure \ref{fig:err_deform_mode_zhao} shows
 the L$_2$ and L$_\infty$ errors as a function of $N_x$.
Again, both the errors are nearly proportional to $N_x^{-1}$, 
 indicating the first-order accuracy for capturing the particle deformation.

\begin{figure}[t]
\begin{center}
\epsfig{file=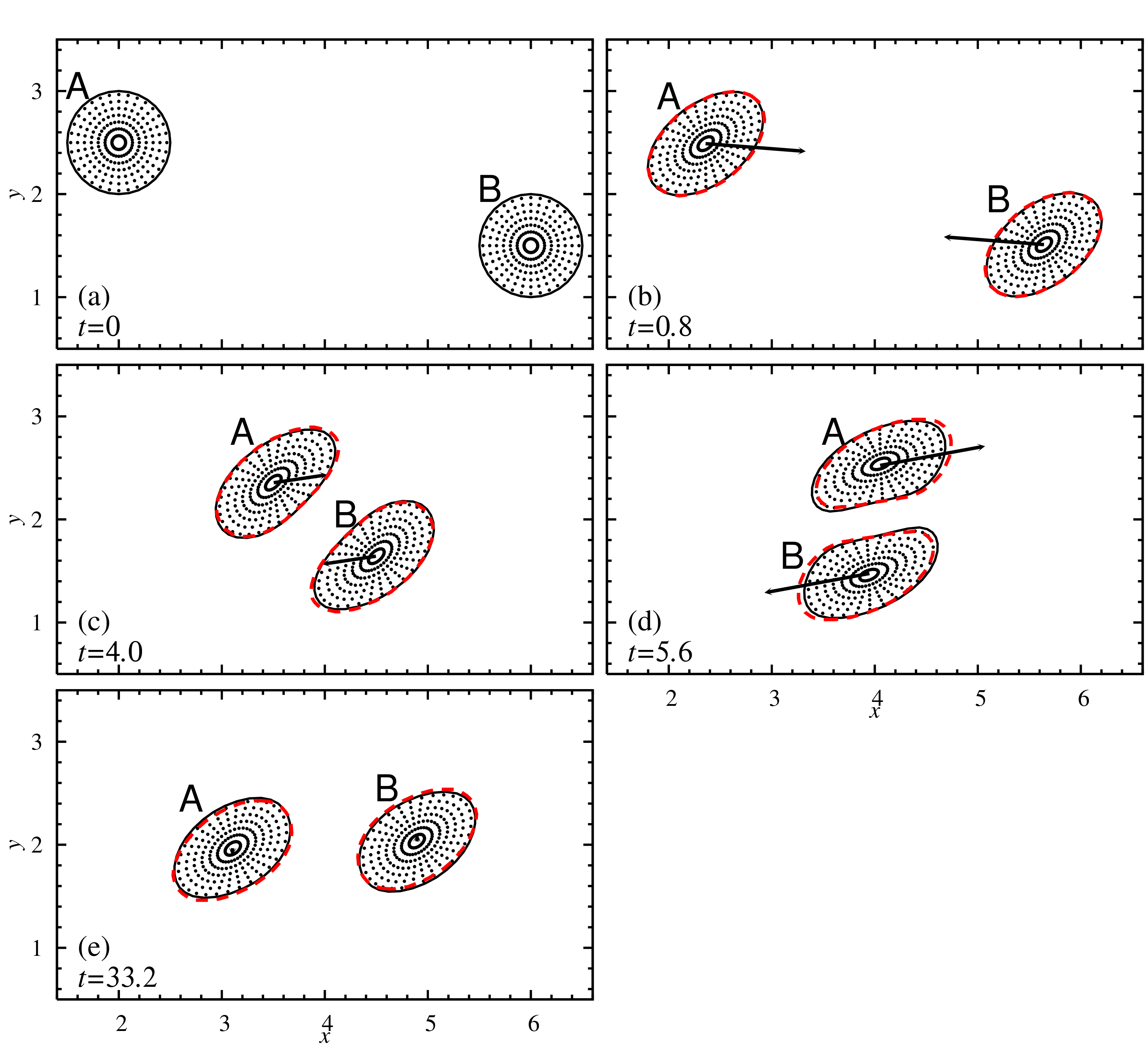,width=12.5cm,angle=270}
\end{center}
\caption{
Comparison of particle-particle interactions in a Couette flow
 with the simulation result \cite{gao2009}. 
The dashed outline represents the result of Gao \& Hu \cite{gao2009},
 in which the body-fit Lagrangian mesh was used
 to solve the FSI problem.
The dotted material points and the solid outline
 correspond to the present simulation results based on the full Eulerian approach
 with a mesh $1024\times 512$. 
At $t=0.8$, initial deformation; 
At $t=4.0$ and $t=5.6$, ``roll over'' interacting mode; 
At $t=33.2$, ``bounce back'' interacting mode \cite{gao2009}.
}
\label{fig:comp_gao1}
\end{figure}

\begin{figure}[t]
\begin{center}
\epsfig{file=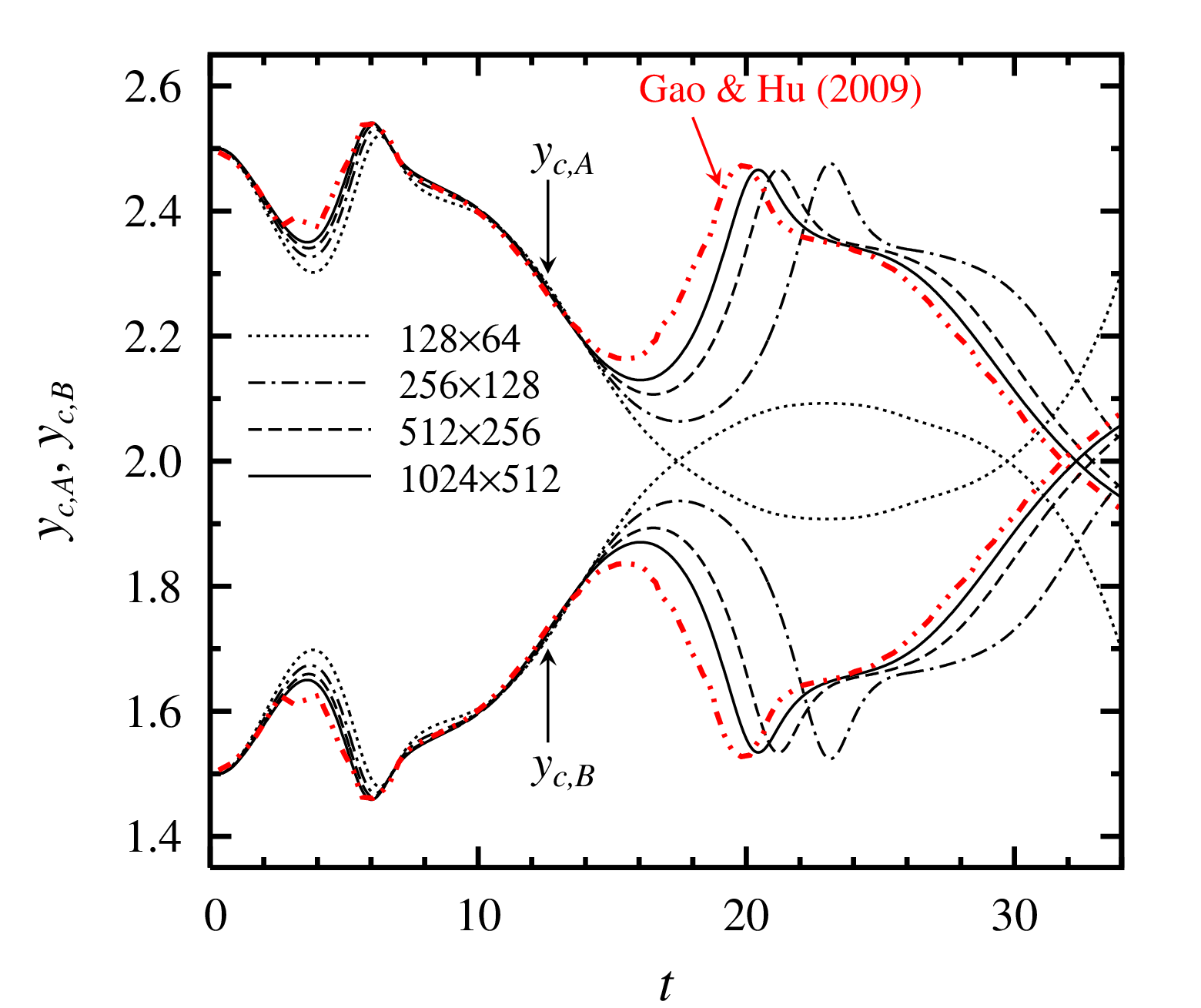,width=9cm,angle=270}
\end{center}
\caption{
Variations of particle $y_c$-position as functions of time for various number of grid points. 
Comparison with the result of Gao \& Hu\cite{gao2009}.
}
\label{fig:comp_gao2}
\end{figure}

\subsubsection{Two particles interaction in a Couette flow}
\label{sec:gao}

We here make a comparison with the available numerical analysis
 of the interaction between two deformable particles in a Couette flow
 performed by Gao \& Hu \cite{gao2009},
 who adopted body-fit Lagrangian mesh.
The computational extent is $L_x\times L_y = 8 \times 4$,
 which is the same as \cite{gao2009}.
Initially, the system is at rest. 
Two unstressed solid particles are initially circular with a radius of $0.5$, 
 and centered at ${\bm x}_{c,A}=(2, 2.5)$ and ${\bm x}_{c,B}=(6, 1.5)$
 as depicted in figure \ref{fig:comp_gao1}(a).
The upper and lower plates located at $y=4$ and $y=0$, respectively,
 start to move impulsively to drive the fluid and solid motions
 at speeds of $V_{W}^{\rm upper}=1$
 and $V_{W}^{\rm lower}=-1$ in $x$ direction. 

The no-slip condition is imposed on the plates,
 while the periodic condition is applied in $x$ direction.
The solid component is purely hyperelastic. 
The material properties are
 $\rho=1$, $\mu_f=20$, $\mu_s=0$, $c_2=40$ and $c_1=c_3=0$.

Figure \ref{fig:comp_gao1} visualizes the two-particle shape 
 for five time instants.
The dotted markers are, again, to represent the solid deformations
 and those markers are not used for computing solid stress or strain.
The arrows at the particle centers are the instantaneous translating
 velocity vectors.
The dashed curve in figure \ref{fig:comp_gao1} 
 represents the outline of the particles obtained 
 by Gao \& Hu \cite{gao2009}.
The particles experience somehow complicated interactions
 involving the ``roll over'' and ``bounce back'' modes 
 as examined in \cite{gao2009}.
The solid shape obtained by the present Eulerian simulation 
 is again in agreement with the well-validated result \cite{gao2009},
 indicating that the particle-particle interaction
 is also reasonably captured by the present approach.

Figure \ref{fig:comp_gao2}
 shows the temporal evolution of the $y_c$-position 
 of the particle centroid, 
 which is evaluated from (\ref{eq:xc}),
 for various grid resolutions
($N_x\times N_y = 128\times 64$, $256\times 128$, $512\times 256$, $1024\times 512$). 
In the full Lagrangian computation \cite{gao2009}, 
 the finite element mesh is refined within the particle-particle gap, 
 whereas in the present Eulerian simulation, the grid size is uniform and fixed. 
When the plot shows peaks around $t=3.0$, $t= 16.0$ and $t=20.0$, 
 the gap between the particles is narrow,
 and the particle undergoes relatively strong hydrodynamic force
 owing to a squeezing effect. 
Such a narrow-gap effect is less resolved by the present method
 than the full Lagrangian method
 especially for the low spatial resolution cases, 
 that is reflected on the larger deviations from the result by Gao \& Hu \cite{gao2009}
 preferentially at the peaks. 
In the higher spatial resolution, 
 the profiles of the present simulation
 get closer to the full Lagrangian result \cite{gao2009}.

\begin{figure}[h]
\begin{center}
\epsfig{file=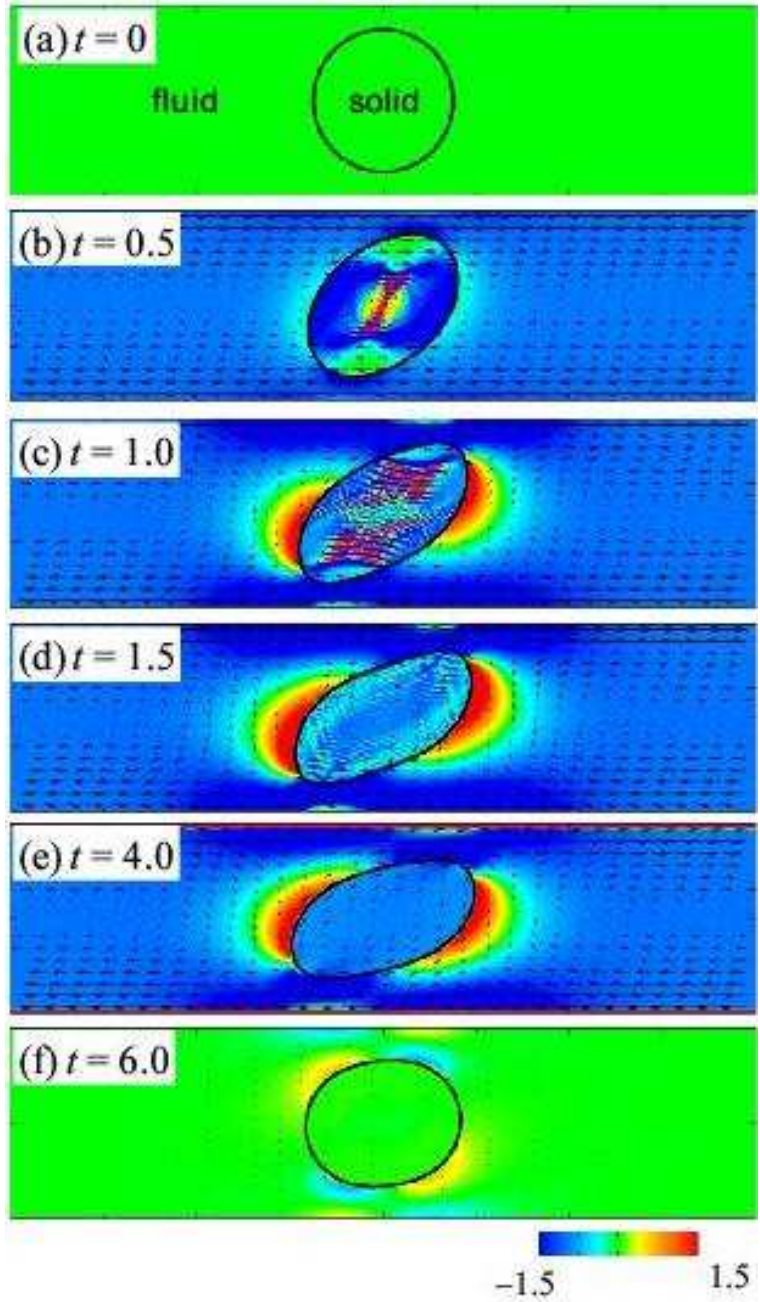,width=10cm,angle=0}
\end{center}
\caption{
Snapshots of the velocity (arrows) and vorticity (color) fields 
 involving a circular particle 
 in the imposing-releasing shear flow 
 between two parallel plates.
The number of grid points is $1024\times 256$.
The upper and lower plates move at speed of $1$ and $-1$ 
 within a period of $t\in [0,4]$, 
 and then stop after $t=4$.
The solid obeys an incompressible Saint Venant-Kirchhoff law
 with $\mu_s=0$, $\lambda_{\mbox{\tiny Lam\'e}}^s=6$
 and $\mu_{\mbox{\tiny Lam\'e}}^s=4$.
}
\label{fig:snap_rev_circ}
\end{figure}

\subsection{Reversibility in shape of hyperelastic material}
The hyperelastic material generally exhibits reversibility in shape
 when it is released from stress. 
In the total Lagrangian method using the finite element mesh, 
 since the tracked material point
 links both the reference and current configurations,
 the reversibility can be captured with little difficulty.
By contrast, the Eulerian fixed grid point
 retains no information on the reference configuration. 
Therefore, one may raise a shortcoming
 that the Eulerian approach is likely to 
 lose the information about the original shape
 once the material is stressed to deform. 
We here perform a reversibility examination. 

\subsubsection{For a circular particle}
\label{sec:rev_circ}

We here deal with a shear flow between two plane plates
 involving a hyperelastic particle.
The distance between the plates is $L_y = 2$.
The computational extent in $x$ direction is set to $L_x = 8$. 
The upper and lower plates are located at $y=1$ and $y=-1$, respectively.
Initially, the system is at rest. 
An unstressed solid particle is initially circular with a radius of $0.75$,
 and centered at the middle position $(0,0)$ between the plates
 as depicted in figure \ref{fig:snap_rev_circ}(a).
The no-slip condition is imposed on the plates,
 whereas the periodic condition is applied in $x$ direction. 
We fix the material properties $\rho=1$, $\mu_f=1$ and $\mu_s=0$.
We consider two kinds of materials:
 one is the linear Mooney-Rivlin material
 with $c_1=4$, $c_2=2$ and $c_3=0$, 
 and the other is the Saint Venant-Kirchhoff material
 with $\lambda_{\mbox{\tiny Lam\'e}}^s=6$ and $\mu_{\mbox{\tiny Lam\'e}}^s=4$
 (i.e., $c_1=4$, $c_2=-2$ and $c_3=1.75$).

The system motion is controlled as follows.
Within a period of $0\leq t \leq 4$,
 the upper and lower plates move at speeds of $V_{W}^{\rm upper}=1$
 and $V_{W}^{\rm lower}=-1$ in $x$ direction, respectively,
 to drive the fluid and solid motions.
After $t=4$, the moving plates stop 
 (i.e. $V_{W}^{\rm upper} = V_{W}^{\rm lower} = 0$) 
 to release the particle from the shearing force.

Figure \ref{fig:snap_rev_circ} visualizes
 the particle deformation
 and the flow field for six consecutive time instants.
As the shear flow is induced by the moving plates,
 the shearing force is imposed on the solid particle,
 and causes the particle elongation
 toward the extensional direction. 
In the transient state during the development of the deformation, 
 it is observed in figure \ref{fig:snap_rev_circ}(b)(c) that
 the transverse elastic waves travel inside the solid,
 and are reflected by the fluid-structure interface.
The wave amplitude is damped through the repetitious reflections with time
 as shown in figure \ref{fig:snap_rev_circ}(d)(e).
As examined in \cite{gao2009}, 
 the elastic wave propagation inside the particle
 may play an important role on the deformation. 
As shown in figure \ref{fig:snap_rev_circ}(e), 
 the vorticity inside the particle at $t=4$ is negative,
 indicating that the particle experiences a tank-treading like motion. 
After the shearing force is released
 by setting the wall velocities to be zero at $t=4$,
 the fluid flow rapidly decays
 and the deformed particle gradually recovers the circular shape. 
At $t=6$ as shown in figure \ref{fig:snap_rev_circ}(f),
 the vorticity in the bulk fluid is almost zero, 
 while the non-zero vorticity forms near the fluid-structure interface, 
 indicating the particle shape is under recovery.

\begin{figure}[h]
\begin{center}
\epsfig{file=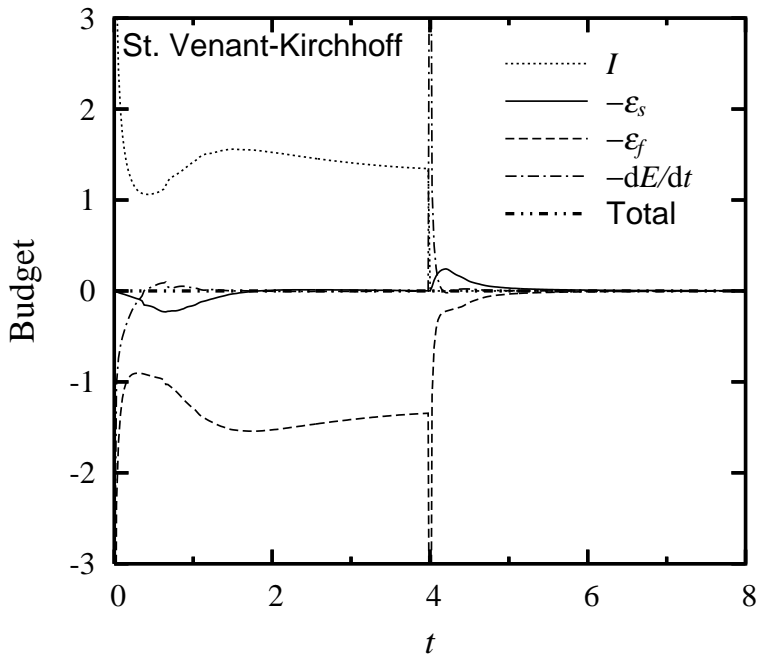,width=9cm,angle=270}

\end{center}
\caption{
The budget of the kinetic-energy transport
 (\ref{eq:en_bud1})
 in the imposing-releasing shear flow.
The conditions are the same as those of figure \ref{fig:snap_rev_circ}.
The dotted, solid, dashed, and dashed-dotted curves 
 correspond to 
 the energy input rate ${\cal I}$, 
 the strain energy rate $-\varepsilon_s$,
 the energy dissipation rate $-\varepsilon_f$, 
 and the kinetic-energy transport $-{\rm d}E/{\rm d}t$, 
 respectively.
The each component is provided in (\ref{eq:en_bud2}).
The dashed-double-dotted curve corresponds to the summation of
 the left-hand-side terms of (\ref{eq:en_bud1}).
}
\label{fig:en_bud}
\end{figure}

\begin{figure}[h]
\begin{center}

\epsfig{file=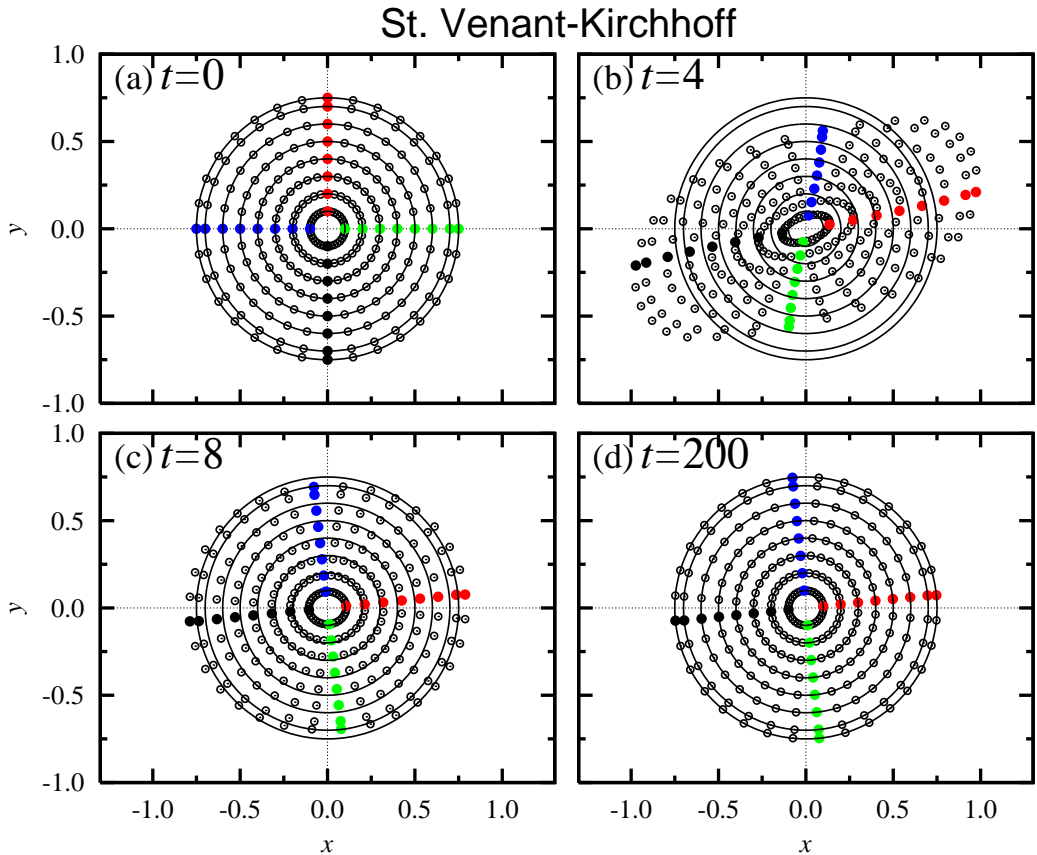,width=11.5cm,angle=270}
\end{center}
\caption{
Material point distribution 
 in the imposing-releasing shear flow
 involving a circular particle between two parallel plates
 with the $1024\times 256$ mesh. 
The conditions are the same as those of figure \ref{fig:snap_rev_circ}.
The colored filled circles are distributed to demonstrate the rotation. 
}
\label{fig:mp_imp_circ_rel}
\end{figure}

The complete recovery in the particle shape is established
 when the hyperelastic strain energy potential
 returns to the initial state.
Therefore, the energy transfer between the fluid and solid phases
 is relevant to it. 
We here check whether the energy transport is numerically conserved 
 during the simulation. 
In the system addressed in this section,
 the budget of the kinetic-energy transport is written as 
\begin{equation}
{\cal I}-\varepsilon_s-\varepsilon_f-\frac{{\rm d}E}{{\rm d}t}=0,
\label{eq:en_bud1}
\end{equation}
where ${\cal I}$, $\varepsilon_s$, $\epsilon_f$, and $E$,
 respectively, denote averaged quantities of 
 the energy input rate, 
 the strain energy rate, 
 the energy dissipation rate, 
 and the kinetic-energy,
 expressed as
\begin{equation}
\begin{split}
{\cal I}=&
\frac{\mu_f}{L_y}
\left(
V_{W}^{\rm upper}\left\langle
\frac{\partial v_x}{\partial y}
\right\rangle_{\Gamma_{W}^{\rm upper}}
\!\!\!\!\!\!-
V_W^{\rm lower}\left\langle
\frac{\partial v_x}{\partial y}
\right\rangle_{\Gamma_{W}^{\rm lower}}\right),\\
\varepsilon_s =&
\left\langle \phi_s {\bm D}^\prime:{\bm \sigma}_s^\prime
\right\rangle_{\Omega},\\
\varepsilon_f =&
2\mu_f\left\langle (1-\phi_s){\bm D}^\prime:{\bm D}^\prime
\right\rangle_{\Omega},\\
E=&
\frac{\rho}{2}
\left\langle {\bm v}\cdot{\bm v}\right\rangle_{\Omega},
\label{eq:en_bud2}
\end{split}
\end{equation}
where $\langle...\rangle_{\Gamma_{W}}$
 stands for the average over the wall, 
 and $\langle...\rangle_{\Omega}$
 for the average over the entire domain $\Omega$.
Figure \ref{fig:en_bud} shows the time history 
 of the each contribution
 in the left-hand-side of (\ref{eq:en_bud1}).
As the flow evolves, the particle deforms,
 and thereby the particle stores the strain energy
 as indicated by the solid curve with negative value
in figure \ref{fig:en_bud}.
After the walls stop at $t=4$, 
 the particle releases the strain energy. 
The double-chained curve in figure \ref{fig:en_bud}
 shows the summation of the left-hand-side terms of (\ref{eq:en_bud1}).
Its absolute value, corresponding to the numerical error, 
 is less than $10^{-5}$, 
 which is much smaller than the variation of the contributions
 of the individual terms.
The system is well conserved during the simulation
 in view of the energy balance,
 because the equality of (\ref{eq:en_bud1}) is almost fulfilled.
It is important to emphasize that
 the numerical energy conservation hinges upon the finite difference schemes, 
 and the method proposed by Kajishima \cite{kaj1994} is employed in the present study.

\begin{figure}[h]
\begin{center}

\epsfig{file=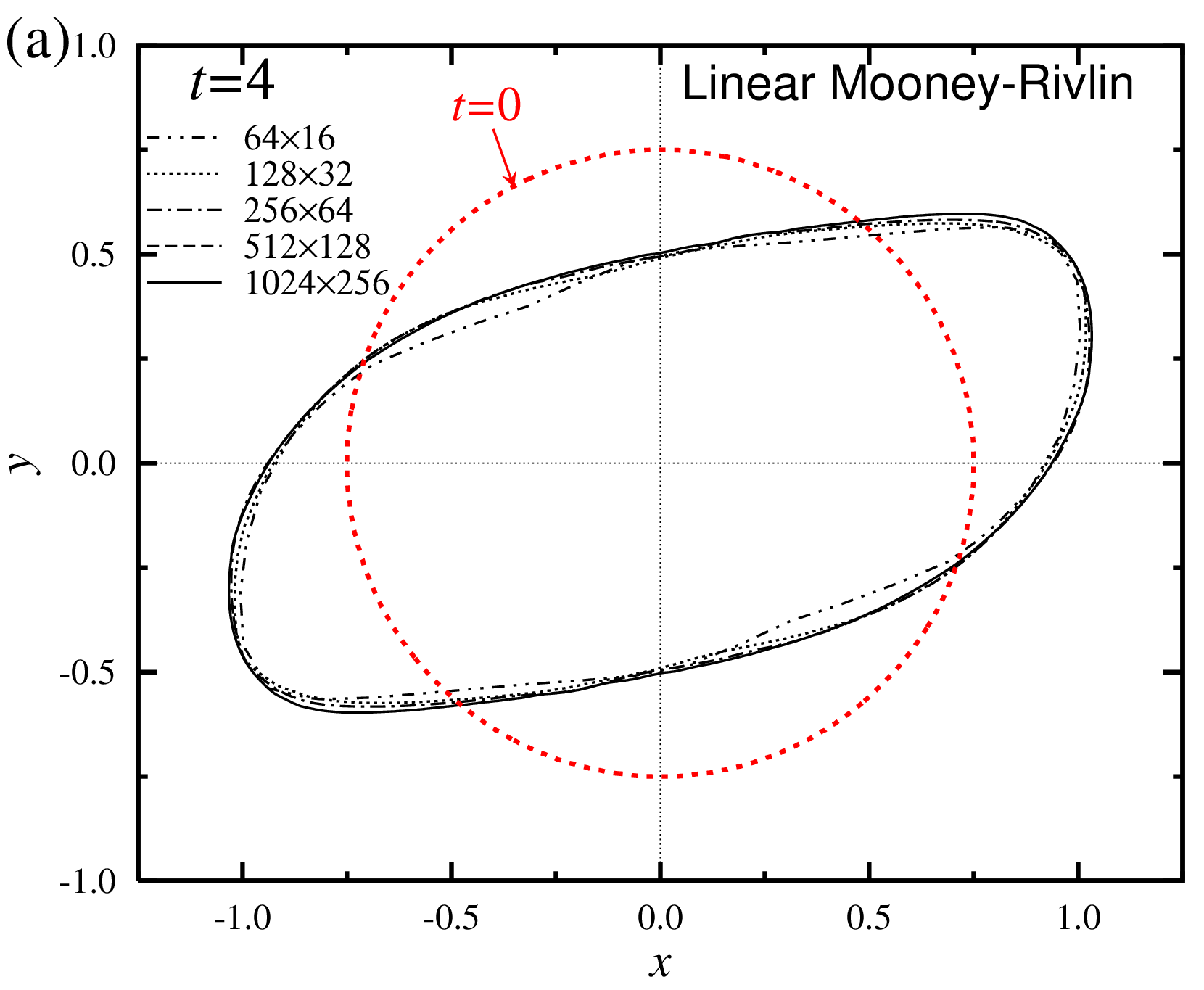,width=5.7cm,angle=270}
\epsfig{file=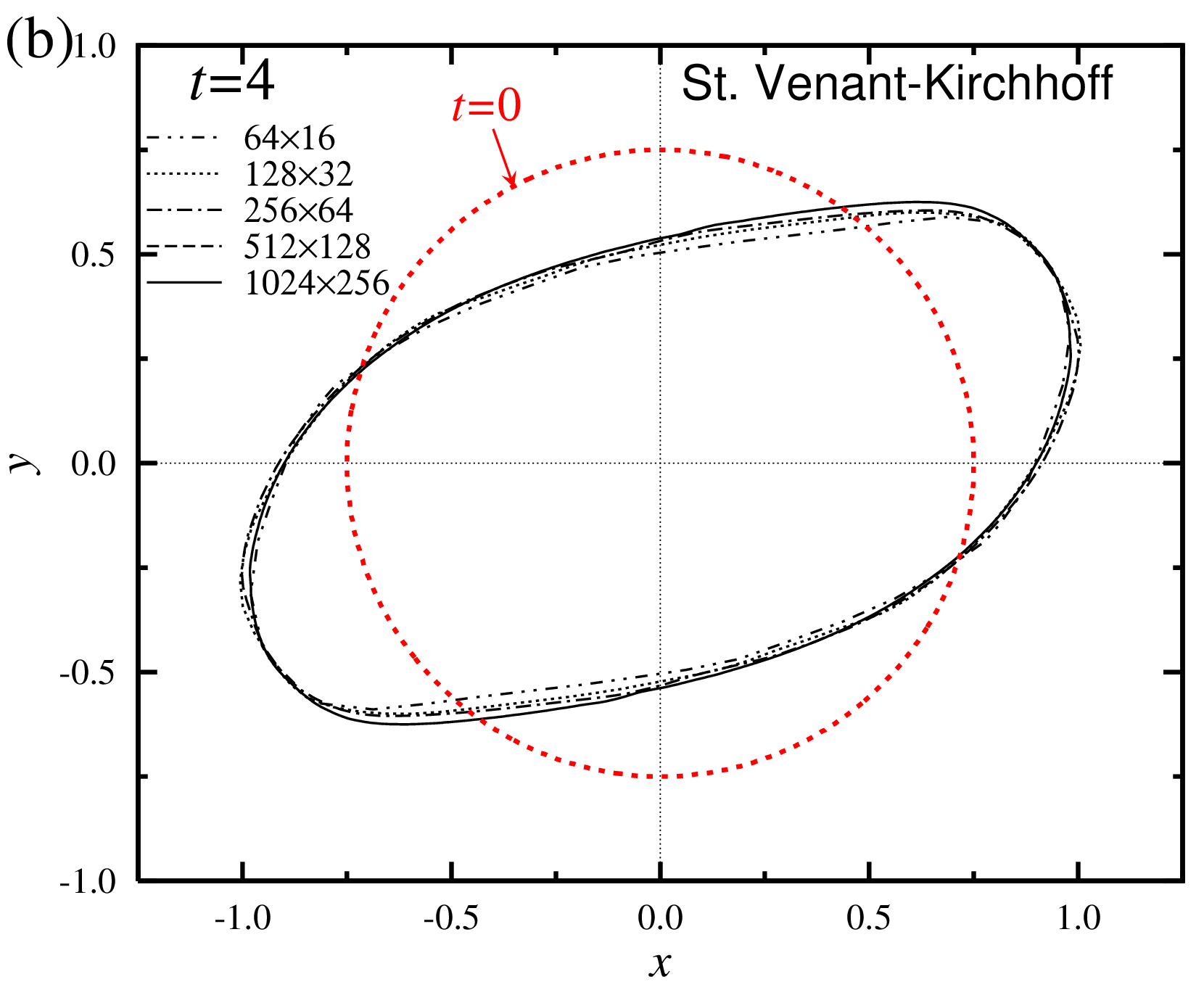,width=5.7cm,angle=270}

\epsfig{file=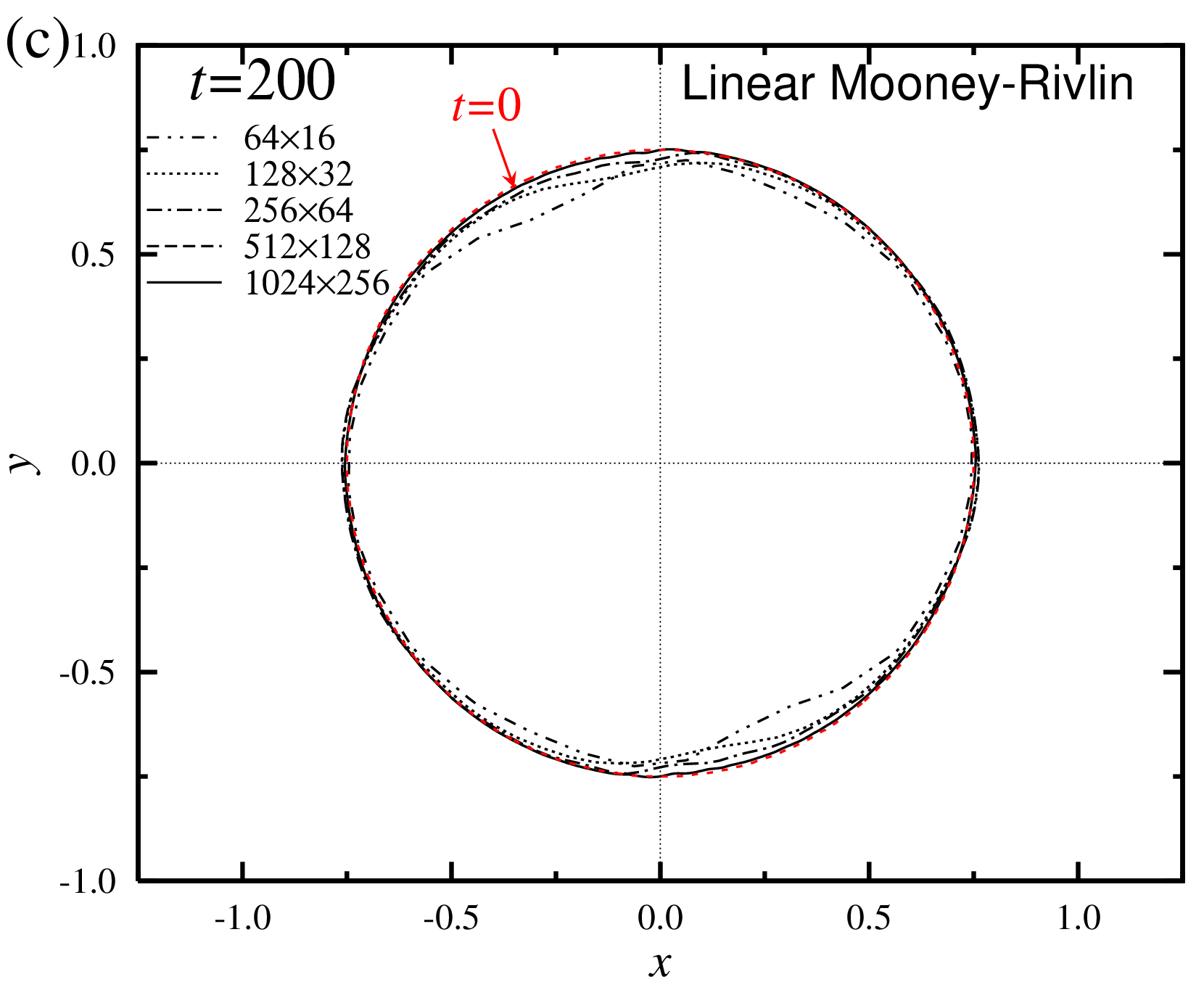,width=5.7cm,angle=270}
\epsfig{file=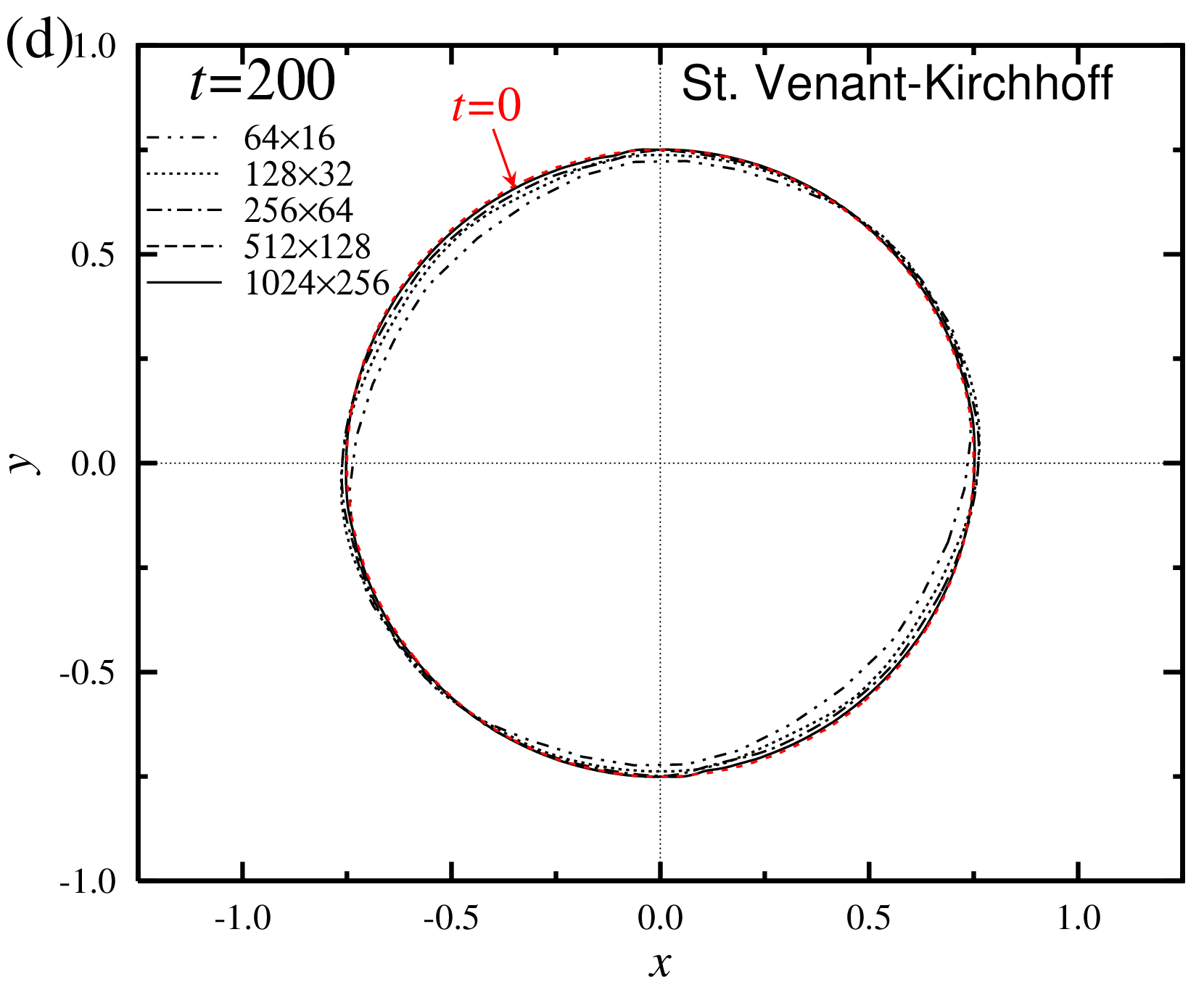,width=5.7cm,angle=270}

\end{center}
\caption{
Outlines of the fluid-structure interface
 in the imposing-releasing shear flow 
 involving a circular particle between two parallel plates
 for various number of grid points
($N_x\times N_y=64\times 16$, $128\times 32$, $256\times 64$,
 $512\times 128$ and $1024\times 256$).
The imposing-releasing shear scheme, the geometry, 
 and the fluid properties are the same as those of figure \ref{fig:snap_rev_circ}.
The left panels: the linear Mooney material 
 with $\mu_s=0$, $c_1=4$, $c_2=2$ and $c_3=0$.
The right panels: the incompressible Saint Venant-Kirchhoff material 
 with $\mu_s=0$, $\lambda_{\mbox{\tiny Lam\'e}}^s=6$
 and $\mu_{\mbox{\tiny Lam\'e}}^s=4$.
The upper panels: at $t=4$.
The lower panels: at $t=200$.
}
\label{fig:outlines_rev_circ}
\end{figure}

To directly demonstrate whether the reversibility can be captured, 
 the distributions of the tracers for four consecutive time instants
 are shown in figure \ref{fig:mp_imp_circ_rel}.
As depicted in figure \ref{fig:mp_imp_circ_rel}(a),
 the tracers are initially seeded on the concentric circles inside the solid
 to demonstrate the local displacements inside the solid.
The bilinear interpolation to the tracer location
 is applied to identifying its velocity, 
 and its position is temporally updated in a Lagrangian way.
Figure \ref{fig:mp_imp_circ_rel}(b) shows the tracer distribution
 at the most deformed instant $t=4$
 when the particle is under the tank-treading like motion.
After the wall velocities is set to be zero at $t=4$, 
 the tracer particles gradually move back toward the initial concentric
 circles with time. 
It should be noted that
 because the degree of freedom corresponding to the rigid rotation
 is allowed,
 the tracer distributions in figure \ref{fig:mp_imp_circ_rel}(c)(d)
 turn in the clockwise directions about 80 degrees with respect to 
 the initial distribution in figure \ref{fig:mp_imp_circ_rel}(a).
At the instant $t=8$,
 when the same period as the shear-imposing stage (four unit time)
 has elapsed after the walls stop,
 the discrepancy between the tracer location and the concentric circle
 is clearly shown in figure \ref{fig:mp_imp_circ_rel}(c),
 indicating that the recovery in the particle shape is still underway.
After a sufficiently long time ($t=200$),
 the tracers are found to be back in the concentric circles 
 as shown in figure \ref{fig:mp_imp_circ_rel}(d).
We may say that
 the present Eulerian approach can capture the reversibility in shape
 under certain right circumstances,
 which will be discussed later.

To assess the effect of the spatial resolution, 
 the outlines of the fluid-structure interface, 
 which are identified as the isolines at $\phi_s=1/2$,
 for various number of grid points for different materials
 are shown in figure \ref{fig:outlines_rev_circ}.
As the spatial resolution is increased, 
 the deformed shape more settles to a convergent curve
 at the instant $t=4$
 when the particle exhibits the most deformed shape,
 and the outline at $t=200$ shows better recovery to the initial curve at $t=0$.

\begin{figure}[h]
\begin{center}

\epsfig{file=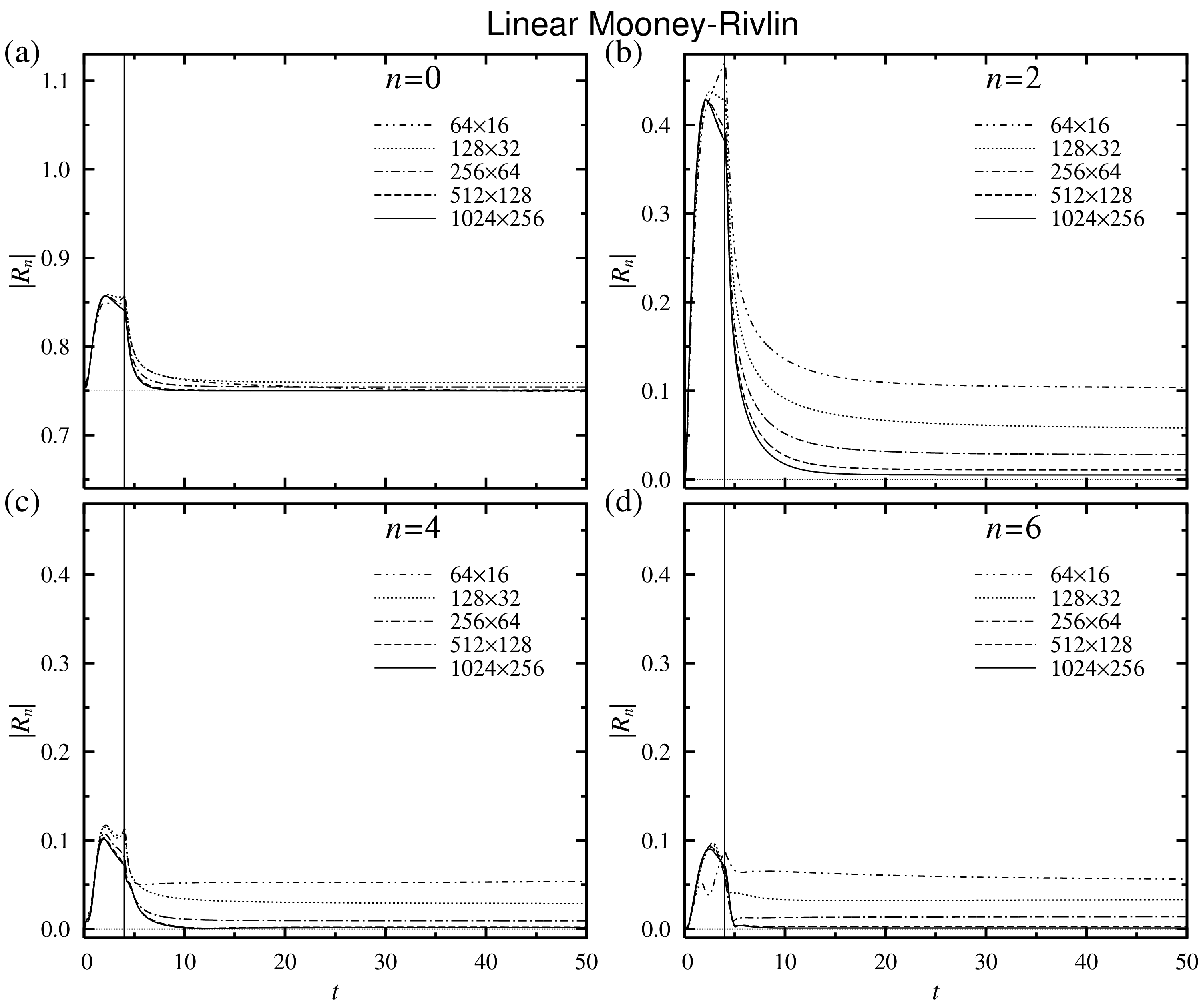,width=11.5cm,angle=270}

\end{center}
\caption{
Time history of the $n$-th order modal amplitude $|R_n|$
 of the particle deformation
 in the imposing-releasing shear flow
 involving a circular particle between two parallel plates
 for various number of grid points
($N_x\times N_y=64\times 16$, $128\times 32$, $256\times 64$,
 $512\times 128$ and $1024\times 256$).
(a) $n=0$, (b) $n=2$, (c) $n=4$, and (d) $n=6$.
The solid obeys a linear Mooney-Rivlin law with
 $\mu_s=0$, $c_1=4$, $c_2=2$ and $c_3=0$.
}
\label{fig:deform_mode_mr}
\end{figure}

\begin{figure}[h]
\begin{center}

\epsfig{file=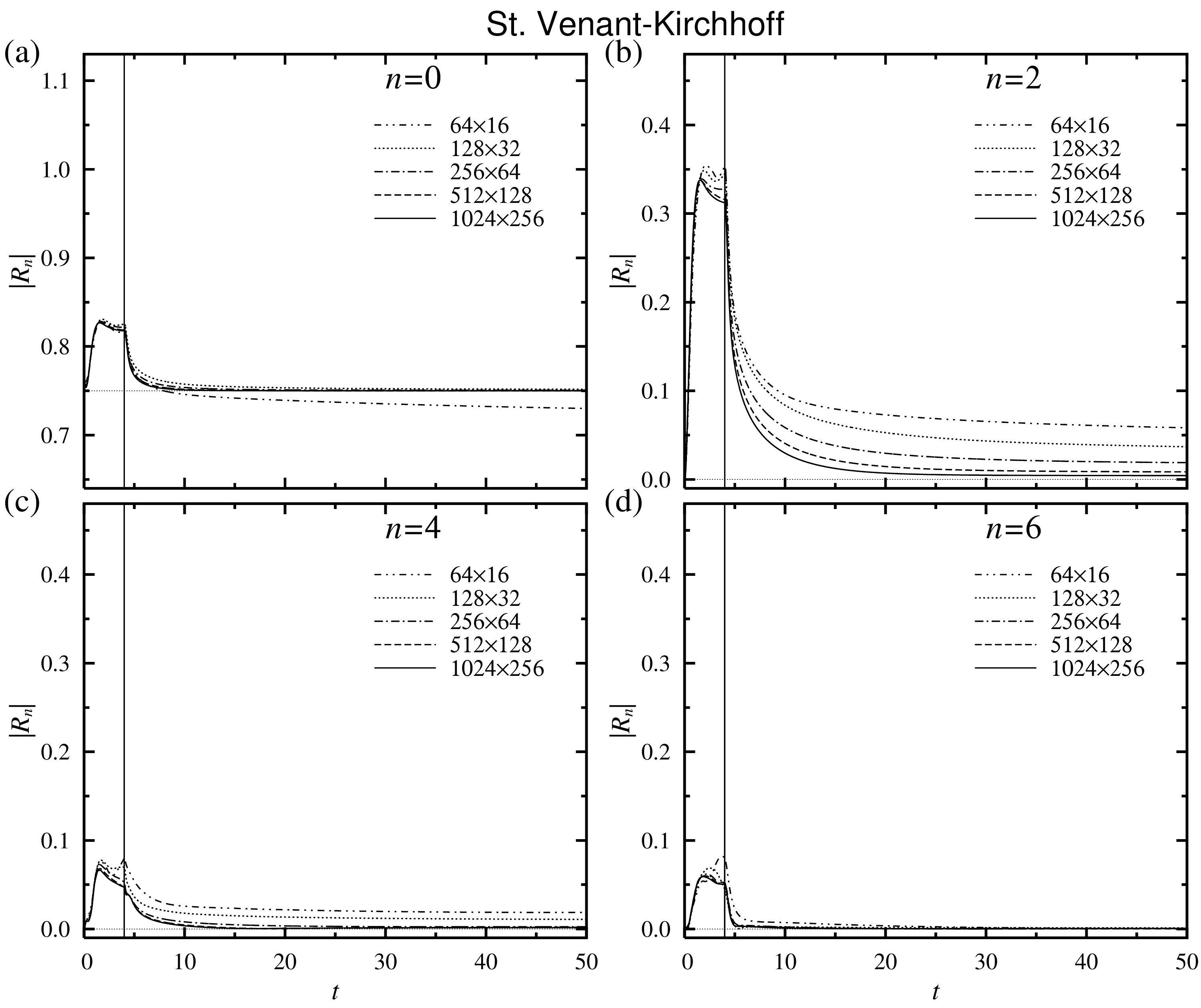,width=11.5cm,angle=270}

\end{center}
\caption{
Same as figure \ref{fig:deform_mode_mr},
 but the solid obeys an incompressible Saint Venant-Kirchhoff law
 with $\mu_s=0$, $\lambda_{\mbox{\tiny Lam\'e}}^s=6$
 and $\mu_{\mbox{\tiny Lam\'e}}^s=4$.
}
\label{fig:deform_mode_svk}
\end{figure}

To further assess the grid convergence behavior in the particle deformation, 
 the deformation mode is here investigated. 
As explained in \S \ref{sec:zhao},
 the deformation modes are determined 
 using (\ref{eq:mode1}), (\ref{eq:mode2}) and (\ref{eq:mode5}).
Due to the symmetry of the system, 
 the odd-number-order modes $R_{c,2n+1}$ and $R_{s,2n+1}$ are identically zero. 
Temporal evolutions of the modal amplitudes
 $|R_n|$  ($n=0$, $2$, $4$ and $4$) of the particle deformation
 for various number of grid points
 are shown in figure \ref{fig:deform_mode_mr}
 (for the linear Mooney-Rivlin material)
 and in figure \ref{fig:deform_mode_svk}
 (for the incompressible Saint Venant-Kirchhoff material).
Both figure \ref{fig:deform_mode_mr} and figure \ref{fig:deform_mode_svk} 
 demonstrate the convergence behavior of the profiles
 with increasing the number of grid points. 
After the walls stop at $t=4$,
 the zeroth-order mode $|R_0|$ approaches $0.75$, 
 corresponding to the unstressed radius,
 in the case of the spatial resolution $N_x\times N_y = 256\times 64$ or higher.
However,
 the higher-order amplitudes $|R_2|$, $|R_4|$ and $|R_6|$
 at the fully developed stage 
 obviously settle to some non-zero values. 
This tendency is more pronounced in the lower grid resolution, 
 indicating that some spurious deformation remains.
However, the deviation from zero for $|R_n|$ ($n\neq 0$)
 vanishes exponentially as the spatial resolution is increased.

\begin{figure}[h]
\begin{center}

\epsfig{file=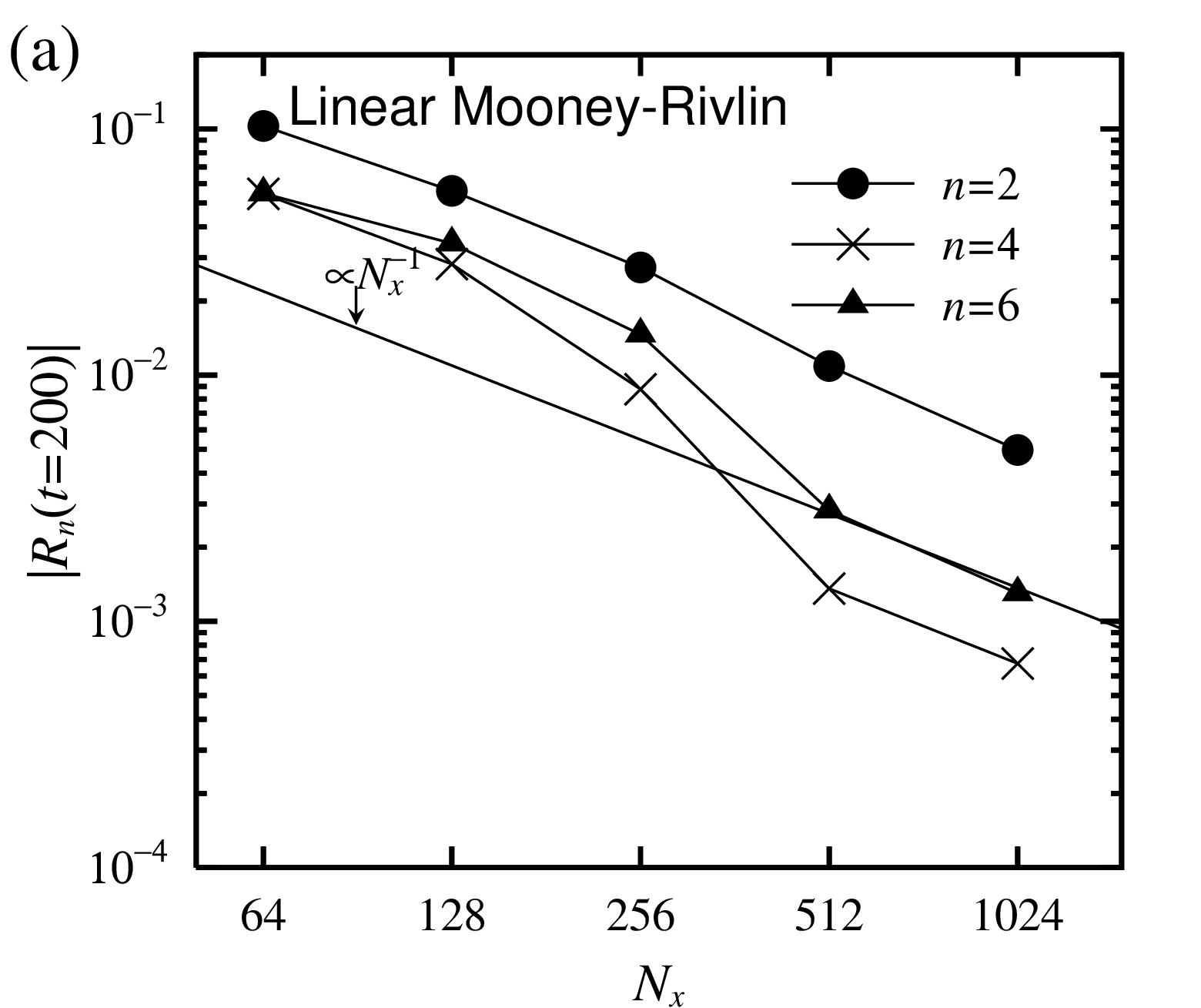,width=5.7cm,angle=270}
\epsfig{file=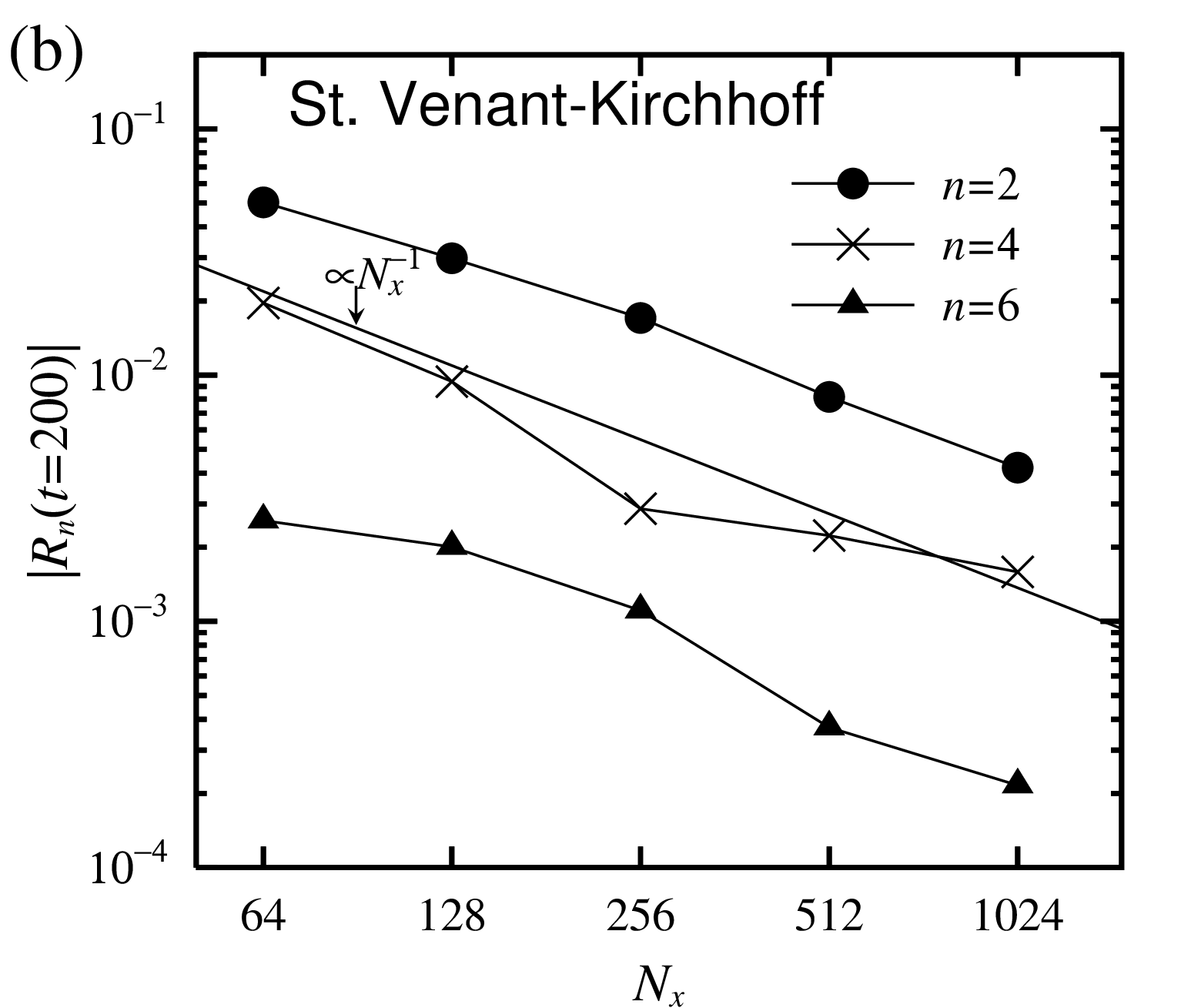,width=5.7cm,angle=270}

\end{center}
\caption{
Residual modal amplitudes $|R_n|$
 of the particle deformation at $t=200$ 
 versus the number $N_x$ of grid points
 in the imposing-releasing shear flow. 
(a): for the results in figure \ref{fig:deform_mode_mr}
 employing the linear Mooney-Rivlin material 
(b): for the results in figure \ref{fig:deform_mode_svk}
 employing the incompressible Saint Venant-Kirchhoff material.
}
\label{fig:res_deform_mode}
\end{figure}

To quantify the spurious residual deformation, 
 the modal amplitudes $|R_n|$ ($n=2$, $4$ and $6$) at $t=200$
 as a function of the number of grid points
 are plotted in figure \ref{fig:res_deform_mode}.
The residual amplitudes are nearly proportional to $N_x^{-1}$,
 indicating the first-order accuracy in capturing unstressed shape. 
As demonstrated in \S \ref{sec:layer} for
 the much simpler system consisting of 
 the fluid-structure layers, 
 the present fluid-structure couping method
 involves the first-order accuracy, 
 which is also reflected on the grid convergence
 of the reversibility in shape. 

\begin{figure}[h]
\begin{center}
\epsfig{file=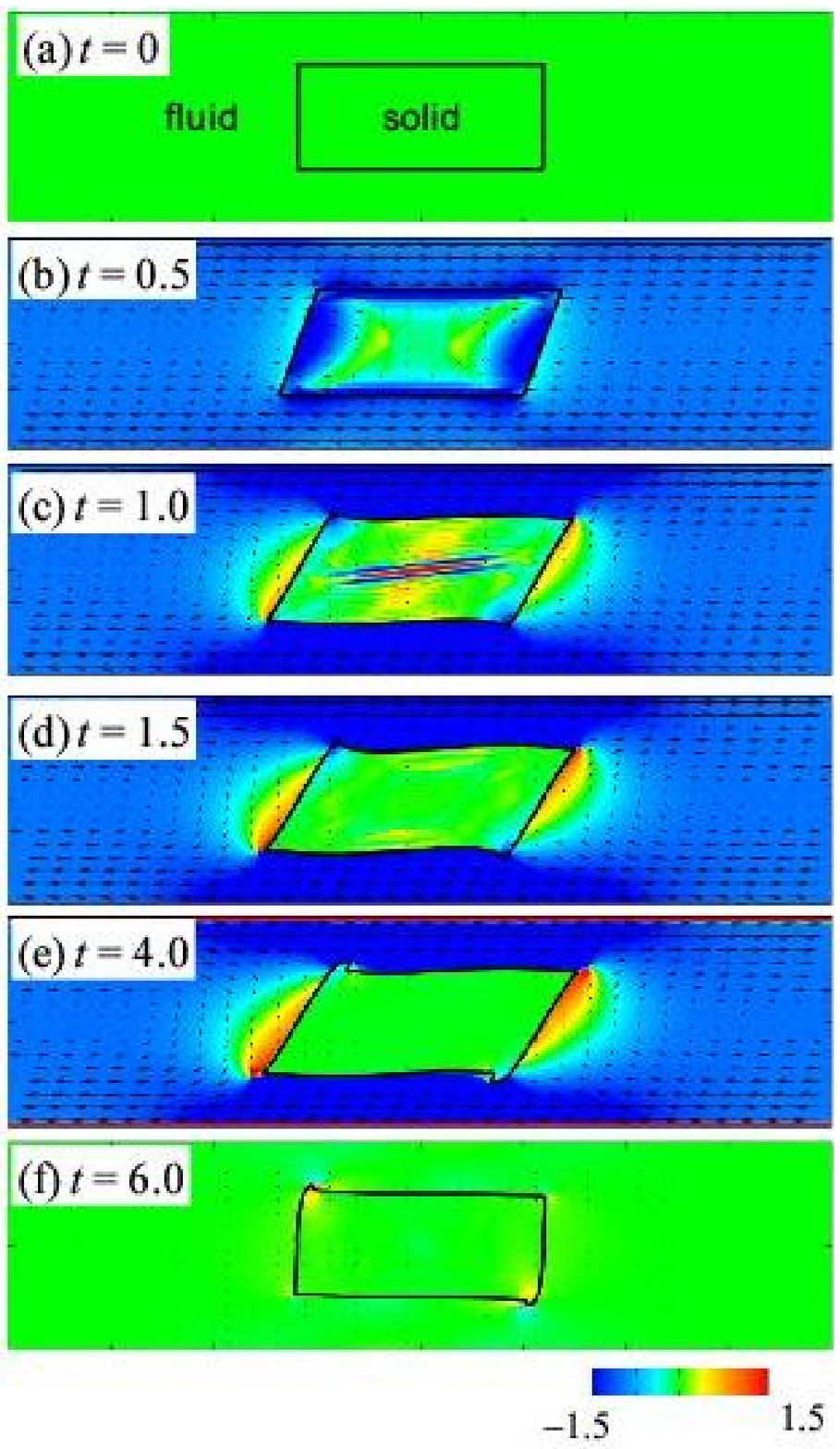,width=10cm,angle=0}
\end{center}
\caption{
Snapshots of the velocity (arrows) and vorticity (color) fields 
 in the imposing-releasing shear flow 
 involving a rectangular particle 
 between two parallel plates. 
The number of grid points is $1024\times 256$.
The upper and lower plates move at speeds of $1$ and $-1$ 
 within a period of $t\in [0,4]$, 
 and then stop after $t=4$.
The solid obeys an incompressible Saint Venant-Kirchhoff law
 with $\mu_s=0$, $\lambda_{\mbox{\tiny Lam\'e}}^s=6$
 and $\mu_{\mbox{\tiny Lam\'e}}^s=4$.
}
\label{fig:snap_rev_rect}
\end{figure}

\begin{figure}[h]
\begin{center}
\epsfig{file=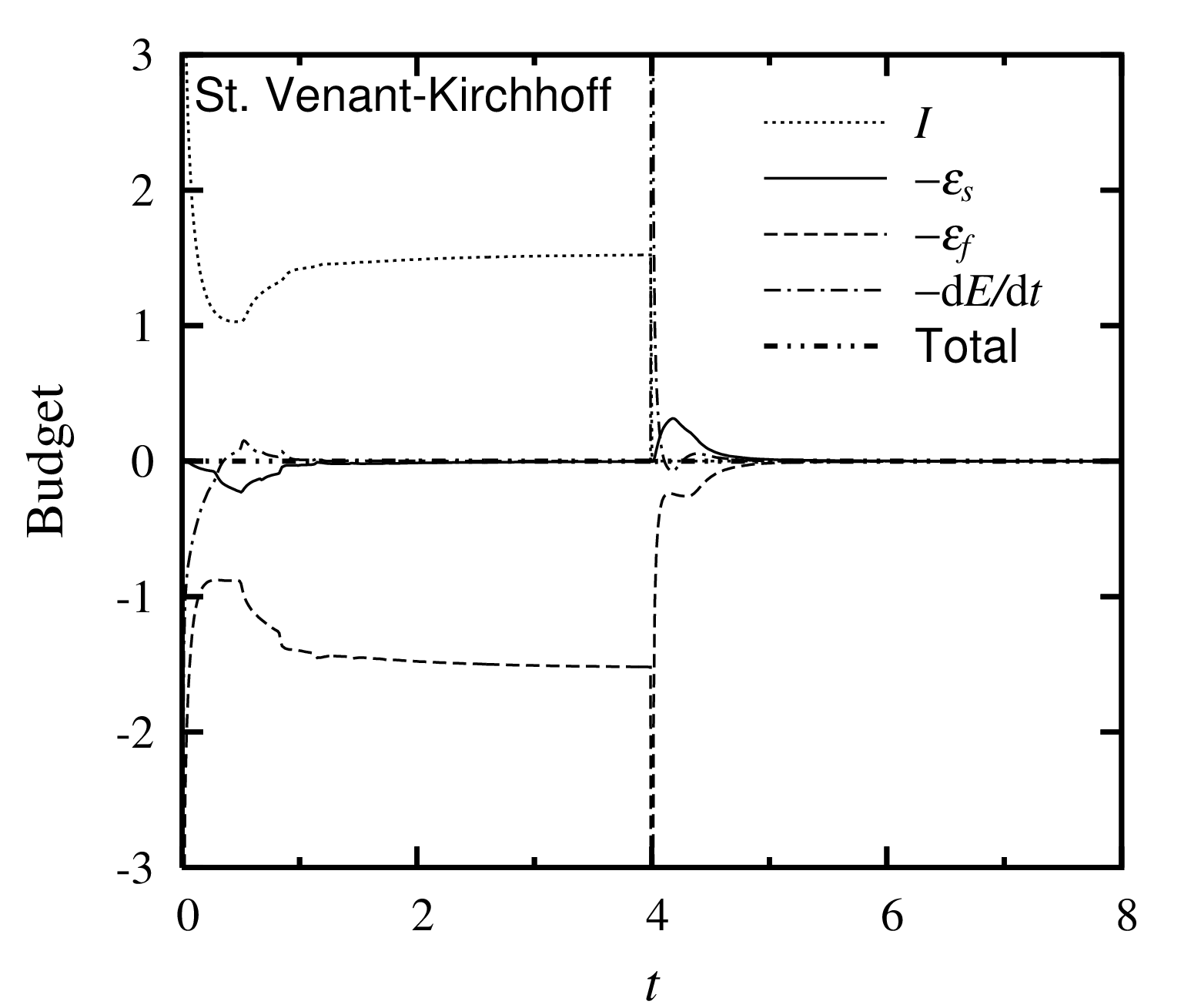,width=9cm,angle=270}

\end{center}
\caption{
The budget of the kinetic-energy transport
 (\ref{eq:en_bud1})
 in the imposing-releasing shear flow.
The conditions are the same as those of figure \ref{fig:snap_rev_rect}.
The dotted, solid, dashed, and dashed-dotted curves 
 correspond to 
 the energy input rate ${\cal I}$, 
 the strain energy rate $-\varepsilon_s$,
 the energy dissipation rate $-\varepsilon_f$, 
 and the kinetic-energy transport $-{\rm d}E/{\rm d}t$, 
 respectively.
The each component is provided in (\ref{eq:en_bud2}).
The dashed-double-dotted curve corresponds to the summation of
 the left-hand-side terms of (\ref{eq:en_bud1}).
}
\label{fig:en_bud_rect}
\end{figure}

\begin{figure}[h]
\begin{center}
\epsfig{file=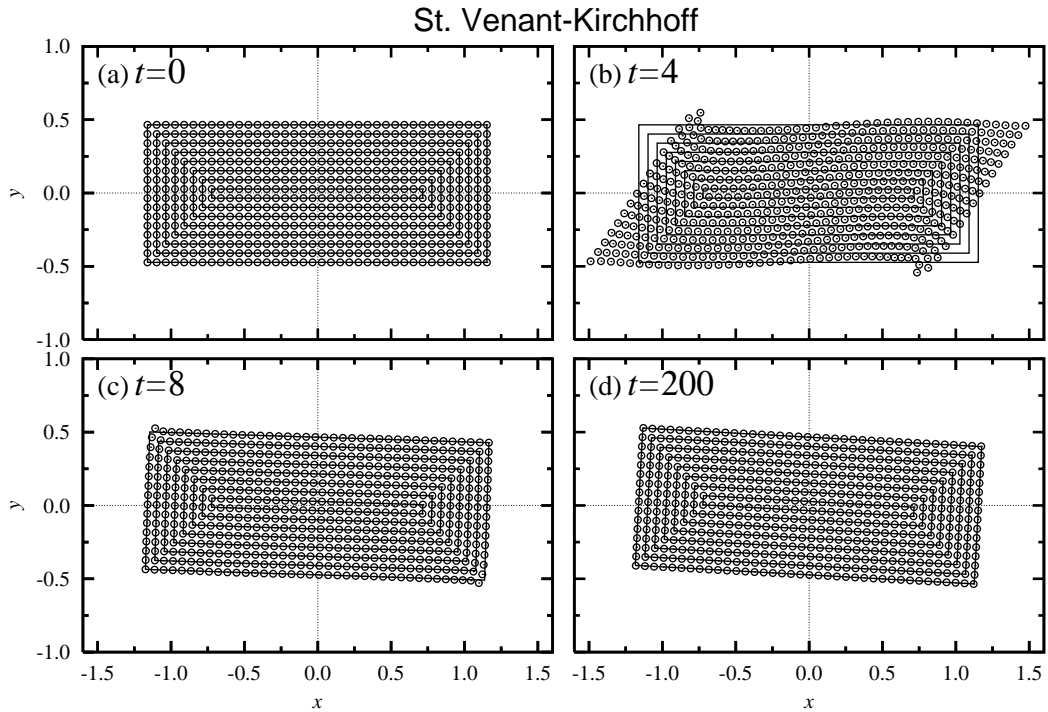,width=9.5cm,angle=270}
\end{center}
\caption{
Material point distribution 
 in the imposing-releasing shear flow
 involving a rectangular particle 
 between two parallel plates. 
 with the $1024\times 256$ mesh. 
The conditions are the same as those of figure \ref{fig:snap_rev_rect}.
}
\label{fig:mp_imp_rect_rel}
\end{figure}

\begin{figure}[h]
\begin{center}

\epsfig{file=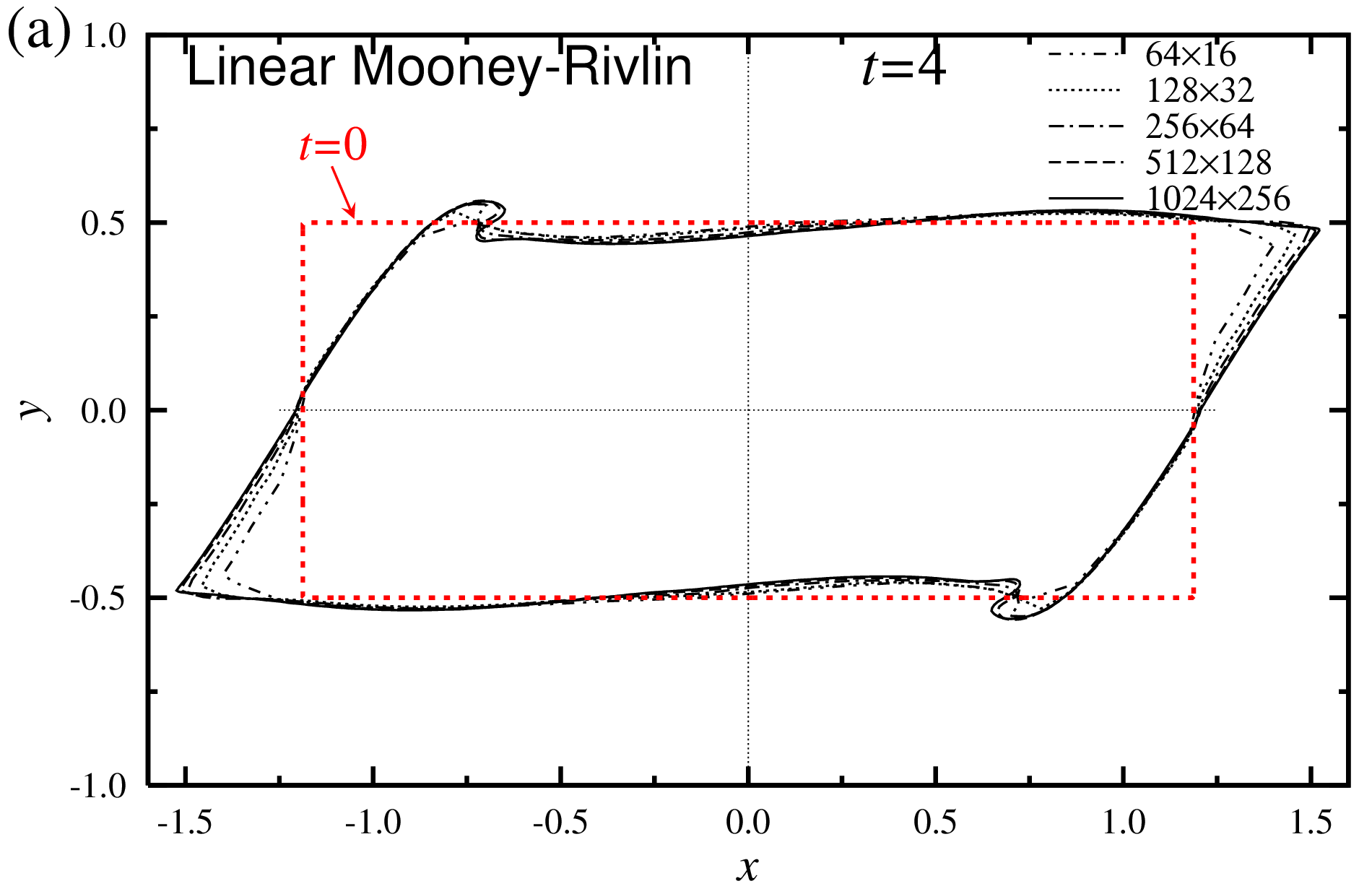,width=4.5cm,angle=270}
\epsfig{file=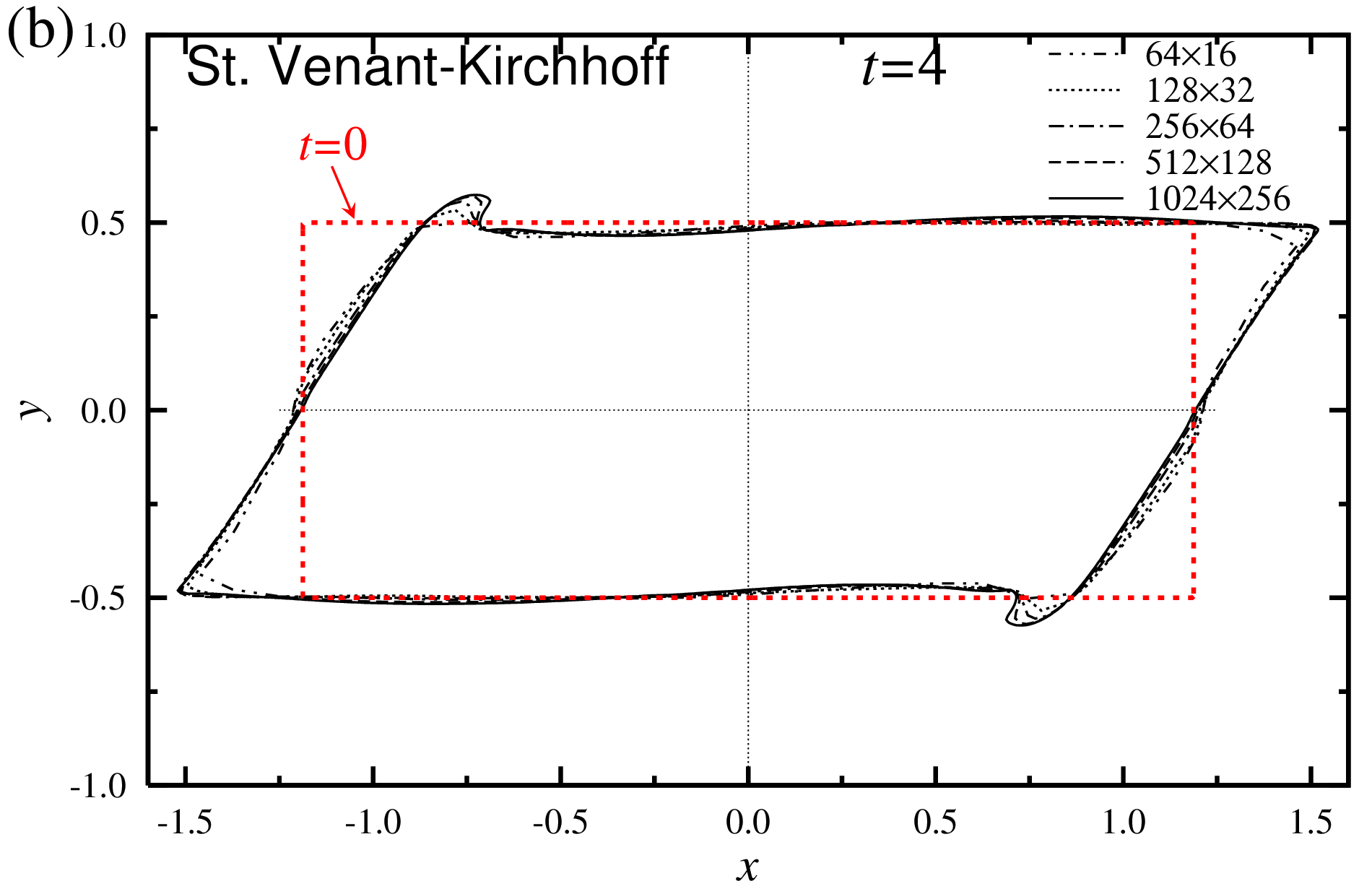,width=4.5cm,angle=270}

\epsfig{file=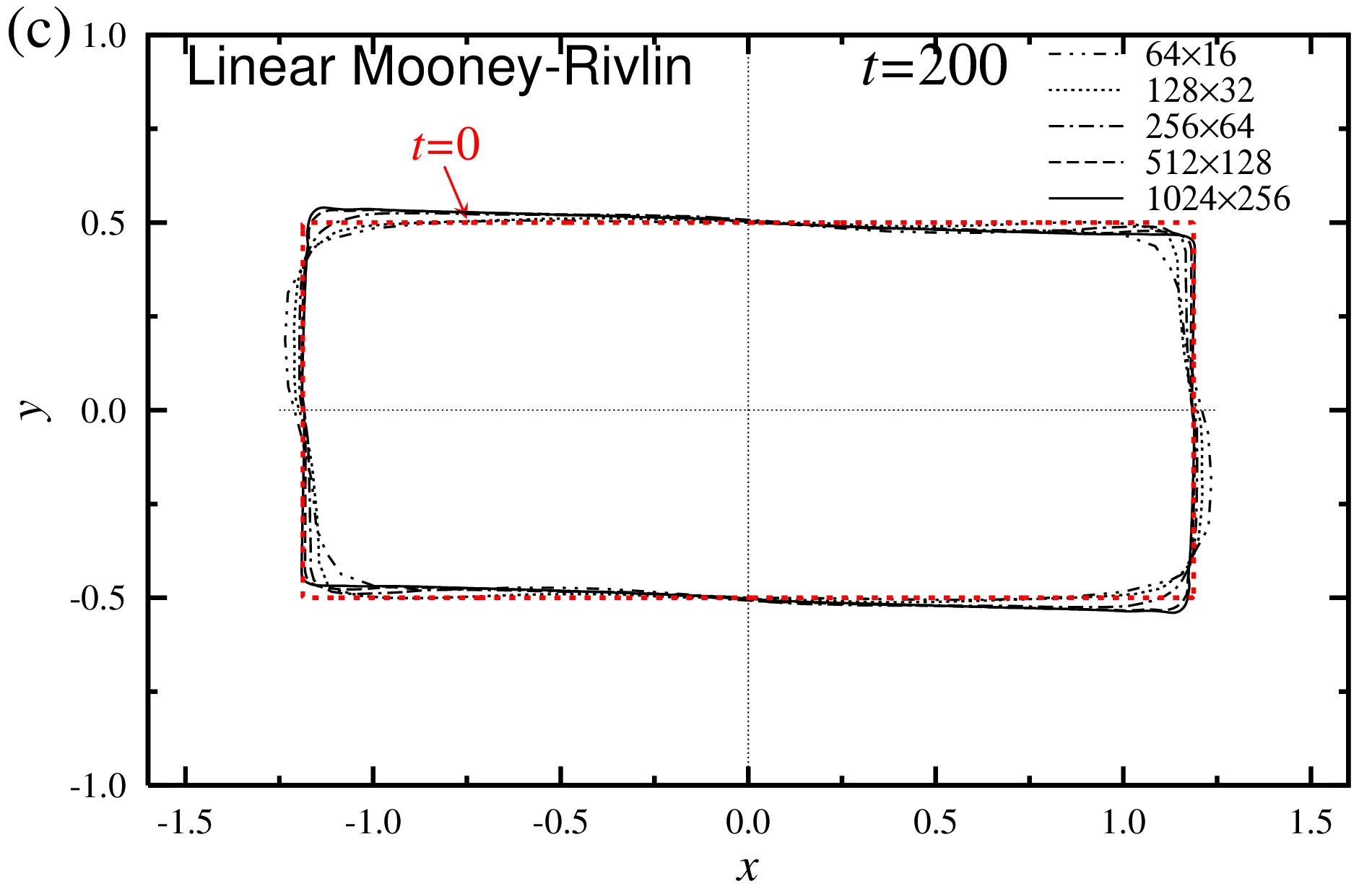,width=4.5cm,angle=270}
\epsfig{file=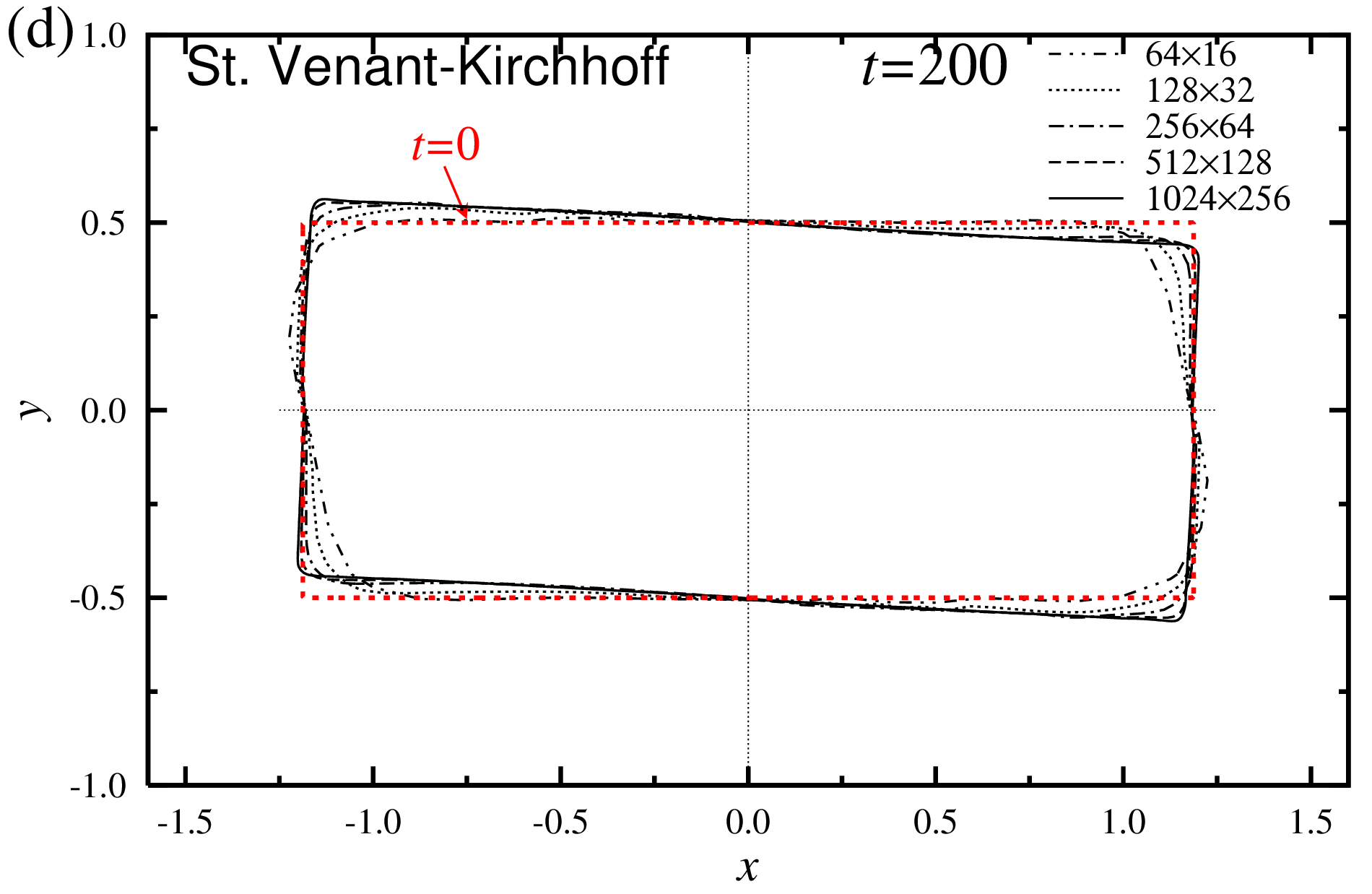,width=4.5cm,angle=270}

\end{center}
\caption{
Outlines of the fluid-structure interface
 in the imposing-releasing shear flow
 involving a rectangular particle 
 between two parallel plates
 for various number of grid points
($N_x\times N_y=64\times 16$, $128\times 32$, $256\times 64$,
 $512\times 128$ and $1024\times 256$).
The imposing-releasing shear scheme, the geometry, 
 and the fluid properties are the same as those of figure \ref{fig:snap_rev_rect}.
The left panels: the linear Mooney material 
 with $\mu_s=0$, $c_1=4$, $c_2=2$ and $c_3=0$.
The right panels: the incompressible Saint Venant-Kirchhoff material 
 with $\mu_s=0$, $\lambda_{\mbox{\tiny Lam\'e}}^s=6$
 and $\mu_{\mbox{\tiny Lam\'e}}^s=4$.
The upper panels: at $t=4$.
The lower panels: at $t=200$.
}
\label{fig:outlines_rev_rect}
\end{figure}

\subsubsection{Shape reversibility of a rectangular particle}

We also perform a reversibility test
 for a rectangular particle with a dimension of $2.375\times 1$
 to demonstrate the applicability of the method to an object
 with a larger aspect ratio and sharp corners.
The initial setup is depicted in figure \ref{fig:snap_rev_rect}(a).
Figure \ref{fig:snap_rev_rect} visualizes
 the particle deformation
 and the velocity and vorticity fields
 for six consecutive time instants.
Similar to the system involving the circular particle
 in figure \ref{fig:snap_rev_circ},
 the elastic wave propagation and its attenuation are observed 
 in figure \ref{fig:snap_rev_rect}(b)(c)(d). 
However, unlike the tank-treading motion
 in figure \ref{fig:snap_rev_circ}(e),
 the vorticity inside the particle at $t=4$
 is not entirely negative in figure \ref{fig:snap_rev_rect}(e), 
 indicating that the particle does not experience
 the tank-treading or tumbling motion.
It is because the rotational motion
 is geometrically suppressed due to the hydrodynamic interaction 
 between the particle and the wall.
As shown in figure \ref{fig:snap_rev_rect}(e), 
 the left-top and right-bottom corners of the object 
 are largely deformed. 
From the subsequent results on
 the shape reversibility in figure \ref{fig:mp_imp_rect_rel}
 and the grid convergence behavior in figure \ref{fig:outlines_rev_rect},
 we strongly envisage that large deformations are resulted from the
 physical mechanism, not induced by numerical errors.
After the shearing force is released at $t=4$, 
 the deformed particle gradually recovers
 the unstressed shape
 as shown in figure \ref{fig:snap_rev_rect}(f) ($t=6$).

Figure \ref{fig:en_bud_rect} shows
 the budget of the kinetic energy transport.
Similar to the results in figure \ref{fig:en_bud}, 
 the numerical error is much smaller than the variation of the contributions
 of the individual terms in (\ref{eq:en_bud1}), 
 indicating that the energy exchange between the fluid and solid phases
 via the solid deformation is reasonably guaranteed.

Figure \ref{fig:mp_imp_rect_rel} shows
 the tracer distributions for four consecutive time instants.
After the particle well deforms at $t=4$, 
 the recovery of the material points
 toward the initial configuration is demonstrated. 
Even though the rotational motion of the particle is suppressed, 
 the object at fully developed state in figure \ref{fig:mp_imp_rect_rel}(d)
 slightly turns in the clockwise direction as compared with
 the initial distribution in figure \ref{fig:mp_imp_rect_rel}(a).
It would be remarkable to note that,
 though the left-top and right-bottom corners of the object
 are strongly stretched at $t=4$, 
 the object gradually resumes the shape of the original corners
 as the time goes on.

Figure \ref{fig:outlines_rev_rect} shows 
 the outlines of the fluid-structure interface
 for various number of grid points and different materials.
The edges of the rectangle are obviously smeared out, 
 which would be the inevitable effect of the numerical dissipation
 involved in the fifth-order WENO scheme,
 which is applied to updating $\phi_s$.
Nevertheless, similar to figure \ref{fig:outlines_rev_circ}, 
 with increasing the number of grid points, 
 the particle shapes at $t=4$ and $t=200$ converge, 
 and the reversibility in shape can be better attained. 

\section{Conclusion and perspectives}
\label{sec:conclusion}

A full Eulerian simulation method 
 for solving Fluid-Structure Interaction (FSI) problems 
 has been developed. 
A volume-of-fluid formulation \cite{hir1981}
 was applied to describing the multi-component geometry. 
The temporal change in the solid deformation 
 was described in the Eulerian frame 
 by updating a left Cauchy-Green deformation tensor, 
 which was used to express the nonlinear Mooney-Rivlin constitutive law.
The validity of the present simulation method 
 was established through comparisons with the analytical solution of the 
 the oscillatory response in fluid-solid parallel layers,
 and also with the available simulation data of
 the solid motion in the lid-driven cavity flow \cite{zha2008}
 and the two-particle interaction in the Couette flow \cite{gao2009}.
We confirmed that the present Eulerian approach
 can capture the reversibility in shape
 as long as the grid resolution is sufficiently high.
Further, we demonstrated that the numerical accuracy 
 due to the fluid-structure coupling is
 of the first-order with respect to the grid size.

The significance of the present full Eulerian simulation method
 may be that the approach showed a feasibility of reducing
 the FSI coupling problem to a simple incompressible fluid flow solvers.
Thus, the conventionally-used efficient computational techniques,
 such as the fast Fourier transform, and multi-grid method, 
 are applicable. 
The present Eulerian method is proved to be
 well-suited for using the voxel-based 
 multi-component geometry on the fixed Cartesian system. 
Once the initial field of the solid volume fraction
 is given over the entire domain,
 the present Eulerian method enables one to carry out 
 the FSI simulation without mesh generation procedure.
The method promises to extend the possibility of the FSI simulation 
 to certain additional classes of problems in the medical field,
 owing to a facility in incorporating the voxel data
 directly converted from medical images.
The practical demonstration is the future subject
 of the present authors.

To improve the accuracy in the present fluid-structure coupling
 to a level available for practical applications,
 it is important to capture the interface more sharply.
We now use the fifth-order WENO method
 for advecting the solid volume fraction field,
 which temporally makes the interface numerically diffusive.
As frequently used in the multiphase flow simulation, 
 to suppress the numerical diffusion, 
 elaborated techniques for the sharp interface advection such as 
 SLIC \cite{noh1976}, PLIC \cite{you1982,gue1999}, and THINC \cite{xia2005} methods
 would be applicable.
As an alternative of the VOF function, the level set function
 \cite{osh2003,set2003} is another option. 
On the dynamic interaction, 
 we now write the stress in a fluid-structure mixture form. 
Although the strain rate has a discontinuity across the fluid-structure interface, 
 it is smoothed out at the grid scale in the present simulation method. 
The ideas of the immersed interface treatment \cite{lev1994,li2006}
 and the localized strain formulation \cite{oka1995} 
 would be effective to improve the accuracy in the fluid-structure coupling. 
It is a challenging task to overcome the multiphysics difficulty
 particularly associated with the difference in constitutive laws
 for fluid and solid. 
Improved accuracy in capturing the interface and
 robust time advancement \cite{ii2009,ii2010}
 are the ongoing subject of the present authors.

\section*{Acknowledgments}
The authors thank Toshiaki Hisada, Robert M. Miura, Huaxiong Huang,
 Lucy T. Zhang, Yoichiro Mori, 
 Shigeho Noda, Teruo Matsuzawa, Hiroshi Okada,
 and Shigenobu Okazawa for fruitful discussions. 
K.S. is grateful to Tong Gao for providing
 the detailed simulation conditions in \S \ref{sec:gao}.
The authors thank the referees for their useful comments and suggestions.
This research was supported by
 Research and Development of the Next-Generation Integrated Simulation
 of Living Matter, a part of the Development
 and Use of the Next-Generation Supercomputer Project
 of the Ministry of Education, Culture, Sports, Science and Technology (MEXT),
 and by the Grant-in-Aid for Young Scientist (B) (No.21760120) of MEXT.




\appendix

\section{Finite difference descriptions}
\label{sec:fdm}

\subsection{For mass conservation equation (\ref{eq:cont2})
and derivatives of the incremental pressure involved
 in (\ref{eq:update_p})$-$(\ref{eq:pres_eq})}

Using the operators in (\ref{eqa:fdop}), we describe 
\begin{equation}
(\nabla\cdot{\bm v})_{i,j}=
\frac{\delta_i(v_x)|_{i,j}}{\Delta_x}+
\frac{\delta_j(v_y)|_{i,j}}{\Delta_y}.
\end{equation}
\begin{equation}
(\partial_x \varphi)_{i+\frac{1}{2},j}=
\frac{\delta_i(\varphi)|_{i+\frac{1}{2},j}}{\Delta_x},\ \ \ 
(\partial_y \varphi)_{i,j+\frac{1}{2}}=
\frac{\delta_j(\varphi)|_{i,j+\frac{1}{2}}}{\Delta_y},
\end{equation}
\begin{equation}
\begin{split}
&(\partial_x^2 \varphi)_{i,j}=
\frac{\delta_i(\varphi)|_{i+\frac{1}{2},j}-\delta_i(\varphi)|_{i-\frac{1}{2},j}
}{\Delta_x^2},
\ \ \ 
(\partial_y^2 \varphi)_{i,j}=
\frac{\delta_j(\varphi)|_{i,j+\frac{1}{2}}-\delta_j(\varphi)|_{i,j-\frac{1}{2}}
}{\Delta_y^2},
\\&
(\partial_x \partial_y \varphi)_{i+\frac{1}{2},j+\frac{1}{2}}=
\frac{\varphi_{i+1,j+1}-\varphi_{i,j+1}-\varphi_{i+1,j}+\varphi_{i,j}}{\Delta_x\Delta_y}.
\end{split}
\end{equation}

\subsection{For momentum conservation equation (\ref{eq:mom3})}

Here, we show the discretization for each term involved only
 in the $x$-momentum equation.
The permutations $i\leftrightarrow j$ and $x\leftrightarrow y$
 lead to the corresponding discretization in the $y$-momentum equation. 
For a quantity $q$, we here introduce interpolation operators denoted by overlines such as
\begin{equation}
\overline{q}^i|_{i,j}=\frac{q_{i+\frac{1}{2},j}+q_{i-\frac{1}{2},j}}{2},
\ \ \ 
\overline{q}^j|_{i,j}=\frac{q_{i,j+\frac{1}{2}}+q_{i,j-\frac{1}{2}}}{2}.
\end{equation}
The advection terms \cite{kaj1994}:
\begin{equation}
\begin{split}
(v_x\partial_x v_x)_{i+\frac{1}{2},j}=&
\frac{
\overline{v_x}^i|_{i,j}\delta_{i}(v_x)|_{i,j}+
\overline{v_x}^i|_{i+1,j}\delta_{i}(v_x)|_{i+1,j}
}{2\Delta_x},\\
(v_y\partial_y v_x)_{i+\frac{1}{2},j}=&
\frac{
\overline{v_y}^j|_{i+\frac{1}{2},j-\frac{1}{2}}\delta_{j}(v_x)|_{i+\frac{1}{2},j-\frac{1}{2}}+
\overline{v_y}^j|_{i+\frac{1}{2},j+\frac{1}{2}}\delta_{j}(v_x)|_{i+\frac{1}{2},j+\frac{1}{2}}
}{2\Delta_y}.
\end{split}
\end{equation}
The pressure gradient and the divergence of the deviatoric stress tensors:
\begin{equation}
\begin{split}
&(\partial_x \tilde{p})_{i+\frac{1}{2},j}=
\frac{\delta_i(\tilde{p})|_{i+\frac{1}{2},j}}{\Delta_x},
\\&
(\partial_x \tilde{\sigma}_{xx})_{i+\frac{1}{2},j}=
\frac{\delta_i(\tilde{\sigma}_{xx})|_{i+\frac{1}{2},j}}{\Delta_x},\ \ 
(\partial_y \tilde{\sigma}_{xy})_{i+\frac{1}{2},j}=
\frac{\delta_j(\tilde{\sigma}_{xy})|_{i+\frac{1}{2},j}}{\Delta_y},
\end{split}
\end{equation}
where  
$$
(\tilde{\sigma}_{xx})_{i,j}=
2(\mu_f+(\mu_s-\mu_f)\phi_{s,i,j})L_{xx,i,j}+
(\phi_s\tilde{\sigma}_{sh,xx})_{i,j},
$$
$$
(\tilde{\sigma}_{xy})_{i+\frac{1}{2},j+\frac{1}{2}}=
\bigl(\mu_f+(\mu_s-\mu_f)\overline{\overline{\phi_{s}}}|_{i+\frac{1}{2},j+\frac{1}{2}}\bigr)
(L_{xy,i+\frac{1}{2},j+\frac{1}{2}}
+L_{yx,i+\frac{1}{2},j+\frac{1}{2}})+
(\phi_s\tilde{\sigma}_{sh,xy})_{i+\frac{1}{2},j+\frac{1}{2}},
$$
\begin{equation*}
\begin{split}
(\phi_{s}\tilde{\sigma}_{sh,xx})_{i,j}
=&
\biggl\{
(2c_1-12c_3)\phi_{s,i,j}^{\frac{1}{2}}
+(2c_2+4c_3){\rm tr}(\tilde{\bm B})_{i,j}
-2c_2\tilde{B}_{xx,i,j}
\biggr\}\tilde{B}_{xx,i,j}
-2c_2\overline{\overline{\tilde{B}_{xy}^2}}\bigr|_{i,j},
\end{split}
\end{equation*}
\begin{equation*}
\begin{split}
(\phi_{s}\tilde{\sigma}_{sh,xy})_{i+\frac{1}{2},j+\frac{1}{2}}
=&
\biggl\{
(2c_1+2c_2-12c_3)\overline{\overline{\phi_{s}^{\frac{1}{2}}}}\bigr|_{i+\frac{1}{2},j+\frac{1}{2}}
+4c_3\overline{\overline{{\rm tr}(\tilde{\bm B})}}\bigr|_{i+\frac{1}{2},j+\frac{1}{2}}
\biggr\}\tilde{B}_{xy,i+\frac{1}{2},j+\frac{1}{2}}.
\end{split}
\end{equation*}
Considering $B_{zz}=1$, we write the trace of $\tilde{\bm B}$ as
$$
{\rm tr}(\tilde{\bm B})_{i,j}=
\tilde{B}_{xx,i,j}+\tilde{B}_{yy,i,j}+\phi_{s,i,j}^{\frac{1}{2}}.
$$

\subsection{For the advection terms in 
(\ref{eq:transphi}) and (\ref{eq:transb02})}
For a quantity $q$ (corresponding to $\phi_s$, $\tilde{B}_{xx}$, or $\tilde{B}_{yy}$)
 defined at the cell centroid $(i,j)$, 
 we apply the fifth-order WENO scheme \cite{liu1994, jia1996}
 to the advection terms in (\ref{eq:transphi}) and (\ref{eq:transb02}). 
The advection term $v_x\partial_x q$ is written as
\begin{equation}
\begin{split}
(v_{x}\partial_x q)_{i,j}^{\rm WENO}=&
\frac{1}{12\Delta_x}
\Biggl\{
\frac{(\overline{v}_{x}^i|_{i,j}+\bigl|\overline{v}_{x}^i|_{i,j}\bigr|)
(a_1^{(-)} g_1^{(-)} + a_2^{(-)} g_2^{(-)} + a_3^{(-)} g_3^{(-)})}
{a_1^{(-)}+a_2^{(-)}+a_3^{(-)}+\epsilon}
\\&+
\frac{(\overline{v}_{x}^i|_{i,j}-\bigl|\overline{v}_{x}^i|_{i,j}\bigr|)
(a_1^{(+)} g_1^{(+)} + a_2^{(+)} g_2^{(+)} + a_3^{(+)} g_3^{(+)})}
{a_1^{(+)}+a_2^{(+)}+a_3^{(+)}+\epsilon}
\Biggr\},
\end{split}
\end{equation}
where $\epsilon$ is a positive tiny number to avoid division by
zero, and 
$$
a_1^{(\pm)} =  (s_2^{(\pm)} s_3^{(\pm)})^2,\ \ 
a_2^{(\pm)} = 6(s_1^{(\pm)} s_3^{(\pm)})^2,\ \ 
a_3^{(\pm)} = 3(s_1^{(\pm)} s_2^{(\pm)})^2,
$$
\begin{equation*}
\begin{split}
g_1^{(\pm)} =& 2\delta_i(q)_{i \pm \frac{5}{2},j}-7\delta_i(q)_{i \pm \frac{3}{2},j}+11\delta_i(q)_{i \pm \frac{1}{2},j},\\
g_2^{(\pm)} =& -\delta_i(q)_{i \pm \frac{3}{2},j}+5\delta_i(q)_{i \pm \frac{1}{2},j}+ 2\delta_i(q)_{i \mp \frac{1}{2},j},\\
g_3^{(\pm)} =& 2\delta_i(q)_{i \pm \frac{1}{2},j}+5\delta_i(q)_{i \mp \frac{1}{2},j}-  \delta_i(q)_{i \mp \frac{3}{2},j},
\end{split}
\end{equation*}
\begin{equation*}
\begin{split}
  s_1^{(\pm)}=&
13 \{ \delta_i(q)_{i \pm \frac{5}{2},j}-2\delta_i(q)_{i \pm \frac{3}{2},j}+ \delta_i(q)_{i \pm \frac{1}{2},j}\}^2\\&+
 3 \{ \delta_i(q)_{i \pm \frac{5}{2},j}-4\delta_i(q)_{i \pm \frac{3}{2},j}+3\delta_i(q)_{i \pm \frac{1}{2},j}\}^2,\\s_2^{(\pm)}=&
13 \{ \delta_i(q)_{i \pm \frac{3}{2},j}-2\delta_i(q)_{i \pm \frac{1}{2},j}+ \delta_i(q)_{i \mp \frac{1}{2},j}\}^2\\&+
 3 \{ \delta_i(q)_{i \pm \frac{3}{2},j}                                   - \delta_i(q)_{i \mp \frac{1}{2},j}\}^2,\\s_3^{(\pm)}=&
13 \{ \delta_i(q)_{i \pm \frac{1}{2},j}-2\delta_i(q)_{i \mp \frac{1}{2},j}+ \delta_i(q)_{i \mp \frac{3}{2},j}\}^2\\&+
 3 \{3\delta_i(q)_{i \pm \frac{1}{2},j}-4\delta_i(q)_{i \mp \frac{1}{2},j}+ \delta_i(q)_{i \mp \frac{3}{2},j}\}^2,
\end{split}
\end{equation*}
Likewise, $(v_{y}\partial_y q)_{i,j}^{\rm WENO}$ is computed
using the interpolated advection velocity $\bar{v}_{y}^j|_{i,j}$. 
For $\tilde{B}_{xy}$ defined at the cell apex $(i+\frac{1}{2},j+\frac{1}{2})$,
 using the interpolated velocities
 $\bar{v}_{x}^j|_{i+\frac{1}{2},j+\frac{1}{2}}$ and $\bar{v}_{y}^i|_{i+\frac{1}{2},j+\frac{1}{2}}$,
 we evaluate ${\bm v}\cdot \nabla \tilde{B}_{xy}$ in a similar manner.

\section{Spectral algorithm to find sharp interface solution for the parallel layers problem}
\label{sec:spec}

We here explain the sharp interface approach to solve the one-dimensional fluid-structure
 coupling problem by means of (pseudo) spectral method. 
We obtain accurate solutions, which are used for validating the
 present full Eulerian model by comparisons. 

Due to the symmetry of the system with respect to $y=0$
 illustrated in figure \ref{fig:schem_layer},  
 we consider the upper half region $y\geq 0$ and
 write the fluid and solid velocities $v_f$, $v_s$
 in a Fourier series form
\begin{equation}
v_f(\tilde{y},t)=V_I(t)+\frac{\tilde{y}}{L_f}(V_W(t)-V_I(t))
+\sum_{k=1}^{\infty}v_{f,k}(t)\sin\frac{\pi k\tilde{y}}{L_f},
\label{eq:vf01}
\end{equation}
\begin{equation}
v_s(y,t)=\frac{V_I(t)y}{L_s}
+\sum_{k=1}^{\infty}v_{s,k}(t)\sin\frac{\pi k y}{L_s},
\label{eq:vs01}
\end{equation}
where $V_I$ is the velocity at the fluid-structure interface ($y=L_s$), 
$V_W$ is the given velocity of the upper wall, 
$v_{f,k}$ and $v_{s,k}$ are expansion coefficients, 
and $\tilde{y}=y-L_s$.
The expressions (\ref{eq:vf01}) and (\ref{eq:vs01})
satisfy the continuity of the velocity ($v_f=v_s$) at the interface $y=L_s$, 
the no-slip condition ($v_f=V_W$) on the upper wall $y=L_s+L_f$, 
and the symmetric condition ($v_s=0$) at $y=0$.
From (\ref{eq:vs01}), 
we readily find the solid displacement $u_s$ as
\begin{equation}
u_s(y,t)=\frac{U_I(t)y}{L_s}
+\sum_{k=1}^{\infty}u_{s,k}(t)\sin\frac{\pi k y}{L_s},
\label{eq:us01}
\end{equation}
where $U_I$ and $u_{s,k}$ yield
\begin{equation}
\frac{{\rm d}U_I}{{\rm d}t}=V_I,\ \ \ 
\frac{{\rm d}u_{s,k}}{{\rm d}t}=v_{s,k}.
\label{eq:us02}
\end{equation}
From the momentum equations (\ref{eq:mom}), 
with the stress expressions (\ref{eq:sigf}) and (\ref{eq:sigs_b01}),
we obtain
\begin{equation}
\frac{{\rm d}V_I}{{\rm d}t}
+\frac{\tilde{y}}{L_f}\left(
\frac{{\rm d}V_W}{{\rm d}t}-\frac{{\rm d}V_I}{{\rm d}t}
\right)
+\sum_{k=1}^{\infty}
\left\{
\frac{{\rm d}v_{f,k}}{{\rm d}t}
+\frac{\mu_f}{\rho}\left(\frac{\pi k}{L_f}\right)^2
v_{f,k}\right\}
\sin\frac{\pi k\tilde{y}}{L_f}=0,
\label{eq:vf02}
\end{equation}
\begin{equation}
\frac{y}{L_s}
\frac{{\rm d}V_I}{{\rm d}t}
+\sum_{k=1}^{\infty}
\left\{
\frac{{\rm d}^2u_{s,k}}{{\rm d}t^2}
+\frac{2(c_1+c_2)}{\rho}\left(\frac{\pi k}{L_s}\right)^2
u_{s,k}
+\frac{\pi k}{\rho L_s}
\sigma_{{\rm NL},k}
\right\}
\sin\frac{\pi k y}{L_s}=0,
\label{eq:vs02}
\end{equation}
where $\sigma_{{\rm NL}}$ denotes the nonlinear contribution in 
the solid stress with respect to the displacement. 
The definition of $\sigma_{{\rm NL}}$
 and the relation with the expansion coefficients $\sigma_{{\rm NL},k}$ are
\begin{equation}
\sigma_{{\rm NL}}\equiv
4c_3\left(
\frac{\partial u_s}{\partial y}
\right)^3
=\sum_{k=0}^\infty
\sigma_{{\rm NL},k}\cos\frac{\pi k y}{L_s}.
\label{eq:signl01}
\end{equation}
From the orthogonality in the sine function, 
 (\ref{eq:vf02}) and (\ref{eq:vs02}) are reduced to
 the modal relations
\begin{equation}
\frac{2}{\pi k}\left\{
\frac{{\rm d}V_I}{{\rm d}t}
-(-1)^k
\frac{{\rm d}V_W}{{\rm d}t}
\right\}
+
\frac{{\rm d}v_{f,k}}{{\rm d}t}
+\frac{\mu_f}{\rho}\left(\frac{\pi k}{L_f}\right)^2
v_{f,k}=0,
\label{eq:vf03}
\end{equation}
\begin{equation}
-\frac{2(-1)^k}{\pi k}
\frac{{\rm d}V_I}{{\rm d}t}
+
\frac{{\rm d}^2u_{s,k}}{{\rm d}t^2}
+\frac{2(c_1+c_2)}{\rho}\left(\frac{\pi k}{L_s}\right)^2
u_{s,k}
+\frac{\pi k}{\rho L_s}
\sigma_{{\rm NL},k}
=0,
\label{eq:vs03}
\end{equation}
for $1\leq k<\infty$.
The continuity of the shear stress at the interface $y=L_s$ is 
\begin{equation}
\begin{split}
&\frac{\mu_f(V_W-V_I)}{L_f}
-\frac{2(c_1+c_2)U_I}{L_s}-\sigma_{{\rm NL},0}
\\&
+\sum_{k=1}^{\infty}
\left[
\frac{\mu_f \pi k v_{f,k}}{L_f}
-
(-1)^k
\left\{
\frac{2(c_1+c_2)\pi k u_{s,k}}{L_s}
+\sigma_{{\rm NL},k}
\right\}\right]
=0.
\label{eq:vs04}
\end{split}
\end{equation}
The equation set to be solved consists of 
(\ref{eq:signl01})$-$(\ref{eq:vs04}).
In the numerical determination of the coefficients $v_{f,k}$, $u_{s,k}$, $V_I$ and $U_I$, 
 we truncate the number of the modes appeared
 in the infinite series summation of (\ref{eq:vs04}) up to $k=K-1$. 
If $K$ is chosen as an integer power of $2$, 
 the fast Fourier sine transform can be applied to efficiently evaluating 
 $v_f$ and $u_s$ respectively given in (\ref{eq:vf01}) and (\ref{eq:us01}),
 and the fast Fourier cosine transforms determine 
 the nonlinear part of the solid stress in a pseudo-spectral way
\begin{equation}
\sigma_{{\rm NL},k}\approx
\frac{8c_3}{N(1+\delta_{k0})}
\sum_{j=0}^{K-1}
\left\{
\frac{U_I}{L_s}
+\sum_{l=0}^{K-1}
\frac{\pi n u_{s,l}}{L_s}
\cos\frac{\pi l (j+\frac{1}{2})}{K}
\right\}^3
\cos\frac{\pi k(j+\frac{1}{2})}{K},
\end{equation}
where $\delta$ is the Kronecker delta.

In the case of $c_3=0$ (the linear Mooney-Rivlin material), 
 the system is linear since $\sigma_{{\rm NL}}$ vanishes. 
Considering the wall velocity is $V_W(t)={\rm Im}(\hat{V}_W\exp(i\omega t))$, 
 we may apply the separation of variable to the velocities and the displacement
\begin{equation}
\begin{split}
V_I(t)=&{\rm Im}(\hat{V}_I\exp(i\omega t)),\\
v_{f,k}(t)=&{\rm Im}(\hat{v}_{f,k}\exp(i\omega t)),\\
u_{s,k}(t)=&{\rm Im}(\hat{u}_{s,k}\exp(i\omega t)),
\end{split}
\end{equation}
which reduce the differential equations 
 (\ref{eq:vf03}) and (\ref{eq:vs03}) with respect to $t$
 into the algebraic ones. 
We readily find the expansion coefficients
\begin{equation}
\hat{v}_{f,k}
=\frac{\left\{(-1)^k\hat{V}_W-\hat{V}_I\right\}\alpha_k}{\pi k},
\end{equation}
\begin{equation}
\hat{u}_{s,k}
=\frac{i(-1)^{k}\hat{V}_I\beta_k}{\pi \omega k},
\end{equation}
\begin{equation}
\hat{V}_I=
\frac{\displaystyle
\frac{\mu_f\hat{V}_W}{L_f}
\left(1+\sum_{k=1}^{K-1}(-1)^k\alpha_k\right)
}{\displaystyle \frac{\mu_f}{L_f}
\left(1+\sum_{k=1}^{K-1}\alpha_k\right)
+\frac{2(c_1+c_2)}{i\omega L_s}
\left(1-\sum_{k=1}^{K-1}\beta_k\right)},
\end{equation}
where 
$$
\alpha_k=
\frac{2i\omega}{\displaystyle \frac{\mu_f\pi^2 k^2}{\rho L_f^2}+i\omega},
\ \ \ 
\beta_k=
\frac{2\omega^2}{\displaystyle \frac{2(c_1+c_2)\pi^2 k^2}{\rho L_s^2}-\omega^2}.
$$

In the case of $c_3\neq 0$ (e.g., the incompressible Saint Venant-Kirchhoff material),
 the system is nonlinear since $\sigma_{{\rm NL}}\neq 0$, 
 and thus the numerical time integration is needed. 
Here, it is carried out 
 using the second-order Adams-Bashforth and Crank-Nicolson schemes. 
We here put an superscript $(n)$ to a quantity
 to indicate the $n$-th time level ($t=n(\Delta t)$). 
If all the variables at the $n$-th and $(n-1)$-th time levels are known,
 together with the prescribed wall velocity
$$
V_W^{(n+1)}\equiv
V_W((n+1)(\Delta t))={\rm Im}(\hat{V}_W\exp(i\omega (n+1)(\Delta t))),
$$
 we update $U_I$, $v_{f,k}$, $u_{s,k}$, and $V_I$
 at the $(n+1)$-th time level:
\begin{equation}
U_I^{(n+1)}=U_I^{(n)}+\frac{(\Delta t)}{2}(V_I^{(n+1)}+V_I^{(n)}),
\end{equation}
\begin{equation}
v_{f,k}^{(n+1)}=
\frac{E_{vf,k}}{\displaystyle
1+\frac{(\Delta t)\mu_f\pi^2 k^2}{2\rho L_f^2}},
\end{equation}
\begin{equation}
\begin{split}
u_{s,k}^{(n+1)}=&
\frac{(-1)^k(\Delta t)(V_I^{(n+1)}-V_I^{(n-1)})}{\pi k}
+2u_{s,k}^{(n)}-u_{s,k}^{(n-1)}
\\&
-(\Delta t)^2\left(
\frac{2(c_1+c_2)\pi^2 k^2}{\rho L_s^2}u_{s,k}^{(n)}
+\frac{\pi k}{\rho L_s}\sigma_{{\rm NL},k}^{(n)}
\right),
\end{split}
\end{equation}
\begin{equation}
V_I^{(n+1)}=\frac{E_{VI}}{\displaystyle
\frac{\mu_f}{L_f}+\left(N-\frac{1}{2}\right)
\frac{(c_1+c_2)(\Delta t)}{L_s}
+\sum_{k=0}^{K-1}
\frac{4\rho \mu_f L_f}{2\rho L_f^2+(\Delta t)\mu_f\pi^2 k^2}
},
\end{equation}
where 
\begin{equation*}
\begin{split}
E_{vf,k}=&
\left(1-\frac{(\Delta t)\mu_f\pi^2 k^2}{2\rho L_f^2}\right)v_{f,k}^{(n)}
-\frac{2\{V_{I}^{(n+1)}-V_{I}^{(n)}-(-1)^k\delta V_W\}}{\pi k},
\end{split}
\end{equation*}
\begin{equation*}
\begin{split}
&E_{VI}=
\frac{\mu_f V_W^{(n+1)}}{L_f}
-\frac{(c_1+c_2)\{2U_I^{(n)}+(\Delta t)V_I^{(n)}\}}{L_s}
-2\sigma_{{\rm NL},0}^{(n)}+\sigma_{{\rm NL},0}^{(n-1)}
\\&
+\sum_{k=1}^{K-1}\Biggl[
\frac{\mu_f}{L_f}\left\{
\frac{\pi k\{2\rho L_f^2-(\Delta t)\mu_f\pi^2 n^2\}v_{f,n}^{N}
+4\rho L_f^2\{V_I^{(n)}+(-1)^k\delta V_W\}
}{2\rho L_f^2+(\Delta t)\mu_f\pi^2 n^2}
\right\}
\\&+\frac{2(c_1+c_2)}{L_s}\Biggl\{
(\Delta t)V_I^{(n-1)}+(-1)^k\pi k\left(
\gamma_k u_{s,k}^{(n)}
+u_{s,k}^{(n-1)}\right)\Biggr\}
\\&
+(-1)^k\left(
\gamma_k \sigma_{{\rm NL},k}^{(n)}+\sigma_{{\rm NL},k}^{(n-1)}
\right)
\Biggr],
\end{split}
\end{equation*}
$$
\delta V_W=V_W^{(n+1)}-V_W^{(n)},\ \ \ 
\gamma_k=\frac{2(c_1+c_2)(\Delta t)^2\pi^2 k^2}{\rho L_s^2}-2.
$$
After a sufficiently long computation, we obtain temporally periodic solutions. 

We checked the convergence of the solution as a function of the truncated mode $K$.
We confirmed that within the parameter range shown in figure \ref{fig:comp_tau},
 the results with $K=2048$ are accurate enough to be
 regarded as the reference solutions for comparison.


\end{document}